\documentclass[12pt]{article}

\usepackage{subfiles}
\usepackage{placeins}
\usepackage{amsthm}
\usepackage{amsfonts}
\usepackage{amssymb}
\usepackage{amsmath}
\usepackage{graphicx}
\usepackage{subfig}
\usepackage{float}
\usepackage{lipsum}
\usepackage{listings}
\usepackage{titling}
\usepackage[bookmarks=true]{hyperref}
\usepackage{mathtools}
\usepackage{xr}
\usepackage{color,soul}

\usepackage[
giveninits=true, backend=biber,maxcitenames=2,doi=false,isbn=false,url=false,eprint=false]{biblatex}
\renewbibmacro{in:}{} 

\renewbibmacro*{volume+number+eid}{
    \printfield{volume}
    \setunit*{\addnbspace}
    \printfield{number}
    \setunit{\addcomma\space}
    \printfield{eid}}
\DeclareFieldFormat[article]{number}{\mkbibparens{#1}}
\AtEveryBibitem{\clearfield{month}}
\AtEveryCitekey{\clearfield{month}}
\renewcommand{\theequation}{
\arabic{equation}%
}

\usepackage{color}
\usepackage{subfig}
\usepackage[]{mcode}
\usepackage{tikz}
\usepackage{parskip}
\usepackage{longtable}
\usepackage{lscape}
\usepackage[]{hyphenat}
\usepackage[toc,page]{appendix}
\usepackage{float, lscape}
\usepackage{array,multirow}
\usepackage{booktabs}
\usepackage{rotating}
\usepackage{afterpage}
\usepackage{tabto}
\usepackage{enumitem}
\usepackage{mathrsfs}
\usepackage{titlesec, blindtext, color}
\usepackage[ruled,linesnumbered]{algorithm2e}

\newtheorem{proposition}{Proposition}
\newtheorem{theorem}{Theorem}
\newtheorem{remark}{Remark}
\newtheorem{corollary}{Corollary}
\renewcommand{\theequation}{
\arabic{equation}%
}

\usepackage{setspace}

\def\spacingset#1{\renewcommand{\baselinestretch}%
{#1}\small\normalsize} \spacingset{1}

\usepackage{xcolor}
\hypersetup{
    colorlinks,
    linkcolor={black},
    citecolor={black},
    urlcolor={black}
}

\newcommand{\blind}{1}

\begin{document}
\spacingset{1.2}
\begin{refsection}

\if1\blind
{
  \title{\bf Large Scale Partial Correlation Screening \\ with Uncertainty Quantification
    \vspace{0.2in}}
    \date{}
  \author{Emily Neo, Peter Radchenko, Bala Rajaratnam \hspace{.2cm}\\
    University of Sydney}
  \maketitle
} \fi

\if0\blind
{
  \bigskip
  \bigskip
  \bigskip
  \begin{center}
    {\LARGE\bf Large Scale Partial Correlation Screening with Uncertainty Quantification}
\end{center}
  \medskip
} \fi

\bigskip



\pagestyle{plain}

\begin{abstract}

Identifying multivariate dependencies in high-dimensional data is an important problem in large-scale inference.  This problem has motivated recent advances in mining (partial) correlations, which focus on the challenging ultra-high dimensional setting where the sample size, $n$, is fixed, while the number of features, $p$, grows without bound.  The state-of-the-art method for partial correlation screening can lead to undesirable results.
This paper introduces a novel principled framework for partial correlation screening with error control (PARSEC), which leverages the connection between partial correlations and regression coefficients.  We establish the inferential properties of PARSEC when~$n$ is fixed and~$p$ grows super-exponentially.  First, we provide “fixed-$n$-large-$p$” asymptotic expressions for the familywise error rate (FWER) and $k$-FWER.  Equally importantly, our analysis leads to a novel discovery which permits the calculation of exact marginal p-values for controlling the false discovery rate (FDR), and also the positive FDR (pFDR). To our knowledge, no other competing approach in the “fixed-$n$-large-$p$” setting allows for error control across the spectrum of multiple hypothesis testing metrics. We establish the computational complexity of PARSEC and rigorously demonstrate its scalability to the large $p$ setting.  The theory and methods are successfully validated on simulated and real data, and PARSEC is shown to outperform the current state-of-the-art.

\end{abstract}
\vspace{0.2in}

\spacingset{1.2}

\section{Introduction}

Modern massive datasets have emerged out of our improved ability to collect, store and analyze data.  Extracting and identifying complex associations and dependencies in such data, while addressing the unique challenges inherent to the increasing number of features, is a critically important task in statistical inference and machine learning.  In the modern ultra-high dimensional setting where the sample size, $n$, is fixed and the number of features, $p$, tends to infinity, spurious associations will be wrongly identified due to random chance alone.  Indeed, it is well understood that an alarming proportion of scientific results reported in leading journals are not reproducible \parencite{Ioannidis:2005}. Hence, there is a compelling need for a principled framework to identify the strongest and most statistically significant associations in the ultra-high dimensional setting, while providing theoretical safeguards against identifying spurious associations.  The resulting significant associations can subsequently be represented as a correlation or partial correlation graph.

For a given set of features, sample correlation coefficients are a basic measure of bivariate or marginal (linear) dependency.  Alternatively, partial correlation coefficients provide a multivariate measure of dependency, and have found widespread use.  Notably, in the Gaussian setting, zero partial correlation coefficients imply conditional independence.  In this setting, the partial correlation matrix is typically estimated using the inverse of the sample covariance matrix \parencite{Dempster:1972,Anderson:2003}. However, this approach is problematic in high dimensions, as the inverse is not well-defined when $p>n$.

A number of methods for estimating sparse partial correlation matrices have been proposed for the high-dimensional asymptotic regime where both $p$ and $n$ tend to infinity.  The corresponding approaches are often based on the $\ell_1$-penalized likelihood framework, which adds an $\ell_1$ penalty to either the Gaussian likelihood or some pseudo-likelihood - see, for example, \textcite{BanerjeeEtAl:2008}, \textcite{KhareOhRajaratnam:2015}, and the references therein. These $\ell_1$-based approaches are appropriate when both~$n$ and~$p$ are large and provide a framework for model selection by shrinking small partial correlations to zero.
However, the focus of $\ell_1$ approaches is on variable selection rather than hypothesis testing.
Moreover, they also require the determination of the tuning/penalty parameter.  Such $\ell_1$-based approaches are also computationally expensive in large~$p$ data regimes due to the predominant use of iterative optimization algorithms.  Furthermore, theoretical safeguards for the $\ell_1$-penalized methods are established in the setting when both~$n$ and $p$ go to infinity.  For these reasons, the $\ell_1$-penalized framework is not always amenable to the modern ultra-high dimensional setting where $n$ is fixed and $p$ goes to infinity - from both a statistical and a computational perspective.

Limited order partial correlation methods provide an alternative approach for estimating partial correlation matrices.  These methods first employ preliminary statistical testing on bivariate marginal correlations or ``$q$-order" partial correlations (where $q<p-2$) to reduce dimensionality of the conditioning set -- see, for example, the useful works of \textcite{MagweneKim:2004}, \textcite{CasteloRoverato:2009} and references therein. \textcite{LiangEtAl:2015} notes that these methods do not evaluate full-order partial correlation coefficients, and as such they may result in estimates that are closer to marginal correlations.  We note that the pre-processing step of other methods seem to also have the disadvantage that it does not readily facilitate \textit{statistical inference}.
 \textcite{HeroRajaratnam:2012} developed an ultra-high dimensional approach with uncertainty quantification for the purpose of screening variables which are highly partially correlated with many others.  Their approach is based on quantifying the distribution of the number of exceedances of the generalized inverse of the sample correlation matrix in the ultra-high dimensional setting where $n$ is fixed and $p\rightarrow\infty$.
While the emphasis of their approach is on screening highly connected variables (and not parameters), a natural adaptation of their approach, which we refer to as PCS-Hub, can be used for screening edges in the corresponding partial correlation graph. However, our empirical analysis demonstrates that PCS-Hub can be deficient in ultra-high dimensional settings.
To our knowledge, this important problem has remained largely unsolved for more than 10 years.

In this paper, we introduce a principled framework for PARtial correlation Screening with Error Control (PARSEC), which moves beyond the model selection framework at the center of most existing high-dimensional partial correlation learning methods.
PARSEC provides stable estimates of partial correlation coefficients by breaking down the partial correlation estimation problem into a series of simpler regression problems.  PARSEC is inferentially sound, easy to use, computationally efficient and thus can be immediately deployed.

In Section~\ref{sec:methodology} we introduce the specifics of the PARSEC method, and in Section~\ref{sec:theory} we establish theoretical properties of PARSEC and use these for inference in the ultra-high dimensional setting, thus going beyond model selection.
In Section \ref{sec:Algorithms} we analyze the computational complexity of our approach, demonstrating its scalability.  Section~\ref{sec:simulated_data} provides extensive simulations to illustrate PARSEC's computational and inferential performance.  Lastly, the efficacy of PARSEC is demonstrated in Section \ref{sec:real_applications} on real applications, including breast cancer gene screening and a portfolio selection problem.



\section{Methodology} \label{sec:methodology}

In this section, we introduce PARSEC, a novel scalable method for partial correlation screening with uncertainty quantification in ultra-high dimensional data regimes.  Section \ref{sec:methods_preliminaries} provides preliminary information and notation, and in Section~\ref{sec:parsec_method} we motivate and describe the PARSEC method in detail.

\subsection{Preliminaries} \label{sec:methods_preliminaries}

We let $\mathbf{X} = [X_1,...,X_p]^\top \in \mathbb{R}^p$ be a vector of random variables with mean $\boldsymbol{\mu}$, covariance~$\mathbf{\Sigma}_{p \times p}$, and let $\mathbf{\Omega}=\mathbf{\Sigma}^{-1}$.  We define the corresponding (marginal) correlations as $\rho_{ij}=\sigma_{ij}/\sqrt{\sigma_{ii}\sigma_{jj}}$, and partial correlations as $\rho^{ij}=-\omega_{ij}/\sqrt{\omega_{ii}\omega_{jj}}$.
Assuming a sample size of~$n$, let $\mathbb{X}$ denote the corresponding $n \times p$ data matrix, where column vector~$\mathbf{X}_j$ and row vector~$\mathbf{X}_{(i)}$
represent the $j$-th column and the $i$-row, respectively:
$\mathbb{X} = [\mathbf{X}_1,...,\mathbf{X}_p]= [\mathbf{X}^\top_{(1)},..., \mathbf{X}^\top_{(p)}]^\top$.
We define the sample mean of the $j$-th column as $\bar{X}_j = n^{-1} \sum^n_{i=1} X_{ij}$, write $\bar{\mathbf{X}} = [\bar{X}_1,...,\bar{X}_p]$ for the vector of sample means, and define the sample covariance matrix as $\mathbf{S} = \sum^n_{i=1} (\mathbf{X}_{(i)} - \bar{\mathbf{X}})^\top (\mathbf{X}_{(i)} - \bar{\mathbf{X}})/(n-1)$.  Now let $\mathbf{1}$ denote a column vector consisting of $1$s. We calculate multivariate Z-scores by standardizing the columns of the data matrix: $\mathbf{Z}_j = (\mathbf{X}_j - \bar{\mathbf{X}}_j \textbf{1})/\sqrt{\mathbf{S}_{jj} (n-1)}$, which permits the following representation of the sample correlation matrix: $\mathbf{R} = \mathbb{Z}^\top\mathbb{Z}$.  We say that a random matrix $\mathbb{X}\in \mathbb{R}^{n \times p}$ has a \emph{vector-elliptical} distribution with location parameter $\boldsymbol{\mu} \in \mathbb{R}^p$ and positive definite covariance parameter $\mathbf{\Sigma} \in \mathbb{R}^{p\times p}$ if its density can be expressed as follows \parencite{Anderson:2003}:


\begin{equation}
f_{\mathbb{X}}(\mathbb{X}) =
\det(\mathbf{\Sigma})^{-n/2}g(\text{tr}((\mathbb{X}-\boldsymbol{\mu}\mathbf{1}^\top)\mathbf{\Sigma}^{-1}(\mathbb{X}-\boldsymbol{\mu}\mathbf{1}^\top)^\top)),
\label{eqn:hello}
\end{equation}


\noindent where the shape function $g: \mathbb{R}\to [0,\infty)$ is such that $\int f_{\mathbb{X}}(\mathbb{X}) \; d\mathbb{X} =1$.

\begin{figure}[t!]
\centering
  \includegraphics[width=50mm]{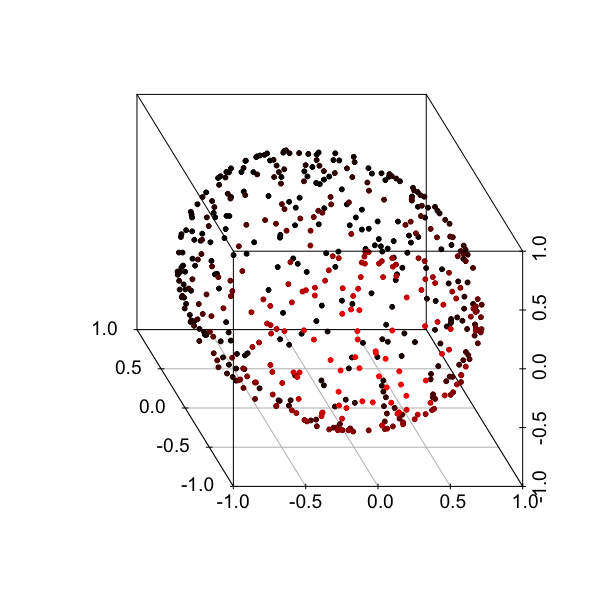}
  \includegraphics[width=50mm]{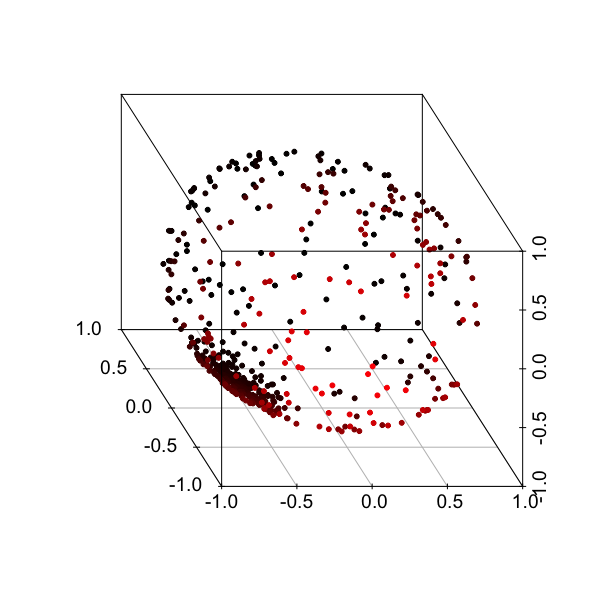}
    \includegraphics[width=50mm]{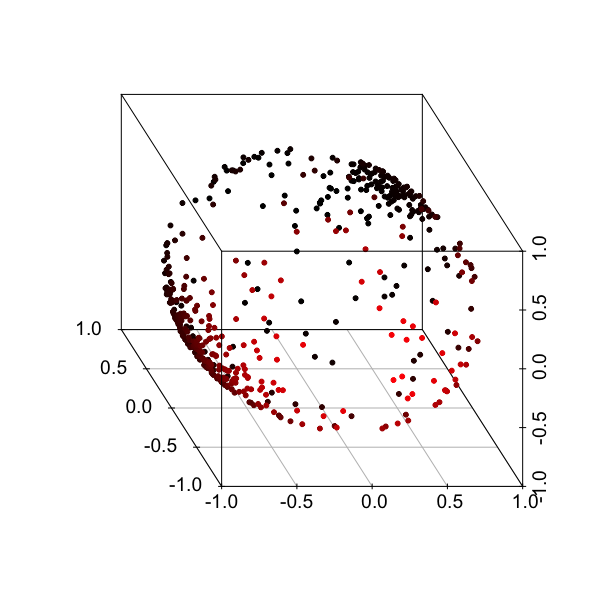}
\caption{In the Gaussian setting when $p=500$ and $n=4$: U-scores associated with a diagonal covariance matrix (left); block-diagonal covariance with one block of size 200 (center); block-diagonal covariance with two blocks of size 200 and 100 (right).  In the diagonal covariance setting, the U-scores are uniformly distributed on the unit sphere.  However, in the block-diagonal settings, clusters of U-scores are evident. \label{fig:U_score_demo}}
\end{figure}

In our theoretical analysis of partial correlation screening in ultra-high dimensions, we will exploit the so-called U-score representation: $\mathbf{R} = \mathbb{Z}^\top\mathbb{Z} = \mathbb{U}^\top\mathbb{U}$. Here, $\mathbb{U}$ is an $n-1$ by~$p$ matrix whose $j$-th column corresponds to the $j$-th feature.  In particular, each feature is represented by a U-score, and all U-scores lie on the unit sphere, $S_{n-2}$, providing a lower-dimensional geometric representation of high-dimensional data and the dependencies therein.  To achieve mean centering, U-scores project away components of~$\mathbf{X}_j$ that are orthogonal to the $n-1$ dimensional hyperplane $\{\mathbf{u}\in\mathbb{R}^n: \mathbf{1}^\top\mathbf{u}=0\}$.  This mean centering changes the effective sample size from $n$ to $n-1$.
Furthermore, U-scores that are close to each other correspond to highly positively correlated features. Importantly, when the distribution of~$\mathbb{X}$ is vector-elliptical and $\mathbf{\Sigma}$ is diagonal, the U-scores are uniformly distributed on the unit sphere. A deviation from uniformity in the distribution of the U-scores reveals the existence of dependency. U-scores have the desirable property of preserving correlations between features (as they are constructed through normalization and centering using an orthogonal matrix as described below).
More formally,
let ~$\mathbb{T}_{n\times n}$ be an orthogonal matrix of the form $[n^{-1/2}\textbf{1}, \mathbb{T}_{2:n}]$.  Here, $\mathbb{T}_{2:n}$ can be any orthogonal matrix whose columns are orthogonal to~$\textbf{1}$; in practice, we construct~$\mathbb{T}_{2:n}$ using Gramm-Schmidt orthogonalization. The U-score matrix, $\mathbb{U}_{(n-1)\times p}=[\mathbf{U}_1,...,\mathbf{U}_p]$, can then be obtained from the relation $\mathbb{U} = \mathbb{T}^\top_{2:n} \mathbb{Z}$. The sample correlation between~$\mathbf{X}_j$ and~$\mathbf{X}_k$, $r_{jk}$, thus has a simple relationship with the Euclidean distance between the corresponding U-scores:
\begin{equation} \label{eqn:ls_corr}
r_{jk} = \mathbf{U}_j^\top \mathbf{U}_k = 1 - || \mathbf{U}_j - \mathbf{U}_k ||_2^2/2
\end{equation}
\noindent Figure~\ref{fig:U_score_demo} illustrates the distribution of U-scores for different covariance structures when $n=4$ and $p=500$.  In the diagonal covariance setting, U-scores are uniformly distributed on the unit sphere.  In contrast, dependent U-scores corresponding to block-diagonal covariance settings notably coalesce into clusters.

\textcite{HeroRajaratnam:2011} leveraged the U-score representation above to understand the asymptotic behavior of sample correlation coefficients in sample-deficient settings and screen for significant correlations.
They established that in the ultra-high dimensional asymptotic regime where $n$ is fixed and $p$ goes to infinity, the number of false discoveries  
is approximated by a Poisson distribution.
This approximation holds for a wide class of vector-elliptical distributions and permits the derivation of the critical screening level~$\rho$ that can control the family-wise error rate (FWER) at a pre-specified significance level.  \textcite{HeroRajaratnam:2012} extended the correlation screening framework for discovering highly connected vertices in partial correlation networks (which we term as ``PCS-Hub").  They estimated the partial correlation matrix using the generalized inverse of the sample correlation matrix, $\mathbf{R}^\dagger$, by leveraging the following representation:
\begin{equation}
\mathbf{R}^\dagger= \mathbb{U}^\top [\mathbb{U}\mathbb{U}^\top]^{-2} \mathbb{U}.
\label{eqn:R_Dagger}
\end{equation}
\noindent Their resulting estimate of the partial correlation matrix is the standardized quantity {$\mathbf{P}=\mathbf{D}_{\mathbf{R}^\dagger}^{-1/2}\mathbf{R}^\dagger\mathbf{D}_{\mathbf{R}^\dagger}^{-1/2}$}, where $\mathbf{D}_{\mathbf{A}}$ denotes the diagonal matrix corresponding to the matrix~$\mathbf{A}$.
Equivalently,
\begin{equation}
    \mathbf{P}=\mathbb{Y}^\top\mathbb{Y}, \qquad\text{where}\quad \mathbb{Y}_{(n-1)\times p}=[\mathbb{U}\mathbb{U}^\top]^{-1}\mathbb{U}\mathbf{D}^{-1/2}_{\mathbb{U}^\top[\mathbb{U}\mathbb{U}^\top]^{-2}\mathbb{U}}\;,
\label{eqn:partial.z.scores}
\end{equation}
\noindent where the Y-scores provide an analog of the U-score representation for the partial correlation setting. The columns of~$\mathbb{Y}$ are the so-called partial correlation Z-scores, which also lie on the unit sphere~$\mathbf{S}_{n-2}$. This representation has been used to perform partial correlation screening with statistical error control - see \textcite{HeroRajaratnam:2012}.
Despite these advances, the resulting PCS-Hub approach has some shortcomings: i) it can produce partial correlation estimates that are similar to marginal correlation estimates (see Figure \ref{fig:U_Y_V_score_demo}), and ii) it can fail to pick up a collection of strong signals (see Figure \ref{fig:distr_parsec_mpi}).  Moreover, it does not also readily specify the distribution of partial correlations in the fixed $n$, fixed $p$ setting.
Hence, it is not clear how to use PCS-Hub to obtain exact marginal p-values for partial correlations, thus preventing a meaningful False Discovery Rate (FDR) and/or positive FDR (pFDR) analysis.  These fundamental shortcomings have remained unresolved for the better part of the last 10 to 15 years, and to
address these, we introduce the PARSEC method below.


\subsection{The PARSEC method} \label{sec:parsec_method}

The primary goal of the PARSEC method is to extract and ``screen" high partial correlations in the noisy ultra-high dimensional setting.  More specifically, PARSEC constructs stable estimates of scaled partial correlation coefficients, and
quantifies the distribution of the resulting partial correlation estimates and the distribution of the number of exceedances in order to achieve a desired level of error control for a broad spectrum of error metrics (FWER, k-FWER, FDR, pFDR) - even in the challenging modern ultra-high dimensional setting when the sample size $n$ is fixed and the dimension $p$ tends to infinity.

As mentioned in the Introduction, the distinct advantage of the PCS-Hub method \parencite{HeroRajaratnam:2012} is that it quantifies the behavior of the distribution of partial correlations in the ultra-high dimensional regime (unlike $\ell_1$-based methods), thus enabling uncertainty quantification even in this challenging setting.
We show below that despite this important advantage, PCS-Hub can however be highly deficient in the modern fixed-$n$-large-$p$ setting.
Despite this large gap in the literature, to our knowledge, methods which provide more stable partial correlation estimates in the fixed-$n$-large-$p$ setting, while achieving rigorous statistical error control, have not been proposed.  This begs the question of whether there are alternative methods to stably estimate partial correlation coefficients in the ultra-high dimensional setting, which are still sufficiently tractable so that the distribution of partial correlation estimates can be precisely quantified.

In this paper, we note that one strategic and deliberate approach to circumventing the direct estimation of the inverse correlation matrix is to recognize and exploit the important relationship between partial correlation coefficients and regression coefficients. Recall that regression coefficients can be recast as scaled partial correlation coefficients. This connection at the population level provides an alternative approach to partial correlation estimation (at the sample level).
We show in this paper that decoupling the partial correlation estimation problem into a series of regressions adequately addresses the deficiencies of directly estimating the inverse correlation matrix. Simultaneously, it uniquely allows for the precise quantification of the distribution of the partial correlation estimates in the highly challenging fixed-$n$-large-$p$ setting - resulting in a novel and principled framework for partial correlation screening with uncertainty quantification.

We now describe the proposed approach: the PARSEC method estimates matrix~$\mathbf{H}$ of scaled partial correlations row by row, regressing~$\mathbf{X}_j$ on the rest of the features for $j=1,...,p$. In fact, PARSEC performs this regression at the level of the U-scores instead of the original features,~$\mathbf{X}_j$.
More specifically, we treat the $j$-th U-score,~$\mathbf{U}_j$, as the response variable, and the rest of the U-scores as predictors for $j=1,...,p$.  Leveraging the U-score representation in~(\ref{eqn:R_Dagger}) and denoting by~$\mathbb{U}^{-j}$ the U-score matrix with the $j$-th column excluded, we derive the following expression for the vector of regression coefficient estimates:
$((\mathbb{U}^{-j})^\top \mathbb{U}^{-j})^{\dagger} (\mathbb{U}^{-j})^\top \mathbf{U}_j$, which is the analogue of the classical regression formula $(\mathbb{X}^\top \mathbb{X})^{-1}\mathbb{X}^\top \mathbf{Y}$ at the level of U-scores.
Our partial correlation estimates are then obtained by rescaling the estimated regression coefficients.
More formally, we (a) compute  the matrix of partial correlation Z-scores (cf.~equation~\ref{eqn:partial.z.scores}) that excludes the $j$-th feature:
\begin{equation*}
\tilde{\mathbb{U}}^{-j}:=(\mathbb{U}^{-j}(\mathbb{U}^{-j})^\top)^{-1} \mathbb{U}^{-j} \mathbf{D}^{-\frac{1}{2}}_{ {\mathbb{U}^{-j}}^\top[ {\mathbb{U}^{-j}} ({\mathbb{U}^{-j}})^\top]^{-2} {\mathbb{U}^{-j}}};
\end{equation*}
\noindent and (b) define the $j$-th row of the PARSEC $\mathbb{H}$ matrix as follows:
\begin{equation} \label{eqn:direct_eval_H}
\big(H_{j1},\dots,H_{j(j-1)},H_{j(j+1)},\dots,H_{jp}\big)^\top := (\tilde{\mathbb{U}}^{-j})^\top \mathbf{U}_j.
\end{equation}
\noindent Observe that equation~(\ref{eqn:direct_eval_H}) is a vector containing inner products and parallels the correlation formula in~(\ref{eqn:ls_corr}), $r_{jk} = \mathbf{U}_j^\top \mathbf{U}_k$, and the PCS-Hub-based partial correlation formula in~(\ref{eqn:partial.z.scores}), $p_{jk} = \mathbf{Y}_j^\top \mathbf{Y}_k$.
To construct PARSEC's scaled partial correlation matrix~$\mathbf{H}$, we implement~(\ref{eqn:direct_eval_H}) for $j=1,...,p$. We note that matrix $\mathbb{U}^{-j}(\mathbb{U}^{-j})^\top$, which is inverted in the definition of~$\tilde{\mathbb{U}}^{-j}$, is $(n-1)\times(n-1)$ and hence low-dimensional in our setting of interest. Thus, the PARSEC approach is readily scalable and is ideally suited to the sample-starved regime when $p  \gg  n$.

The second component of the PARSEC framework provides a screening method which provides rigorous error control for detecting high partial correlations.  Given a screening level~$\rho$, PARSEC deems an estimate~$H_{jk}$ of the scaled partial correlation between features~$X_j$ and~$X_k$ a (screening) discovery if $|H_{jk}|\ge\rho$. This screening level, $\rho$, is obtained in a principled manner, using PARSEC's inferential properties.  Specifically, in Section \ref{sec:theory}, we derive a Poisson approximation for the number of discoveries and also expressions for the mean number of discoveries in the ultra-high dimensional regime when $n$ is fixed and the dimension $p$ goes to infinity.  These expressions allow for the specification of various multiple hypothesis testing measures for statistical error control, such as FWER, k-FWER \parencite{RomanoWolf:2005}, FDR and pFDR as a function of the screening level, $\rho$. These can then be used to identify the specific screening level, $\rho$, which leads to the desired level of error control.

\begin{figure}[t]
\centering
\begin{tabular}{ccc}
(a) U-scores, $p$=5000 & (b) Y-scores, $p$=5000 &  (c) PARSEC scores, $p$=5000 \\[6pt]
\includegraphics[width=42mm]{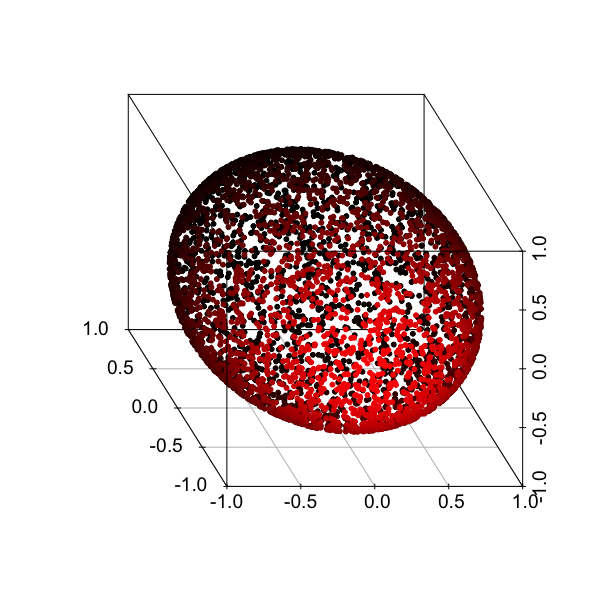}  &
\includegraphics[width=42mm]{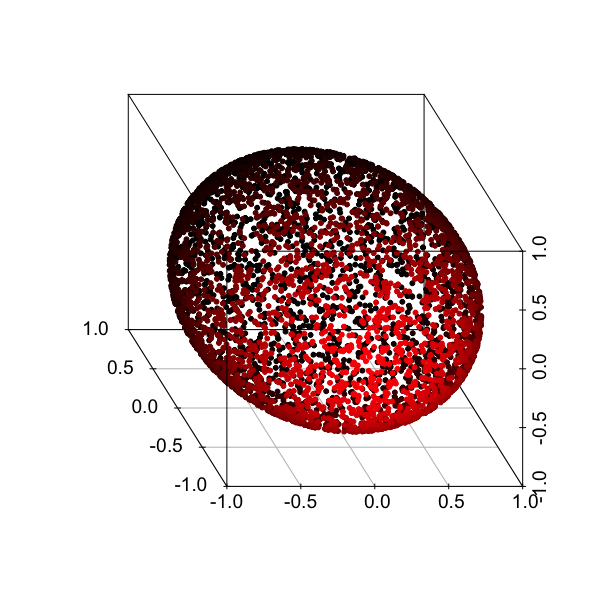} &
\includegraphics[width=42mm]{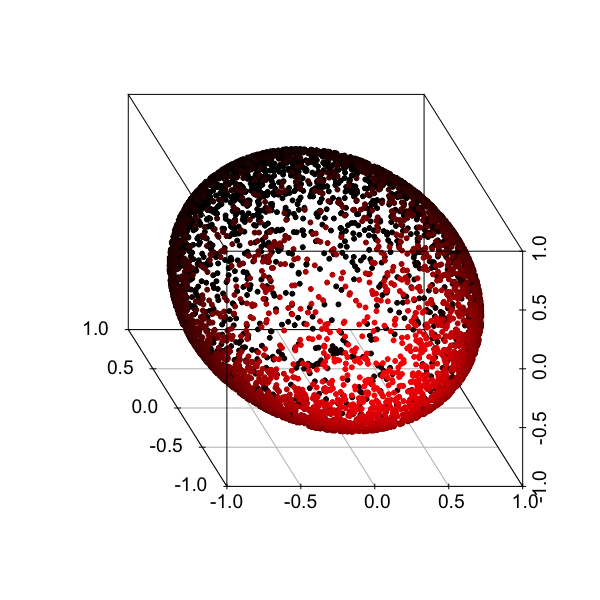} \\
(d) U-scores, $p$=10,000 & (e) Y-scores, $p$=10,000 &  (f) PARSEC scores, $p$=10,000 \\[6pt]
\includegraphics[width=40mm]{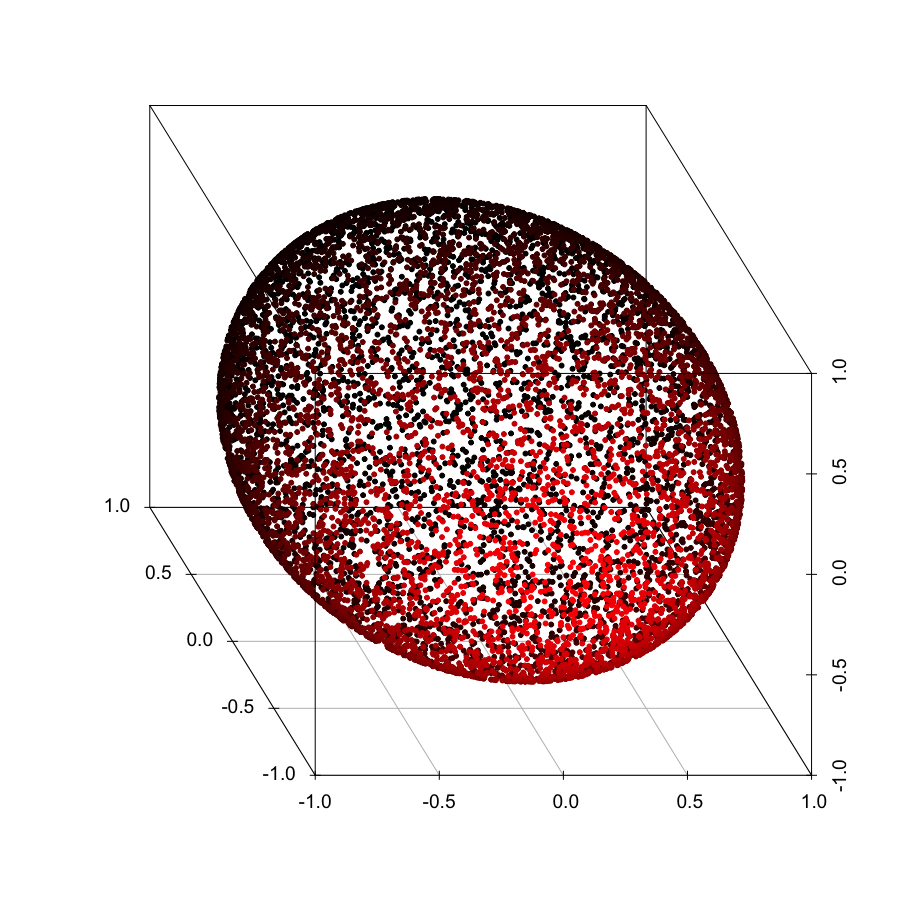}  &
\includegraphics[width=40mm]{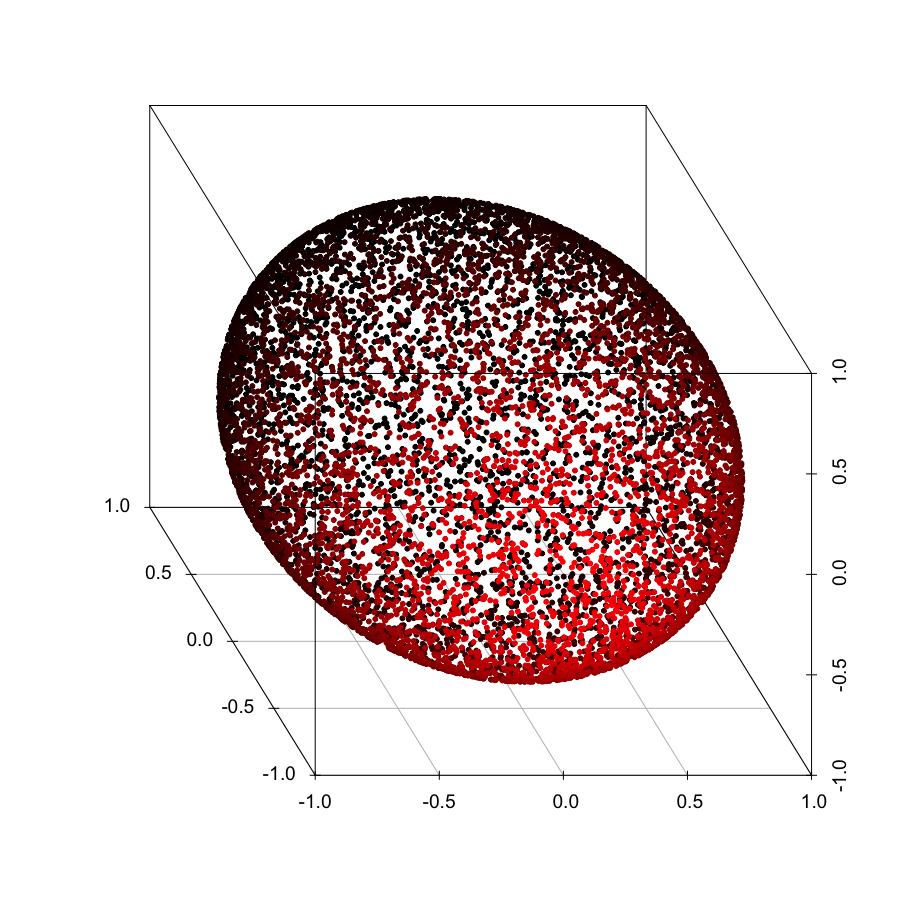} &
\includegraphics[width=40mm]{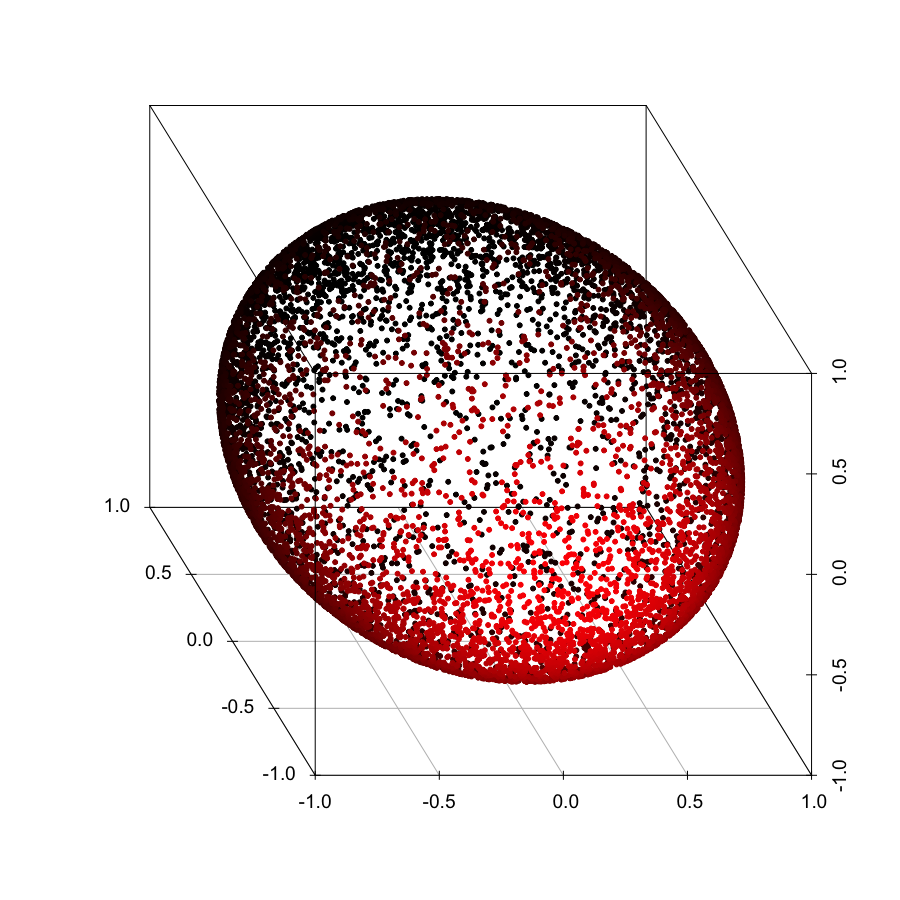} \\
\end{tabular}
\caption{Geometric comparison of Z-type scores used in ultra-high dimensional screening methods for an AR(10) block structure with Gaussian features, block size $=50$,
$n=4$ and varying~$p$.  U-scores are used in large-scale marginal correlation screening and Y-scores are used in partial correlation screening (PCS-Hub).  \label{fig:U_Y_V_score_demo}}
\end{figure}

Through empirical analysis, it can be easily seen that one of the major shortcomings of the PCS-Hub approach is that it can yield partial correlation estimates that are very close to marginal correlation estimates in the high-dimensional sample-starved setting.  Consequently, we find that PCS-Hub does not always produce reliable estimates when partial and marginal correlations are different.  More specifically, the PCS-Hub approach of  \textcite{HeroRajaratnam:2012} uses the Moore-Penrose approach to estimate partial correlations, whereas the PARSEC regression approach in addition also uses the sample marginal correlations, which tend to be more stable (as they do not derive from inverses in high dimensions and can simply be calculated as inner products between features). One of the motivations for the newly proposed PARSEC method is to leverage more stable sample quantities to estimate partial correlations in the challenging ultra-high dimensional setting.

\begin{figure}[t]
\centering
\includegraphics[width=160mm]{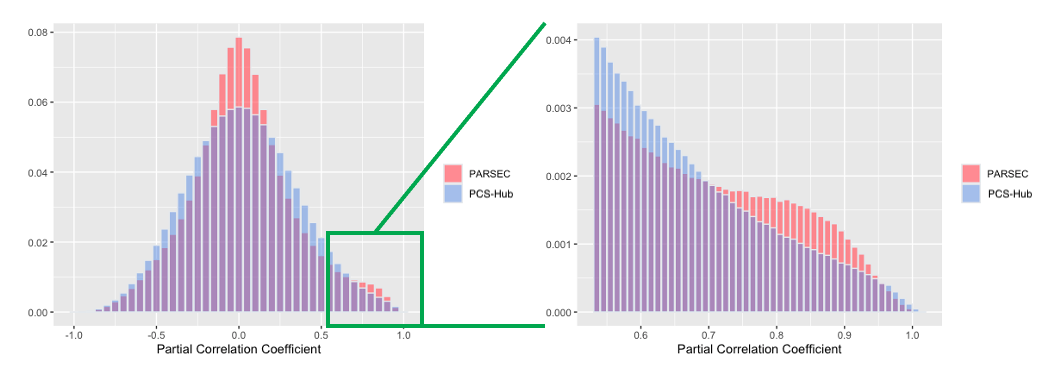}
\caption{A comparison of the distribution of estimated partial correlation coefficients from PARSEC to the PCS-Hub method \parencite{HeroRajaratnam:2012} in an AR(10) setting.  We simulate 1000 replications with Gaussian features where $n=10$, $p=100$, block size=20 and $\phi_1=0.8$ (where $\phi_1$ is the coefficient on the first lag).  Notably, the distribution of PARSEC's partial correlation coefficients is bi-modal.  The mode located closer to 1 corresponds to the true AR(10) model coefficients. We magnify this area of the plot in the right figure. In addition to illustrating PARSEC's statistical power in identifying true signal, the left plot also illustrates PARSEC's effectiveness in controlling  Type I error,  as PARSEC's estimates of the null coefficients are  shrunk closer to zero.  In contrast, the distribution of PCS-Hub partial correlation coefficients is nearly symmetric around zero.  
\label{fig:distr_parsec_mpi}}
\end{figure}

We now provide preliminary evidence that PARSEC can do better at identifying partial correlations than the PCS-Hub approach.
Similar to the U-scores and Y-scores discussed in Section~\ref{sec:methods_preliminaries} for marginal correlation screening and PCS-Hub-based partial correlation screening, the analogous PARSEC scores can be constructed via the corresponding decomposition of the symmetric matrix~$\tilde{\mathbf{H}}$, where~$\tilde{\mathbf{H}}$ is obtained by transposing the upper triangle of PARSEC's $\mathbf{H}$ matrix.  
The closeness of the marginal correlation U-scores and partial correlation Y-scores, as seen in Figure~\ref{fig:U_Y_V_score_demo}, highlights the possible challenges faced by the PCS-Hub-based Y-scores in instances when partial correlations are relatively high but the corresponding marginal correlations are low.  In contrast, PARSEC scores are more distinguishable from marginal correlation U-scores.
To understand whether these differences lead to better inference, we compare the distribution of partial correlation estimates stemming from PCS-Hub and PARSEC, in a setting where the true covariance model has high partial correlations - see Figure \ref{fig:distr_parsec_mpi}.  In contrast to the PCS-Hub-based approach, a bi-modal distribution for PARSEC estimates is immediately identifiable, with a peak closer to 1 corresponding to the large nonzero partial correlations in the underlying model - illustrating PARSEC's improved Type II error control. The second peak closer to zero demonstrates that PARSEC also better shrinks the null coefficients to zero, illustrating PARSEC's improved Type I error control.


\section{Theoretical properties of PARSEC} \label{sec:theory}

In this section we provide the theoretical properties of the PARSEC method in the ultra-high dimensional regime where the sample size~$n$ is fixed while the number of features~$p$ tends to infinity. In particular, we establish the limiting behavior of the number of PARSEC discoveries,
which in turn allows us to derive fixed-$n$-large-$p$ expressions of the corresponding family-wise error rate (FWER) and k-family-wise error rate (k-FWER).
Equally importantly, we also derive the exact marginal p-values for the PARSEC estimates of scaled partial correlation coefficients, which consequently allows us to achieve FDR and pFDR control.

\subsection{Preliminaries}

As our method competes with that of \textcite{HeroRajaratnam:2012}, for comparison purposes we adopt similar notation.
Let $N_{\rho_p}$ denote the number of PARSEC discoveries corresponding to the screening level~$\rho_p\in(0,1)$, i.e., $N_{\rho_p}$ is the number of entries located above the main diagonal in the PARSEC estimate of the scaled partial correlation matrix that exceed~$\rho_p$ in magnitude. We define the spherical cap probability,
\begin{equation}
 P_0=P_0(\rho_{p},n)=a_n\int_{\rho_{p}}^1 (1-u^2)^{(n-4)/2} du, \qquad \text{where} \qquad \text a_n=\frac{2\Gamma([n-1]/2)}{\sqrt{\pi}\Gamma([n-2]/2)},
 \label{eq:spherical.cap.prob}
\end{equation}
\noindent and let~$\|\Delta_{p,q}\|_1$ be the average weak dependency coefficient, defined in \textcite[equation~(A.13)]{HeroRajaratnam:2011}.  We write~$J\big(\overline{f_{\mathbf{U}_{\bullet},\tilde{\mathbf{U}}^{\bullet}_{*-\bullet}}}\big)$ for the normalized integral of the average pairwise density involving a partial and a regular U-score, as formally defined in Appendix~\ref{sec.notation} of this paper.
For the remainder of this section, bounds~$o(\cdot)$ and~$O(\cdot)$, as well as their stochastic counterparts, are understood to hold in the setting where~$n$ is fixed and~$p$ tends to infinity.
In the results that follow, we invoke either or both of the following commonly used assumptions from the literature.
\begin{itemize}
\item[A1.]  The random matrix $\mathbb{X}\in \mathbb{R}^{n \times p}$
has a vector-elliptical distribution with location parameter $\boldsymbol{\mu} \in \mathbb{R}^p$, covariance parameter $\mathbf{\Sigma} \in \mathbb{R}^{p\times p}$, and a differentiable density function that is uniformly bounded in~$p$.
\item[A2.]The elements of $\mathbb{U} \mathbb{U}^\top - E\big(\mathbb{U} \mathbb{U}^\top\big)$ are of order $O_p(s_p)$, where $s_p=o(p)$.
\end{itemize}
As a concrete example, assumption A2 holds with $s_p = p^{1/2}+q_p$ - see \textcite{FirouziEtAl:2017} - if the correlation matrix~$\mathbf{\Omega}$ corresponding to the covariance matrix~$\mathbf{\Sigma}$ is of the form $\mathbf{\Omega}=\mathbf{\Omega}_1+\mathbf{\Omega}_2$, where $\mathbf{\Omega}_1$ is a block-sparse matrix of degree $q_p=o(p)$, and $\mathbf{\Omega}_2=(\omega_{j,k})$ is such that $\omega_{ij}=O\big(f(|j-k|)\big)$ for some function~$f$ that satisfies $\lim_{t\rightarrow\infty}f(t)=0$.

\subsection{Theoretical Results} \label{sec:theoretical_results}

In this section, we establish a sequence of results on the limiting distribution of the number of PARSEC discoveries, $N_{\rho_p}$, as~$p$ tends to infinity but the sample size $n$ remains fixed.  These results allow us to undertake statistical inference in the ultra-high dimensional setting. Our first three theorems provide Poisson approximations for the distribution of the number of discoveries and asymptotic expressions for the corresponding expected value. The assumptions of these theorems get progressively stronger for reasons specified below.
Finally, in the fourth result, we derive exact marginal p-values for PARSEC estimates of scaled partial correlation coefficients.  All the proofs are provided in Appendix~\ref{sec.proofs}.

\vspace{-14pt}

\paragraph{Result 1.} The following theorem
establishes a general result on the limiting behavior of the number of PARSEC discoveries.  Employing assumption A1 leads to an approximation for the expected number of discoveries, and additionally assuming A2 leads to the Poisson approximation given below.
\begin{theorem}
\label{thm.gen}
Let $\eta_p=p(p-1)P_0/2$ and $\delta_p=(s_p+1)/p$. If assumption $A1$ holds, then the expected number of discoveries is given as follows:
\begin{equation*}
E[N_{\rho_{p}}]=\eta_pJ\big(\overline{f_{\mathbf{U}_{\bullet},\tilde{\mathbf{U}}^{\bullet}_{*-\bullet}}}\big)] + O\Big(\eta_p\sqrt{1-\rho_{p}}\Big).
\end{equation*}
\noindent Now, let $N^*_{p}$ denote a Poisson random variable with rate $E[N^*_{p}]=\eta_pJ\big(\overline{f_{\mathbf{U}_{\bullet},\tilde{\mathbf{U}}^{\bullet}_{*-\bullet}}}\big)$. If assumption $A2$ also holds, $(p-1)P_0\le1$, $\delta_p=o\big(1-\rho_{p}\big)$ and
\begin{equation} \label{eqn:result1}
\eta_p\Big[\eta_p(l_p/p)^2+\|\Delta_{p,l_p}\|_1+\big[\sqrt{1-\rho_{p}}+\delta_p(1-{\rho}_{p})^{-1}\big]\big(1+E[N^*_p])^{-1/2}\Big]=o(1)
\end{equation}
\noindent for some arbitrary $l_p\in[p]$, which is allowed to depend on~$p$, then
\begin{equation*}
\sup_{k\in\mathbb{N}}\Big|P(N_{\rho_{p}}>k)-P(N^*_{p}>k)\Big| = o(1).
\end{equation*}
\end{theorem}

\begin{remark}
We note that our approximation for the probability of more than~$k$ discoveries is uniform in~$k$, while the corresponding approximation in \textcite{HeroRajaratnam:2012} is stated only for~$k=0$.  The uniformity of our approximation allows us to use it for~$k$ that is a function of~$p$, for example, $k=p^{1/2}$.
\end{remark}
\begin{remark}
Note that the term $\eta_p[\sqrt{1-{\rho}_{p}}+\delta_p(1-{\rho}_{p})^{-1}]\big(1+E[N^*_p])^{-1/2}$ appears in equation (\ref{eqn:result1}), as compared to $p^2P_0[p^{-1}+(1-\rho_{p})^{1/2}]$ used in \textcite{HeroRajaratnam:2012}.
If $\delta_p=O([1-\rho_{p}]^{3/2})$ and $E[N^*_{p}]\rightarrow\infty$, then our condition for obtaining the Poisson approximation is weaker.
\end{remark}
\begin{remark}
Note that Theorem \ref{thm.gen} does not necessarily require sparsity.
\end{remark}


\paragraph{Result 2.}  We now consider the case where~$\mathbf{\Sigma}$ is a block-sparse matrix of degree~$q_p$.  Imposing block-sparsity yields a better approximation for the Poisson rate parameter and leads to a tractable and useful representation in the modern ultra-high dimensional setting.
\begin{theorem}
\label{thm.block.sparse}
Let $\eta_p=p(p-1)P_0/2$.  If assumption $A1$ holds and $\mathbf{\Sigma}$ is block-sparse of degree~$q_p$, then
\begin{equation*}
E[N_{\rho_{p}}]=\eta_p \left[1+ O\Big(\frac{q_p}{p}\Big)\right].
\end{equation*}
\noindent Now, let $N^*_{p}$ denote a Poisson random variable with rate $E[N^*_{p}]=\eta_p$.  If $(p-1)P_0\le1$, $n>2$, $q_p=o(p)$ and $\eta_p^2\Big(\frac{q_p}{p}\Big)^2+\sqrt{\eta_p}\Big[\frac1{\sqrt{p}}+\frac{q_p}{p}\Big](1-{\rho}_{p})^{-1}=o(1)$, then
\begin{equation}
\label{Bl.sp.Pois.apr1} \sup_{k\in\mathbb{N}}\Big|P(N_{\rho_{p}}>k)-P(N^*_p>k)\Big| = o(1), \quad \text{and}
\end{equation}
\begin{equation*}
\sup_{k\in\mathbb{N}}\Big|P(N_{\rho_{p}}>k)-P(N^*_p>k)\Big| = O\Big(\frac1{p}+\eta_p^2\Big(\frac{q_p}{p}\Big)^2+{\sqrt{\eta_p}\Big[\frac{\sqrt{\log(p)}}{\sqrt{p}}+\frac{q_p}{p}\Big]}(1-{\rho}_{p})^{-1}\Big).
\end{equation*}
\end{theorem}

\noindent The above theorem subsumes the special case when the covariance matrix $\mathbf{\Sigma}$ is diagonal. This special case is important as it is required for setting FWER and k-FWER error control.  In particular, the following corollary follows directly from Theorem~\ref{thm.block.sparse} by setting $q_p=0$.

\begin{corollary}
\label{thm.ell.cont}
If assumption $A1$ holds
and~$\mathbf{\Sigma}$ is diagonal, then $E\big[N_{\rho_{p}}\big]=\eta_p.$  Now,
let~$N^*_{p}$ denote a Poisson random variable with rate $E[N^*_{p}]=\eta_p$.  If $\sqrt{\eta_p/p}(1-\rho_{p})^{-1}=o(1)$, then
\begin{equation*}
\sup_{k\in\mathbb{N}}\Big|P(N_{\rho_{p}}>k)-P(N^*_{\rho_{p}}>k)\Big| = o(1).
\end{equation*}
\end{corollary}
\begin{remark}
Note that Corollary~\ref{thm.ell.cont} allows for the expected number of discoveries to go off to infinity.  In particular, as $P_0=O\big([1-\rho_{p}]^{(n-2)/2}\big)$, the condition imposed on $\rho_{p}$ in the above result is satisfied when $1-\rho_{p}=o\big(p^{-2/(n-6)}\big)$ or, equivalently, $E[N^*_{p}]=o\Big(p^{1-[4/(n-6)]}\Big)$.   Thus, this condition allows $E[N^*_{p}]\rightarrow\infty$, provided that $n>10$.
\end{remark}

\begin{figure}[t]
\centering
\begin{tabular}{cc}
(a) $n=10$, varying $p$ & (b) $p=1000$, varying $n$ \\[6pt]
 \includegraphics[width=78mm]{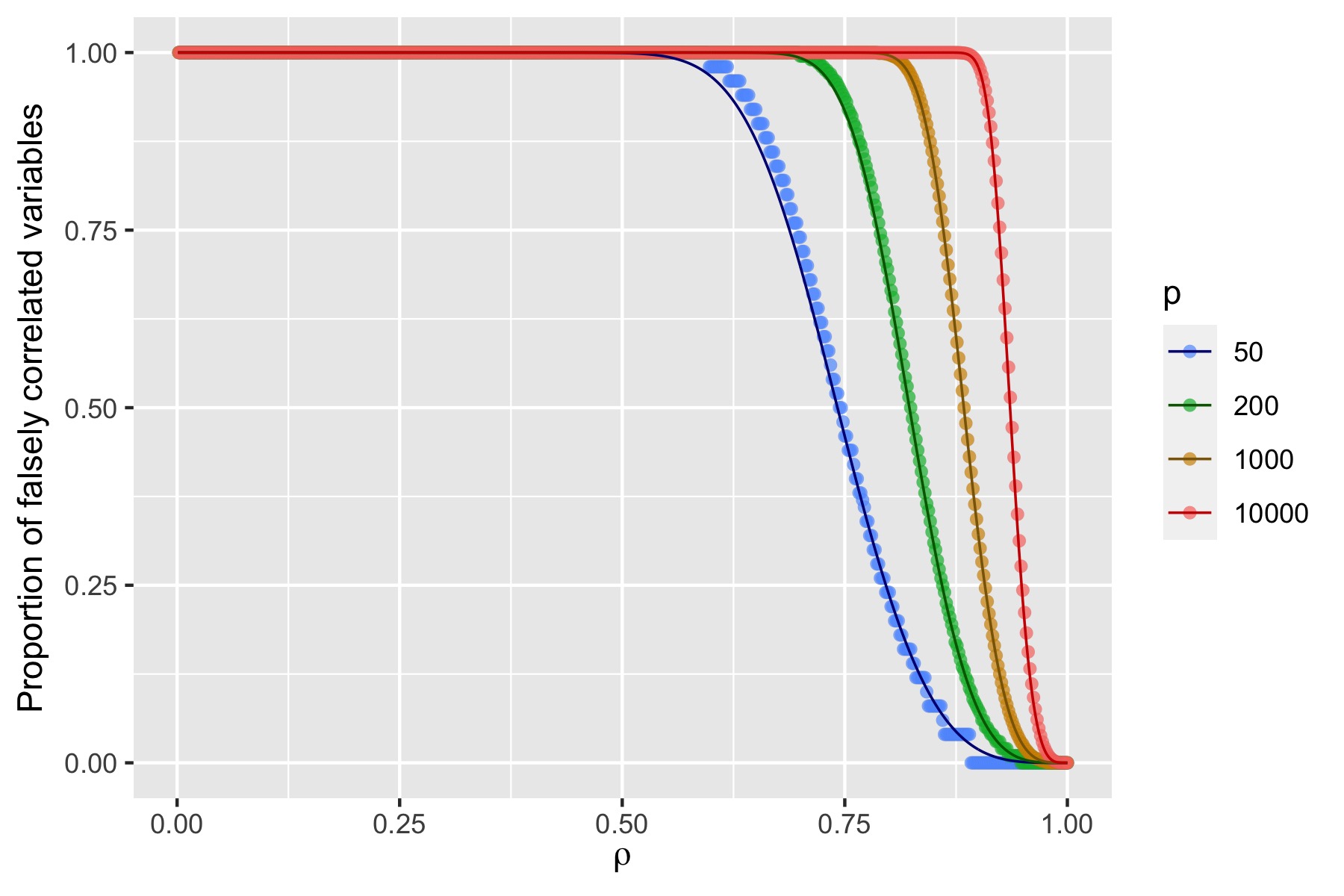}  &
  \includegraphics[width=78mm]{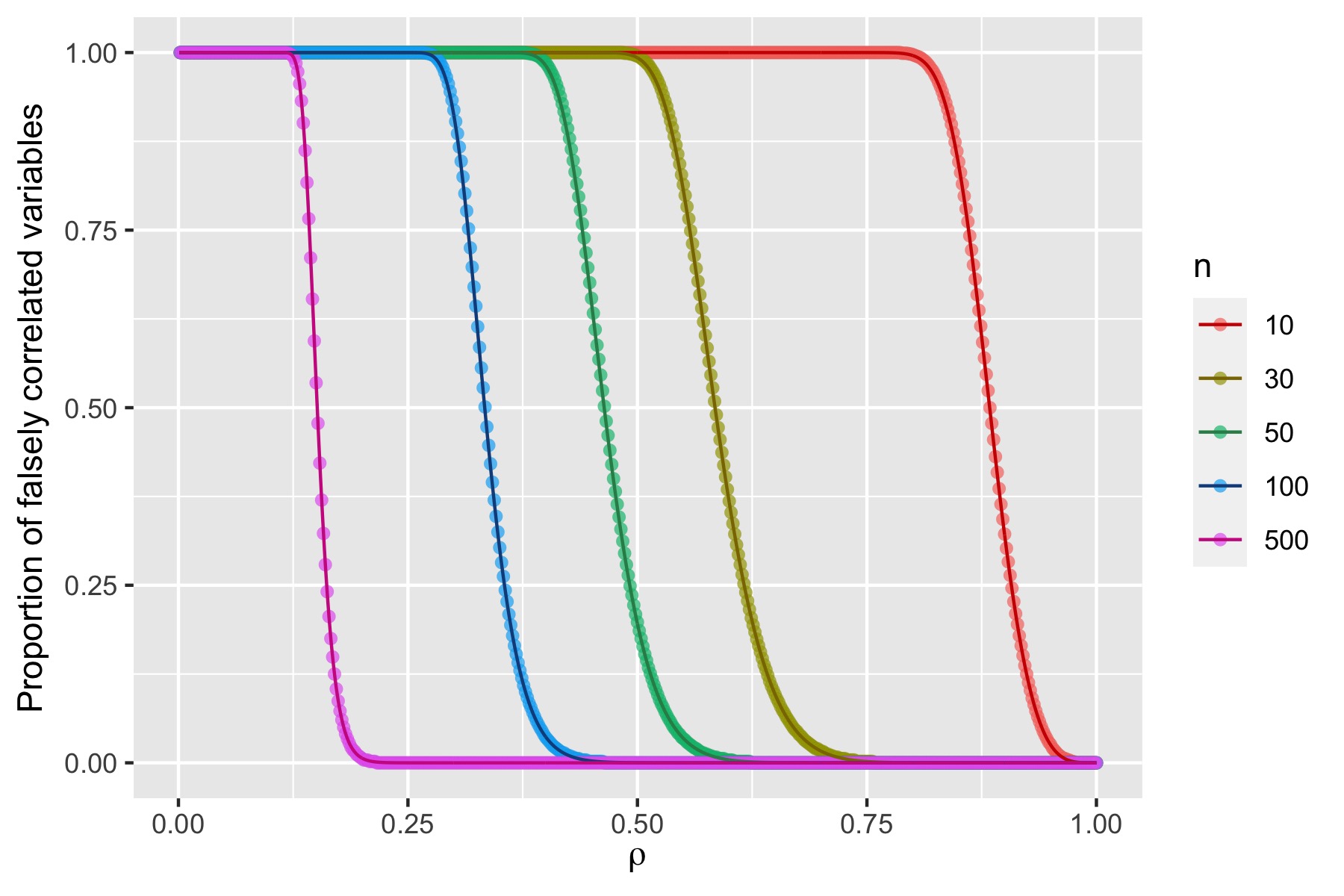}
\end{tabular}
\caption{Illustration of the proportion of features in~$\mathbb{X}$ with at least one partial correlation falsely identified by PARSEC under the null hypothesis of diagonal covariance at screening level $\rho$. In panel~(a), $n=10$ and $p$ is varied; in panel~(b), $p=1000$ and $n$ is varied. We compare the theoretical approximations (displayed as solid curves) with the corresponding median empirical proportions (displayed as circles) over 1000 numerical replications. It is clear that the large $p$ theoretical expressions mirror the true empirical proportion of discovered features, even when the dimension $p$ is small. \label{fig:hub_transitions}}
\end{figure}

\noindent Figure~\ref{fig:hub_transitions} provides numerical validation of Corollary~\ref{thm.ell.cont} in the null model setting, where $\Sigma=\mathbf{I}$.
%
%
The theoretical approximations almost exactly mirror the empirical results, even when~$p$ is as small as~50 (see Appendix~\ref{appendix:phase_transitions} for further details).


\paragraph{Result 3.}  In Appendix Section \ref{appendix:thm.finite.lim}, we complement~Theorem \ref{thm.block.sparse} with a more specialized result, which specifies the limiting behavior~$\rho_p$ that allows the expected number of discoveries to converge to a finite limit as~$p$ goes to infinity.

\paragraph{Result 4.} Our final theorem derives exact marginal p-values for the PARSEC estimates of scaled partial correlation coefficients. In the next subsection, we apply this result to control the FDR and pFDR.
\begin{theorem}


\label{thm.pval}
Let $P_0$ denote the spherical cap probability as given in equation~(\ref{eq:spherical.cap.prob}) and  let $H_{jk}$ denote the PARSEC partial correlation estimate between features~${X}_j$ and~${X}_k$.  Suppose that assumption $A1$ is satisfied
and~$\mathbf{\Sigma}$ is diagonal.  Then for each non-negative~$\rho$,
\begin{equation*}
P(|H_{jk}|>\rho) = P_0(\rho,n).
\end{equation*}
\end{theorem}
\begin{remark}
Note that the above result is an ``exact" result and holds for any fixed~$n$ and any fixed~$p$. It is thus a ``doubly finite" sample-dimension result. To our knowledge, no competing method for large-scale partial correlation screening (such as the gold standard PCS-Hub method) readily provides any type of distinct distribution for the estimated {\it partial} correlation in the large $p$ setting, let alone an exact one. The above exact result for the PARSEC approach facilitates the calculation of exact marginal p-values.
\end{remark}

\vspace{0.5cm}

\begin{remark}We note that the assumptions invoked in order to prove our results are not restrictive for a number of compelling reasons. First, the main theorem (please see Theorem 1) does not assume sparsity, and even if sparsity is imposed, the block sparse matrix is allowed to grow in dimension as $p \rightarrow \infty$ and the result will still hold true (please see Theorem 2). Hence the assumptions are not strong in this regard, especially compared to work in high dimensional inference where sparsity is almost always assumed. In this sense, the results in the paper are quite general. Second, Assumptions A1 and A2 used in the paper are also invoked in the PCS-Hub paper and in \textcite{FirouziEtAl:2017} respectively, both of which are screening papers that appear in {\textit{IEEE Transactions on Information Theory}}. Thus, the assumptions invoked for the above results to hold are quite general and are not any more restrictive than those invoked in the literature. Third, neither assumption A1 nor assumption A2 are required to obtain the exact distribution of the distribution of the PARSEC partial correlation co-efficient (please see Theorem 3).
\end{remark}

\subsection{Ultra high-dimensional statistical inference with PARSEC}
\label{uncertainty.subsec}

We now demonstrate how the theoretical results stated above can be used for statistical inference (and not only for model selection) in the ultra high-dimensional settings where~$n$ is fixed and~$p$ tends to infinity. In particular, we use these results to control the FWER, $k$-FWER, FDR and pFDR metrics for our proposed method.


\paragraph{$k$-FWER:} Because FWER is a special case of $k$-FWER when $k=0$, we will focus on the more general $k$-FWER metric.  Let $\alpha$ be a pre-specified level of error control, $k$ the allowable number of false discoveries, $p$ the  dimension, and $n$ the sample size. By Corollary~\ref{thm.ell.cont}, if we choose $\rho_p$ so that~$k$ is a $(1-\alpha)$-level quantile of the Poisson distribution with rate $p(p-1)P_0(\rho_p,n)/2$, using formula~(\ref{eq:spherical.cap.prob}) for the spherical cap probability~$P_0$, then $P(N_{\rho_{p}}>k)\approx\alpha$ for large~$p$, thus allowing us to control the $k$-FWER at level~$\alpha$.  The results of Corollary~\ref{thm.ell.cont} can also be employed in other useful ways.  First, Corollary~\ref{thm.ell.cont} allows us to quantify the true level of error control achieved when a user pre-specifies a value for the screening level~$\rho_p$ (i.e., a type of generalization of the p-value in the multiple hypothesis testing context).  Second, note that sometimes it may be difficult to specify the value of $k$.  In such cases, given $n$, $p$, $\rho_p$ and $\alpha$, Corollary~\ref{thm.ell.cont} allows us to quantify the number of false discoveries $k$ that are being implicitly tolerated at that level of error control.



\paragraph{FDR:} To control the FDR, we shall compute marginal p-values as follows.  Recall that $H_{jk}$ is the PARSEC estimate of the partial correlation between features~${X}_j$ and~${X}_k$.  Under the null hypothesis that~$\mathbf{\Sigma}$ is diagonal, i.e. each partial correlation coefficient is zero, by Theorem~\ref{thm.pval}, $P(|H_{jk}|>c) = P_0(c,n)$ for~$c>0$.  Thus, the \emph{exact} non-asymptotic marginal p-value corresponding to $\hat{\rho}^{jk} = H_{jk}$ is given by $P_0(|H_{jk}|,n)$. These marginal p-values for off-diagonal elements, together with the Benjamini-Hochberg (BH) and/or Benjamini-Yakuteli (BY) method, can be used to control the FDR.
In addition to FDR, PARSEC is also readily amenable to controlling the pFDR \parencite{Storey:2002}.  We note that FDR control requires the calculation and ordering of $\binom{p}{2}$ p-values.
To address this challenge, we implement a scalable FDR procedure described in Appendix \ref{appendix:scalable_FDR_algorithm}.


Finally, we note that our inferential approach is rich enough to cover Gaussian graphical modelling as a special case.  In addition to learning the Gaussian graphical model structure, our approach can quantify the uncertainty associated with the estimated model - a property that is lacking in contemporary ultra-high dimensional methods.  One of the other advantages of the PARSEC approach is that it can also be used to identify a conditional set for each feature, similar to useful methods proposed in the context of high dimensional regression (see \textcite{ChoFryzlewicz:2012} and other similar methods).


\section{Algorithms and computational properties}\label{sec:Algorithms}

We now provide a suite of algorithms which facilitate the implementation of PARSEC, especially in ultra-high dimensions.  In particular, Section~\ref{sec:base_parsec} provides an algorithm for the base implementation of PARSEC, followed by its highly scalable version.
Lastly, we derive PARSEC's computational complexity in Section~\ref{sec:computational_complexity}, illustrating PARSEC's scalability in large~$p$ settings.

\subsection{The base and scalable implementation of PARSEC} \label{sec:base_parsec}

The PARSEC approach described in Section \ref{sec:methodology} can also be expressed as a formal algorithm.
 This ``base PARSEC algorithm" directly evaluates scaled partial correlation coefficients in matrix $\mathbf{H}$, through~$p$ parallelizable operations, and is provided in Algorithm~\ref{alg:directeval_H} in Appendix \ref{appendix:base_algorithm}.
As the base PARSEC algorithm requires $p$ matrix inversions (Step 8 in Algorithm 1), it can thus be computationally expensive in ultra-high dimensional regimes.  To address this challenge, we now propose an alternative and more scalable algorithm which exploits Sherman-Morrison-Woodbury rank one updates to speed up the evaluation of $\mathbf{H}$. Recall some of the previously used notation and formulas. Given a matrix $\mathbb{M}$, we write $\mathbf{M}_k$ for the $k$-th column of $\mathbb{M}$ and write $M_{jk}$ for the element of~$\mathbb{M}$ located in the $j$-th row and $k$-th column.  The off-diagonal elements in the $j$-th row of matrix $\mathbf{H}$ are contained in the vector $\big(\tilde{\mathbb{U}}^{-j}\big)^\top\mathbf{U}_j$, where $\tilde{\mathbb{U}}^{-j}=\mathbb{C}_j\mathbb{D}_j^{-1/2}$ with $\mathbb{C}_j=\big(\mathbb{U}^{-j} (\mathbb{U}^{-j})^\top\big)^{-1}\mathbb{U}^{-j}$, and $\mathbb{D}_j$ is a diagonal matrix whose diagonal equals to that of the matrix $(\mathbb{U}^{-j})^\top\big(\mathbb{U}^{-j} (\mathbb{U}^{-j})^\top\big)^{-2}\mathbb{U}^{-j}$.  Recall the matrix $\mathbb{U}$ which involves all the U-scores.  We can compute $\big(\mathbb{U}^{-j} (\mathbb{U}^{-j})^\top\big)^{-1}$ for each~$j$ using simple Sherman-Morrison-Woodbury rank one updates as follows:
\begin{equation*}
\Big(\mathbb{U}^{-j} (\mathbb{U}^{-j})^\top\Big)^{-1} = \Big(\mathbb{U} \mathbb{U}^\top - \mathbf{U}_j{\mathbf{U}_j}^\top \Big)^{-1} = \mathbf{A}+\Big[\frac1{1-\mathbf{U}_j^\top\mathbf{A}\mathbf{U}_j}\Big]\mathbf{A}\mathbf{U}_j\mathbf{U}_j^\top\mathbf{A},
\end{equation*}
where $\mathbf{A} = \big(\mathbb{U} \mathbb{U}^\top\big)^{-1}$.  Defining $\mathbf{B} = \mathbb{U}^\top \mathbf{A} \mathbb{U}$ and $\mathbf{F} = \mathbf{A} \mathbb{U}$, we can simplify the updates to
\begin{equation}
H_{jk} = \Big(\frac{B_{kj}}{1-B_{jj}}\Big)\Big\|\mathbf{F}_k+\frac{B_{kj}}{1-B_{jj}}\mathbf{F}_j \Big\|^{-1}_2
\label{eqn:rank_1_speedup}
\end{equation}
or, equivalently, $H_{jk} = \Big\|(1-B_{jj})B_{kj}^{-1}\mathbf{F}_k+\mathbf{F}_j \Big\|^{-1}_2$ - see Appendix \ref{appendix:comp_complexity} for more details.
The procedure using equation~(\ref{eqn:rank_1_speedup}), summarized in Algorithm \ref{alg:scalable_eval_H}, leads to significant computational savings.

{\begin{algorithm}[t]
\small
 \caption{{\bf Scalable implementation of PARSEC} \label{alg:scalable_eval_H}}
\KwIn{$\mathbb{X}_{n\times p}$}
\KwOut{$\mathbf{H}_{p\times p}$}
\Begin{
\For{$j=1$ \KwTo p}{
$ \displaystyle \mathbf{Z}_j = \dfrac{\mathbf{X}_j - \bar{X}_j\mathbf{1}}{\sqrt{\mathbf{S}_{jj}(n-1)}}$}
Define $\mathbb{T}= [n^{-1/2}\mathbf{1}, \mathbf{T}_{2:n}]$, where $\mathbf{1}^\top \mathbf{T} = [\sqrt{n},0,...,0]$ and $\textbf{T}_{2:n}^\top \textbf{T}_{2:n}=\textbf{I}_{n-1}$ \;
 Obtain $ \displaystyle \mathbb{U}_{n-1 \times p} =\mathbf{T}^\top_{2:n} \mathbb{Z}$ \;
 Obtain $ \displaystyle \mathbf{A} = (\mathbb{U} \mathbb{U}^\top)^{-1}$ \;
 Obtain $ \displaystyle \mathbf{B} = \mathbb{U}^\top \mathbf{A} \mathbb{U} $ \;
 Obtain $ \displaystyle \mathbf{F} = \mathbf{A} \mathbb{U}$ \;
 \For{$k=1$ \KwTo p}{
  \For{$j=1$ \KwTo p except $j=k$ }{
  $ \displaystyle H_{kj} =  \Big( \dfrac{\mathbf{B}_{jk}}{1-\mathbf{B}_{kk}} \Big) \Big\lVert \mathbf{F}_j + \Big( \dfrac{\mathbf{B}_{kj}}{1-\mathbf{B}_{kk}} \Big) \mathbf{F}_k \Big\rVert^{-1}_2 $ \;}}}
\end{algorithm}}

\subsection{Computational Complexity} \label{sec:computational_complexity}

The following theorem quantifies the computational complexity of the scalable implementation of PARSEC (see Appendix \ref{appendix:comp_complexity} for the proof).

\vspace{0.2cm}

\begin{theorem} \label{thm:comp_complexity}
The computational complexity of the unparallelized version of Algorithm \ref{alg:scalable_eval_H} is $O(np^2+n^2p+n^3)$, i.e., quadratic in $p$. The $p$-core computational complexity of the parallelized version of Algorithm \ref{alg:scalable_eval_H} is $O(n^2p+n^3)$, i.e., linear in $p$.
\end{theorem}

 \vspace{-6pt}

\paragraph{Computational Complexity:} Table \ref{tbl:computation_complexity_comparision} compares PARSEC's computational complexity with that of state-of-the-art Gaussian likelihood and pseudo-likelihood methods. PARSEC's computational complexity is superior to both $\ell_1$-based pseudo-likelihood and $\ell_1$-based Gaussian likelihood approaches, irrespective of parallelization. In particular, unparallelized PARSEC's computational complexity is of order $O(p^2)$ and is therefore better than existing $\ell_1$-penalized likelihood based methods because it is non-iterative. In addition and more importantly, PARSEC is amenable to parallelization:  the computational complexity of the parallelized implementation of PARSEC is linear in~$p$. This improvement stems from two sources: i) the parallelizable nature of PARSEC; and ii) the non-iterative nature of PARSEC as it yields a closed-form solution that does not require optimization.   Parallelization also allows PARSEC's computational complexity to be competitive with the state-of-the-art PCS-Hub approach.  As such, PARSEC is highly efficient.

\begin{table}[t]
\centering
\small
\begin{tabular}{| p{2.7cm} | p{3.8cm}  | p{3.6cm} | p{2.2cm} | p{2cm} |} \cline{2-5}
 \multicolumn{1}{l|}{}  & \multicolumn{4}{c|}{\textbf{Method}} \\ \cline{1-5}
 \multicolumn{1}{|l|}{}  & \textbf{Gaussian likelihood}  & \textbf{Pseudo-likelihood}  & \multicolumn{2}{c|}{\textbf{PARSEC} (Algorithm \ref{alg:scalable_eval_H})}  \\ \cline{4-5}
 \textbf{Property} &  \parencite{BanerjeeEtAl:2008} & \parencite{KhareOhRajaratnam:2015} & \multicolumn{1}{c|}{\textbf{unparallelized}} & \multicolumn{1}{c|}{\textbf{parallelized}} \\ \cline{1-5}
Computational complexity &  \multicolumn{1}{c|}{$O(tp^4)$} & \multicolumn{1}{c|}{$O(\min\{tnp^2,tp^3\})$} & \multicolumn{1}{c|}{$O(np^2+n^2p+n^3)$}& \multicolumn{1}{c|}{$O(n^2p+n^3)$} \\ \cline{1-5}
Order in $p$ & \multicolumn{1}{c|}{$O(p^4)$} & \multicolumn{1}{c|}{$O(p^2)$} & \multicolumn{1}{c|}{$O(p^2)$} & \multicolumn{1}{c|}{$O(p)$} \\ \cline{1-5}
Order in $t$ & \multicolumn{1}{c|}{$O(t)$} & \multicolumn{1}{c|}{$O(t)$} & \multicolumn{1}{c|}{$O(1)$} & \multicolumn{1}{c|}{$O(1)$} \\ \cline{1-5}
\end{tabular}
\caption[Computational complexity comparison for different methods]{Computational complexity comparison with competing sparse partial correlation methods ($t$ denotes the number of iterations required).
\label{tbl:computation_complexity_comparision}}
\end{table}

\vspace{-12pt}

\paragraph{Storage and Memory Complexity:} In addition to improved computational complexity, PARSEC is also amenable to alternative data structures, which provide storage gains (improved ``memory/storage complexity"). Concrete details are provided in Appendix \ref{appendix:storage_complexity}.


\section{Numerical validation of statistical properties} \label{sec:simulated_data}

We now proceed to validate the theoretical guarantees developed for PARSEC via numerical simulations and compare our approach to its competitor, PCS-Hub, in terms of both a) inferential and b) computational performance in the ultra-high dimensional setting.

\subsection{Inferential properties and performance} \label{sec:simulation_inferential}

Our assessment of PARSEC's inferential performance has two main components.  We first validate the accuracy of PARSEC's theory by evaluating the error control achieved in the null model where $\Omega=\mathbf{I}$ (Type I error control).  Thereafter, we assess PARSEC's screening performance in structured models, in terms of statistical power and ability to recover true underlying signals (Type II error control).

\vspace{-12pt}

\paragraph{Assessing error control in the null model where $\Omega=\mathbf{I}$:}
First, we proceed to understand the level of error control afforded by PARSEC and PCS-Hub.   To this end, we evaluate the accuracy of the asymptotic expressions developed for PARSEC and PCS-Hub in the null setting, i.e., when the true partial correlation matrix is the identity matrix.
To reflect the ultra-high dimensional data regime, we simulate data with varying $p$ and fixed $n$, where $n=30$ and $p=1,000$, $p=10,000$ and even higher, all the way up to $p=100,000$.
To assess FWER and k-FWER error control, we specify partial correlation coefficient screening levels, $\rho_p$, for both PARSEC and PCS-Hub using the inferential procedures in Section~\ref{uncertainty.subsec}.  We also assess PARSEC's FDR control, using exact marginal p-values as outlined in Theorem \ref{thm.pval}.  Recall from Section \ref{sec:methods_preliminaries} that the PCS-Hub method does not provide a means to achieve FDR control in the partial correlation setting.  
As such, a direct comparison with PCS-Hub is not possible.  The results in Table \ref{tbl:sim_ec_null} show that PARSEC is competitive with PCS-Hub with respect to the highly conservative error control measure FWER.  In contemporary applications where FWER may be too conservative, measures such as k-FWER and FDR can often be highly beneficial. Table \ref{tbl:sim_ec_null} indicates that when  k-FWER is used, PCS-Hub displays severe shortcomings whereas PARSEC maintains its performance: the proportion of PARSEC discoveries above $k$ remains consistent with $\alpha$ when using k-FWER.  Importantly, when considering FDR, the average number of PARSEC false discoveries is also consistently controlled by $\alpha$, whereas PCS-Hub does not have a means to control FDR in the partial correlation setting.  In addition to FDR, we note that PARSEC is able to control the pFDR.
 In Appendix Section \ref{appendix:pfdr_null} we provide  additional assessment of PARSEC's pFDR error control.  In summary, PARSEC's compelling performance across multiple error control measures highlights its versatility as a reliable inferential method in the modern ultra-high dimensional setting.

{
\begin{table}[t]
\centering
\small
\begin{tabular}{ll rr  rr  rr }
\toprule
& &  \multicolumn{2}{c}{\textbf{p=$10^3$}} &  \multicolumn{2}{c}{\textbf{p=$10^4$}} & \multicolumn{2}{c}{\textbf{p=$10^5$}} \\ \cmidrule(r){3-4} \cmidrule(r){5-6} \cmidrule(r){7-8}
 & &  PARSEC & PCS-Hub &  PARSEC & PCS-Hub &  PARSEC & PCS-Hub  \\
\midrule
\multicolumn{8}{l}{\textbf{k-FWER (\% of replications $>$ k) }} \\ \cmidrule(r){2-8}
FWER & $\alpha=0.01$  & 0.009  & 0.004 & 0.009 & 0.01  & 0.013 & 0.014    \\
& $\alpha=0.05$ & 0.065 & 0.041 &  0.06 &  0.06  & 0.043 &  0.046 \\ \cmidrule(r){3-8}
k-FWER & $\alpha=0.01,k=1\%$ &  0.018 & 0 & 0.013 & 0 &  0.012 & 0  \\
& $\alpha=0.01,k=5\%$ &  0.019 & 0 & 0.012 & 0 &  0.009 & 0 \\
& $\alpha=0.05,k=1\%$ &  0.078 &  0.001 & 0.056 & 0 &  0.047 & 0   \\
& $\alpha=0.05,k=5\%$ &  0.072 & 0 & 0.063 & 0 & 0.053 & 0\\
\midrule
 \multicolumn{8}{l}{\textbf{FDR (averaged over all replications)}} \\ \cmidrule(r){2-8}
FDR-BH & $\alpha=0.01$ & 0.009 &  d.n.e  & 0.009 & d.n.e & 0.013 & d.n.e \\
& $\alpha=0.05$ & 0.066 &  d.n.e  & 0.051  & d.n.e & 0.043 &  d.n.e \\ 
\bottomrule
\end{tabular}
\caption[Simulation performance, error control as p increases]{Error control with fixed $n$ ($n=30$) and increasing $p$ in null models. Performance measures are evaluated over 1000 replications for each dimension. ``d.n.e." denotes `does not exist' as the PCS-Hub approach does not readily provide FDR control.  Note, the value of $k$ used in the k-FWER error control measure is proportional to the dimension $p$.  We set $k$ equal to the specified percentage of $p(p-1)/2$, the total number of all possible partial correlations.
\label{tbl:sim_ec_null}}
\end{table}
}

\vspace{-12pt}

\paragraph{Screening performance in non-null models.}

We now assess PARSEC's screening performance in non-null models and investigate the setting when the~$p$ features are either Gaussian or heavy-tailed. We consider various structures for the covariance matrix~$\mathbf{\Sigma}$, including auto-regressive (AR) block, block covariance, and star structures, which are summarized as follows:
\begin{itemize}
    \item \textbf{AR block}. Features~$X_j$ with $j\le a$ follow an AR model of order~$d$; the rest of the features are independently generated.  The first order coefficient in the AR model equals $\phi_1$; the remaining AR coefficients equal $(1-\phi_1)/(d-1)$.
    \item \textbf{Block}. We set $\Sigma_{jk}=\rho\mathbf{1}{\{j\neq k, j\le a, k\le a\}}+1{\{j=k\}}$.
    \item \textbf{Star}. The partial dependence among the features is represented by~$k$ ``stars'' or ``hubs". Each star has a central feature connected to~$e$ other features, and there are no other connections within the star. All the nonzero elements of the inverse covariance are set equal to~$c$. We consider two types of star structures: (i) the ``Connected Star'' case, where each pair of stars has a connection between them via at least one non-zero inverse covariance element; and (ii) the ``Disconnected Star'' case, where there are no connections between the stars.
\end{itemize}

{
\begin{table}[ht!]
\small
\centering
\begin{tabular}{llcccccc}
\toprule
 &  & \multicolumn{3}{c}{\textbf{PARSEC}} &  \multicolumn{3}{c}{\textbf{PCS-Hub}}\\ \cmidrule(r){3-5} \cmidrule(r){6-8}
\textbf{Structure} & \textbf{Dimensions} & \textbf{AUC} & \textbf{FPR$<$.1} & \textbf{Timing}  & \textbf{AUC} & \textbf{FPR$<$.1} & \textbf{Timing} \\
\midrule
 \multicolumn{5}{l}{\textbf{AR(1) Block, $\phi_1=0.7$}} \\
& a=50, n=20 & \textbf{0.991} &	\textbf{0.925}  & 0.063 &	0.988 & 	0.910 & 0.084\\
& a=100, n=50 & \textbf{0.999} &	\textbf{0.999} & 0.101 &	\textbf{0.999} & 	\textbf{0.999} & 0.117 \\
& a=500, n=100  & \textbf{0.999} &	\textbf{0.999}  & 0.135 &	\textbf{0.999} & 	\textbf{0.999} & 0.168\\ \cmidrule(r){2-8}
 \multicolumn{5}{l}{\textbf{AR(2) Block, $\phi_1=0.7$}} \\
& a=50, n=20 & \textbf{0.981} &	\textbf{0.895} & 0.060 &	0.979 & 	0.870 & 0.080\\
& a=100, n=50 &\textbf{0.999} &	\textbf{0.996} & 0.108 &	0.998 & 	0.982  & 0.118\\
& a=500, n=100  &  \textbf{0.999} &	\textbf{0.999} & 0.157 & 	0.997 & 	0.979 & 0.193 \\ \cmidrule(r){2-8}
 \multicolumn{5}{l}{\textbf{AR(5) Blocks, $\phi_1=0.7$}} \\
& a=50, n=20 & \textbf{0.999} & \textbf{0.999} & 0.073 & \textbf{0.999} & 0.996 & 0.096 \\
& a=100, n=50 & \textbf{0.999} & \textbf{0.999} & 0.142 & \textbf{0.999} & 0.998 & 0.158 \\
& a=500, n=100  & \textbf{0.999} & \textbf{0.992} & 0.240 & 0.998 & 0.985 & 0.273 \\ \cmidrule(r){2-8}
 \multicolumn{5}{l}{\textbf{AR(10) Blocks, $\phi_1=0.7$}} \\
& a=50, n=20 & \textbf{0.849} & \textbf{0.542} & 0.118 & 0.819 & 0.485 & 0.145 \\
& a=100, n=50 & \textbf{0.892} & \textbf{0.646} & 0.170 & 0.806 & 0.489 & 0.184 \\
& a=500, n=100 & \textbf{0.884} & \textbf{0.638} & 0.186 & 0.695 & 0.337 & 0.217 \\ \cmidrule(r){2-8}
 \multicolumn{5}{l}{\textbf{Block, $\sigma=0.7$}} \\
& a=5, n=20 & \textbf{0.998} & \textbf{0.975} & 0.051 & 0.996 & 0.961 & 0.069 \\
& a=30, n=60 & \textbf{0.998} & \textbf{0.983} & 0.086 & 0.920 & 0.606 & 0.103 \\
& a=50, n=100 & \textbf{0.970} & \textbf{0.756} & 0.122 & 0.589 & 0.124 & 0.152 \\ \cmidrule(r){2-8}
 \multicolumn{5}{l}{\textbf{Connected Star, $c=-0.35$, $n=30$}} \\
& k=5, e=2 & \textbf{0.917} & \textbf{0.612} & 0.049  & 0.904 & 0.576 & 0.059 \\
& k=10, e=4 & \textbf{0.974} & \textbf{0.839} & 0.049 & 0.968 & 0.808 & 0.061 \\   \cmidrule(r){2-8}
 \multicolumn{5}{l}{\textbf{Disconnected Star, $c=-0.35$, $n=30$}} \\
& k=5, e=2 & \textbf{0.868} & \textbf{0.498} & 0.054 & 0.853 & 0.468 & 0.074 \\
& k=20, e=2 &\textbf{ 0.868} & \textbf{0.487} & 0.054 & 0.851 & 0.470 & 0.074 \\
& k=20, e=4 & \textbf{0.917} & \textbf{0.630} & 0.053 & 0.907 & 0.596 & 0.070 \\
& k=50, e=4 & \textbf{0.922} & \textbf{0.639} & 0.052 & 0.911 & 0.604 & 0.071 \\
\bottomrule
\end{tabular}
\caption[Simulation performance, p=1000]{AUC values for the various covariance structures with $p=1000$ and varying~$n$.  Each setting is replicated 1000 times.  $\sigma$ denotes the coefficients of non-zero elements simulated from a covariance matrix in block settings, $\phi_1$ is the coefficient of the first order lag in AR block settings, and $a$ provides the block size. $c$ denotes the value of non-zero elements in the inverse covariance matrix in star structures. Performance is measured using median AUC, and median AUC where the FPR range is limited to less than 0.1. We also report median wall-times (in seconds).   We highlight in bold the best method (highest AUC) in each setting.  In all instances, PARSEC provides equal or superior performance.
\label{tbl:sim_p1000_various_aucs}}
\end{table}
}

\noindent In what follows, we undertake a comprehensive assessment of PARSEC's ability to identify true partial correlations while controlling the number of false discoveries using multiple commonly-used metrics which balance the false positive and false negative rates.

Table~\ref{tbl:sim_p1000_various_aucs} compares the screening performance of PARSEC and PCS-Hub for the above covariance structures using Area-Under-the-Curve (AUC), when $p=1000$ and the sample size~$n$ is varied.  We provide the median AUC over the entire ROC curve and also the median AUC for the part of the curve where the False Positive Rate (FPR) is less than~0.1. The latter measure assesses performance for more realistic values of FPR.  In summary, PARSEC produces results which are always as good as and often much better than PCS-Hub across both measures and over all covariance structures considered, providing evidence of its superior screening performance.  PARSEC also exhibits wall-times that are competitive with PCS-Hub.  These timings are lower in all instances considered in Table \ref{tbl:sim_p1000_various_aucs}.  Moreover, we observe that when~$n$ is varied but~$p$ is fixed, PCS-Hub wall-times stay relatively constant.  However, wall-times for PARSEC decrease for smaller~$n$, that is the sample-starved settings PARSEC is designed for.  Table \ref{tbl:sim_p1000_MCC} in Appendix Section \ref{appendix:type_2_error} also compares PARSEC and PCS-Hub in terms of another metric: Matthews Correlation Coefficient (MCC), a consolidated performance measure which integrates both the false positive and false negative rates.  We observe that PARSEC uniformly obtains higher MCC values across all four error control measures, across all models, and across all signal strengths.  Note that the AR block cases and the ``connected star" model in Table~\ref{tbl:sim_p1000_various_aucs} do not assume sparsity in the correlation matrix, highlighting the technical strengths of the PARSEC approach.
Though the composite measures AUC and MCC affirm PARSEC's strong consolidated performance, they do not readily assess statistical power when an error metric (such as FWER/k-FWER/FDR/pFDR) is fixed at a pre-specified level $\alpha$ (i.e., at a practically relevant point on the ROC curve).  As an inferential method, PARSEC has the key advantage of being able to target a desired point on the ROC curve - it does not inadvertently  quantify performance in regions of the curve which are not relevant to an inferential task.
We investigate this distinct inferential question in Appendix Section \ref{appendix:type_2_error} where PARSEC is shown to achieve consistently higher sensitivity rates (i.e., statistical power) than PCS-Hub at pre-specified levels of error control.
We note that PARSEC provides improved identification of true partial correlation coefficients and favorably shrinks null coefficients, as illustrated in Figure~\ref{fig:distr_parsec_mpi} (from Section \ref{sec:methods_preliminaries}) - see Appendix \ref{appendix:type_2_error} for further illustrations.  Note also that the theoretical results for PARSEC hold for a wider class of distributions than the Gaussian (i.e., the class of vector-elliptical distributions).  In an effort to understand PARSEC's performance in more general settings, we investigated its performance in the context of heavier tail distributions such as the multivariate T (see Appendix \ref{appendix:heavier_simulated}).  The results therein convincingly demonstrate PARSEC's superior performance in a broader class of distributions.

\subsection{Computational performance} \label{sec:comp_sim_performance}

\begin{table}[t]
\centering
\small
\begin{tabular}{cc}
(a) Screening algorithm wall-times & (b) FDR algorithm wall-times \\[6pt]
\begin{tabular}{lccc}
\toprule
   &  \multicolumn{2}{c}{\textbf{PARSEC}} & \textbf{PCS-Hub}  \\ \cmidrule(r){2-3}
&  \textbf{Base} & \textbf{Scalable} \\
\midrule
$n$=30, $p$=$10^3$ & 3.92 & 0.08 & 0.11 \\
$n$=30, $p$=$10^4$ & 389.08 & 53.07 & 57.33 \\
$n$=30, $p$=$10^5$ & - & 7376.76 & 11972.08\\
\bottomrule
\end{tabular} &
\begin{tabular}{lcc}
\toprule
  &  \multicolumn{2}{c}{\textbf{FDR}} \\ \cmidrule(r){2-3}
& \textbf{Original} & \textbf{Iterative} \\
\midrule
$n$=30, $p$=$10^3$ & 6.86 & 0.06 \\
$n$=30, $p$=$10^4$ & 1442.10 & 20.63 \\
$n$=30, $p$=$10^5$ & - & 2150.09 \\
\bottomrule
\end{tabular}
\end{tabular}
\caption[Timing of original vs. speed-up computations]{Timing comparison of a) base versus scalable PARSEC algorithm and PCS-Hub (left), and b) the original versus the iterative FDR method (right), in the null setting where $\Omega=\mathbf{I}$.  Timings provided are median wall-times over 100 simulated datasets for each structure, reported in seconds, limited to a maximum of 24 hours. Note, in the case when $p=100,000$ the base PARSEC algorithm and FDR original algorithm do not yield a solution within 24 hours, affirming the benefits of the respective scalable versions. \label{tbl:timing_fdr_rank_1}}
\end{table}

Table \ref{tbl:timing_fdr_rank_1} (a) compares wall-times for PARSEC's base implementation (Algorithm \ref{alg:directeval_H}), PARSEC's scalable approach (Algorithm \ref{alg:scalable_eval_H}) and the PCS-Hub approach.  We report the median wall-times (in seconds), calculated over 100 simulated datasets.  It is clear that PARSEC's scalable implementation (Algorithm \ref{alg:scalable_eval_H}) exhibits competitive computational performance relative to the PCS-Hub approach, with significantly superior wall-times in higher dimensions such as  $p=100,000$.
We further investigate the source of PARSEC's superior computational performance and demonstrate that it stems from all three aspects of computing: i) computational speed (i.e., processing speed), ii) storage needs, and iii) memory allocation requirements (see Appendix Section \ref{appendix:computational_performance}).

\section{Real Applications} \label{sec:real_applications}

We now proceed to demonstrate PARSEC's efficacy on modern high-dimensional applications, starting with Section \ref{sec:breast_cancer_gene_screening} which illustrates how PARSEC's scalability to large~$p$ problems enables novel insights in breast cancer gene screening.  Thereafter, Section~\ref{sec:finance_application_mvp} demonstrates how PARSEC can be used for down-stream applications of covariance estimation in finance.

%
\subsection{Breast Cancer Gene Screening} \label{sec:breast_cancer_gene_screening}

Partial correlation graphs and network analysis are popular in cancer research as they can potentially be used to identify influential biomarkers for targeted treatment through the detection of highly-connected hubs  \parencite{GrimesPotterDatta:2019}.  Discovery of genetic alterations responsible for tumor growth or survival has revolutionized and driven much recent cancer research, and hence the identification of hub genes can provide breakthrough knowledge on how breast cancer can be better treated \parencite{GiorgioHancockBrancolini:2018}.  In particular, the reliable identification of novel genes that are pivotal in the development or prevention of the disease provides the basis for future research into potential gene therapeutic targets or treatment \parencite{LiEtAl:2018}.  The sample-starved nature of gene expression data however presents significant challenges when constructing gene dependency networks.

In the context of gene expression data, one often has to deal with over 20,000 genes in a single analysis, but with sample sizes in the hundreds or even fewer. $\ell_1$-based methods such as CONCORD are less adept at handling this many features.  Hence, the standard approach in the literature is to first reduce the number of features using Cox regression models, and then build a partial correlation graph from the reduced set of genes (see \citeauthor{KhareOhRajaratnam:2015} \citeyear{KhareOhRajaratnam:2015} and the references therein).  This heuristic step via Cox regression serves to reduce the number of features to a more manageable one ($\sim$1,000).  However, as this heuristic step is based on marginal associations between survival rates and individual genes, it may inadvertently eliminate an important gene, or a set of genes, which are jointly critical for understanding the targeted disease or outcome.  In contrast, PARSEC's immediate scalability circumvents any such heuristic reduction step.  This scalability permits us to jointly model all biomarkers simultaneously, and thus potentially discover gene expressions that have not yet been identified.

In addition to the considerable reduction in the number of genes, existing heuristic approaches determine the sparsity of the partial correlation graph by pre-specifying an $\ell_1$-penalty/threshold level or the total number of edges of the resultant graph \parencite{KhareOhRajaratnam:2015}.
 This additional second layer of heuristics associated with $\ell_1$-based methods is required as the ``ground-truth" is often not available to determine the penalty parameter through cross-validation.
Instead, PARSEC is rooted in a rigorous inferential framework which allows us to assess statistical significance - thus making the analysis more precise.  In summary, PARSEC provides three significant advantages over existing approaches in such biomedical settings: i) scalability for handling full gene expression sets, ii) theoretical guarantees in sample-deficient settings, and iii) a principled approach for assessing statistical significance of hub genes.

We now proceed to illustrate the performance of PARSEC on a popular health application adapted from a breast cancer study, on which competing $\ell_1$ methods have also been implemented - see  \textcite{KhareOhRajaratnam:2015} and references therein.  In particular, in previous applications, existing clinical information and univariate Cox regression analysis (p-value $<$ 0.0003) were used to significantly reduce the original sample to approximately 1000 genes.
PARSEC's scalability however allows us to readily consider the original full gene expression set comprised of all 24,481 gene expression levels.  Note that the data contains missing values and thus the final analysis included 15,220 genes expression levels for 174 breast cancer patients (see Appendix \ref{sec:appendix_gene_screening} for additional details).

The hub genes with the 10 highest number of edges identified using a suite of error control metrics are reported in Table \ref{tbl:breastCancerGenes} in Appendix \ref{sec:appendix_gene_screening}.  To identify hubs, we employ three approaches: i) k-FWER screening ($\alpha = 0.05$, k = $1\%$ of $p^2$), ii) FWER screening ($\alpha = 0.05$) and iii) FDR control ($\alpha = 0.05$).  As reported in Appendix \ref{sec:appendix_gene_screening}, PARSEC identifies a unique set of top hub genes not reported by other covariance estimation methods \parencite[see][]{KhareOhRajaratnam:2015}.  These top hub genes identified by PARSEC also feature prominently in recent (independent) biological studies as either being prognostic or as potential therapeutic targets.  For example, there is growing evidence around the promise of HSPG2 as a target for treatment due to its negative association with survival for patients diagnosed with Triple Negative Breast Cancer  \parencite[TNBC,][]{KalscheuerEtAl:2019}.

\noindent
\begin{minipage}{0.53\linewidth}
\begin{figure}[H]
\centering
\includegraphics[width=58mm]{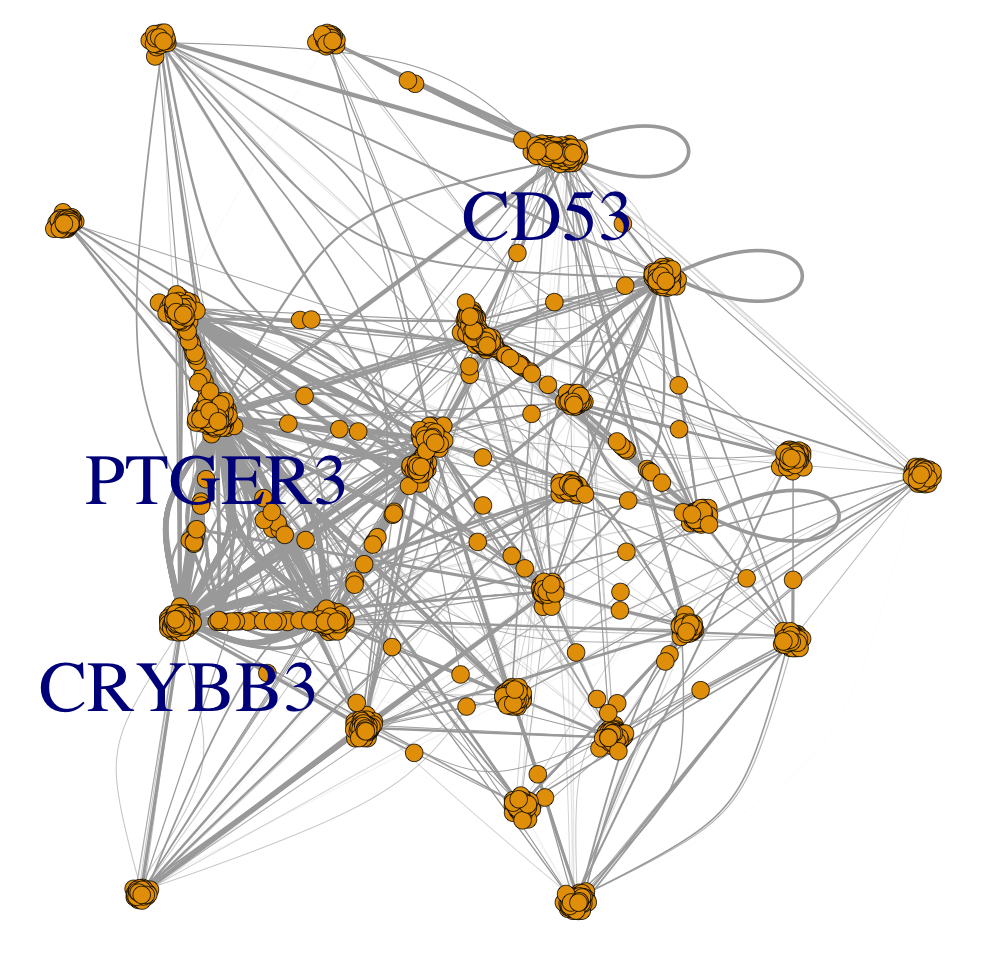}
\caption[Breast cancer gene screening network]{Network of genes screened, using k-FWER screening ($\alpha = 0.05$ and k = $1\%$ of $p^2$).  Edge weights (thickness) reflect the 1 minus p-values of partial correlation coefficients which exceed the screening level.  \label{fig:breastCancerGenesNetwork}}
\end{figure}
\end{minipage}\hfill
\begin{minipage}{0.42\linewidth}
In Figure \ref{fig:breastCancerGenesNetwork}, we present the inferred network using the k-FWER screening approach with $\alpha = 0.05$ and k = $1\%$ of $p^2$, where the edge weights represent the p-values of partial correlation coefficients which exceed the screening level.  A number of centralized gene hubs are immediately visible, such as the cluster centralized around PTGER3 (EP3).  The gene EP3 has been identified as a prognostic marker to breast cancer,  and hence the nearest genes associated with this hub have been hypothesized to repre- 

\end{minipage}

\noindent
sent strong candidates for further research into potential therapeutic targets \parencite{SemmlingerEtAl:2018}.

In summary, the biomedical significance of the top hub genes identified by PARSEC provides empirical evidence of the method's efficacy.
PARSEC's ability to simultaneously model the full gene expression set leads to the discovery of an entirely new set of hub genes previously not identified by $\ell_1$ methods (recall that such methods employ a heuristic step to achieve variable reduction).  In addition, PARSEC also successfully manages to identify the set of hub genes obtained via $\ell_1$ methods.
Appendix \ref{sec:appendix_gene_screening} provides a detailed analysis of these identified hub genes, including assessments of their statistical significance, which $\ell_1$ methods are not able to quantify.  As PARSEC decomposes the partial correlation learning problem into $p$ separate regression problems, it also allows us to undertake uncertainty quantification at the level of each hub gene and not necessarily the entire graph. In particular, we can assess the statistical significance of the strongest partial correlations for each of these biomarkers individually.

\subsection{Minimum Variance Portfolio Selection} \label{sec:finance_application_mvp}

We now demonstrate how PARSEC can be leveraged in financial portfolio selection, where a stable down-stream estimate of the (inverse) covariance matrix is a critical input in determining optimal weights.  For the sake of brevity a concise exposition is provided here - a detailed and comprehensive analysis can be found in Appendix \ref{appendix:finance}.
We consider the minimum variance portfolio framework as implemented by \textcite{WonEtAl:2013}.
Let $\mathbf{\Sigma}_t$ denote the covariance matrix of the daily returns for period~$t$. The minimum variance portfolio problem is defined as: $\text{min} \; \mathbf{w}^\top_t \mathbf{\Sigma}_t \mathbf{w}_t \; \text{subject to} \; \textbf{1}^\top  \mathbf{w}_t = 1,$
where $\mathbf{w}_t$ denotes portfolio weights, and has an analytic solution $\boldsymbol{\mathbf{w}}^*_{t} = (\textbf{1}^\top\mathbf{\Sigma}_t^{-1}\textbf{1})^{-1} \mathbf{\Sigma}_t^{-1} \textbf{1}$.  Since the data is non-stationary, a rolling high-dimensional covariance estimate is required as the effective sample size is low.  We re-estimate~$\mathbf{\Sigma}_t$ repeatedly at the beginning of each investment period~$t$, using a sample size of $n$ daily (adjusted) returns preceding the period, addressing the challenge of non-stationarity in financial returns.

We consider securities in the S\&P500 index and use a 20-year investment horizon starting from January 1, 1995 and ending on January 1, 2015. We re-estimate $\Sigma_t$ using past data from the ``estimation horizon" period, which is defined as data from the immediate past consisting of $n$ days.  These covariance estimates are then used to compute the portfolio weights, $\boldsymbol{\mathbf{w}}^*_{t}$, at the beginning of each monthly investment (or ``hold-out") period, with~$\boldsymbol{\mathbf{w}}^*_{t}$ held constant until the next investment period.  We employ the PARSEC and PCS-Hub approaches to identify significant partial correlations by controlling either FWER, k-FWER or FDR.  Once the graph structure is determined, we obtain estimates of the non-zero elements of the inverse covariance matrix by employing either a likelihood-based or a pseudo-likelihood-based estimation approach (see Appendix \ref{appendix:algs}).  We also compare PARSEC to CONCORD, which is a leading $\ell_1$-penalized pseudo-likelihood method.

\begin{table}[t]
\small
\begin{tabular}{ p{1.8cm} p{1.2cm} p{1.05cm} p{1.05cm} p{1.2cm} p{1.05cm} p{1.05cm} p{1.2cm} p{1.2cm} p{1.2cm} p{1.2cm} }
\toprule
&  & \multicolumn{5}{c}{\textbf{PARSEC}} & \multicolumn{4}{c}{\textbf{PCS-Hub}}  \\ \cmidrule(r){3-7} \cmidrule(r){8-11}
$\textbf{N}_{est}$ &  \rotatebox{90}{\parbox{2cm}{\textbf{CONCORD}, \\ CV }}  &  \rotatebox{90}{\parbox{2cm}{\textbf{FWER, \\ $\alpha = 0.05$} }} & \rotatebox{90}{\parbox{2cm}{\textbf{k-FWER, \\ $\alpha = 0.01, \\ k=\approx1000$} }} &  \rotatebox{90}{\parbox{2cm}{\textbf{k-FWER, \\ $\alpha = 0.05, \\ k\approx5000$} }} &  \rotatebox{90}{\parbox{2cm}{\textbf{FDR-BH, \\ $\alpha = 0.05$}}} & \rotatebox{90}{\parbox{2cm}{\textbf{PARSEC} \\ CV }} &  \rotatebox{90}{\parbox{2cm}{\textbf{FWER, \\ $\alpha = 0.05$} }} & \rotatebox{90}{\parbox{2cm}{\textbf{k-FWER, \\ $\alpha = 0.01, \\ k\approx1000$} }} &  \rotatebox{90}{\parbox{2cm}{\textbf{k-FWER, \\ $\alpha = 0.05, \\ k\approx5000$}}} &  \rotatebox{90}{\parbox{2cm}{\textbf{PCS-Hub}\\ CV }}   \\
\midrule
1 month & -1.1979 & \textbf{0.4244} & 0.2212 & -0.8092 & \textbf{0.4239} & 0.1066 & 0.4180 & -0.3369 & -1.1453 & -0.2099 \\  \cmidrule(r){2-11}
2 months & 0.1939 & \textbf{0.4228} & 0.2965 & -0.6791 & \textbf{0.4240} & 0.2321 & \textbf{0.4203} & -0.9000 & -0.3107 & -0.3724 \\  \cmidrule(r){2-11}
3 months & 0.1677 & \textbf{0.4203} & 0.3158 & -0.5948 & \textbf{0.4199} & 0.2404 & \textbf{0.4165} & -0.8645 & -1.0378 & -0.5830 \\  \cmidrule(r){2-11}
6 months & 0.2737 & \textbf{0.4442} & 0.3808 & 0.1275 & \textbf{0.4445} & 0.3517 & \textbf{0.4469} & -0.4426 & -0.7094 & 0.1095 \\  \cmidrule(r){2-11}
12 months & 0.2593 & \textbf{0.4621} & 0.4343 & 0.3368 & \textbf{0.4620} & 0.4226 & 0.4462 & -0.3906 & -0.8427 & -1.0158 \\
\bottomrule
\end{tabular}
\caption[Minimum Variance Portfolio Adjusted Sharpe Ratios]{Comparison of Adjusted Sharpe Ratios across different methods.  Similar to \textcite{KhareOhRajaratnam:2015}, the highest Adjusted Sharpe Ratios for each estimation horizon, and values within 1\% of this maximum, are highlighted in bold. \label{tbl:MVP_Adjusted_SharpeRatio_corr}}
\end{table}


We apply back-testing to compare the behavior of each portfolio using the performance metrics implemented in \textcite{WonEtAl:2013}.  We use the industry-standard measure, the Sharpe Ratio, to assess the effect of turnover and portfolio stability over the entire investment horizon.  The results are reported in Table~\ref{tbl:MVP_Adjusted_SharpeRatio_corr} under different estimation horizons.  It is clear PARSEC  provides uniformly competitive estimates across every estimation horizon.  Superior performance is attained when using FWER screening ($\alpha=0.05$) and FDR-BH screening ($\alpha=0.05$).  In contrast, the PCS-MPI method exhibits volatile performance.

\section{Conclusion}

PARSEC yields a novel ultra-high dimensional method for partial correlation screening with statistical error control.  It provides a highly tractable approach with superior screening and inferential performance.  
PARSEC's strength as a partial correlation screening approach also presents potential avenues for future work.  First, PARSEC's scalability can be further improved by combining PARSEC with other large-scale screening methods such as the marginal correlation screening framework.  Highly-scalable marginal correlation screening approaches could be used as a preliminary step for removing singleton features uncorrelated with the others.  Thereafter, PARSEC can  be exploited to identify significant partial correlations on a reduced subset of features within a shorter total run-time.  Second, PARSEC is ideal for ultra-high dimensional, sample-starved applications, including the climate sciences, biomedical sciences and social sciences.\\

\noindent {\bf Acknowledgments: } The authors gratefully acknowledge funding from the Australian Research Council (ARC)/Discovery Projects (DP).

\printbibliography
\end{refsection}

\begin{appendices}
\renewcommand{\theequation}{A.\arabic{equation}}
\renewcommand{\thealgocf}{A\arabic{algocf}}

\begin{refsection}

\section{Proofs}
\label{sec.proofs}

\subsection{Additional notation}
\label{sec.notation}


When $k>j$, we will slightly abuse the notation and let $\tilde{\mathbf{U}}^{-j}_{k}$ be column $k-1$ (rather than column~$k$) in the matrix~$\tilde{\mathbb{U}}^{-j}$. Such modification of the indexing ensures that~$\tilde{\mathbf{U}}^{-j}_{k}$ is the partial U-score
corresponding to the variable~$X_k$ rather than~$X_{k-1}$. We write $f_{\mathbf{U}_j,\tilde{\mathbf{U}}^{-j}_{k}}$ for the joint density of $(\mathbf{U}_j,\tilde{\mathbf{U}}^{-j}_{k})$ and let
\begin{equation*}
\overline{f_{\mathbf{U}_{\bullet},\tilde{\mathbf{U}}^{\bullet}_{*-\bullet}}}({\bf u},{\bf v}) = \tfrac{2}{p(p-1)}\sum_{j=1}^{p-1}\sum_{k=j+1}^p\big[\tfrac12f_{\mathbf{U}_j,\tilde{\mathbf{U}}^{-j}_{k}}({\bf u},{\bf v}) +\tfrac12f_{\mathbf{U}_j,\tilde{\mathbf{U}}^{-j}_{k}}(-{\bf u},{\bf v}) \big],
\end{equation*}
by analogy with the corresponding definition in \textcite{HeroRajaratnam:2011}, equation~(3.6).  We also let
\begin{equation*}
J\big(\,\overline{f_{\mathbf{U}_{\bullet},\tilde{\mathbf{U}}^{\bullet}_{*-\bullet}}}\,\big) = |S_{n-2}|\int_{S_{n-2}} \overline{f_{\mathbf{U}_{\bullet},\tilde{\mathbf{U}}^{\bullet}_{*-\bullet}}}({\bf v},{\bf v}) d{\bf v}.
\end{equation*}

\subsection{Proof of Theorem~\ref{thm.gen}}
Writing $B(\rho,{\bf v})$ for the union of the spherical cap regions on~$S_{n-2}$ centered at~${\bf v}$ and~$-{\bf v}$ with radius $\sqrt{2(1-\rho)}$, we note that $E[N_{\rho_p}]=\sum_{ j<k}P(|\mathbf{U}_j\tilde{\mathbf{U}}_k^{-j}|>\rho_p )$ and
\begin{equation}
\label{cap.prob.int}
P\big(|\mathbf{U}_j\tilde{\mathbf{U}}_k^{-j}|>\rho_p \big) = \int_{S_{n-2}} \int_{B(\rho_p,{\bf v})}f_{\mathbf{U}_j,\tilde{\mathbf{U}}^{-j}_{k}}({\bf u},{\bf v})d{\bf u} d{\bf v}.
\end{equation}
As noted in \textcite{HeroRajaratnam:2011}, $P_0$ is the proportional surface area of $B(\rho,{\bf v})$ on $S_{n-2}$, i.e., $P_0 = |B(\rho,{\bf v})|/|S_{n-2}|$.
To control the probability in display~(\ref{cap.prob.int}), we use bound~(48) in the proof of Proposition~1 in \textcite{HeroRajaratnam:2012}, which applies the mean value theorem to the corresponding integral.  This bound yields
\begin{equation}
\label{cap.prob.appr}
P\big(|\mathbf{U}_j\tilde{\mathbf{U}}_k^{-j}|>\rho_p \big)=P_0J\big(\,\overline{f_{\mathbf{U}_{j},\tilde{\mathbf{U}}^{-j}_{k}}}\big)+O\big(P_0\sqrt{1-\rho_p}\,\big),
\end{equation}
where the approximation is uniform over~$j$ and~$k$. Consequently,
\begin{equation*}
E\big[N_{\rho_p}\big]=p(p-1)P_0J\big(\,\overline{f_{\mathbf{U}_{\bullet},\tilde{\mathbf{U}}^{\bullet}_{*-\bullet}}}\,\big)+O\big(p^2P_0\sqrt{1-\rho_p}\big),
\end{equation*}
which establishes the stated bound on the expected number of discoveries. Recalling the definition of $N^*_p$, we can also write the above bound as
\begin{equation}
\label{th1.prf.exp.bnd.Nst}
E\big[N_{\rho_p}\big]=E\big[N^*_{p}\big]+O\big(p^2P_0\sqrt{1-\rho_p}\big).
\end{equation}

We write $\mathbb{I}_{n-1}$ for the $(n-1)\times(n-1)$ identity matrix. Because the marginal distribution of each U-score~$\mathbf{U}_{k}$ is uniform on the sphere $S_{n-2}$, we derive
\begin{equation*}
E\big(\mathbb{U} {\mathbb{U}}^\top\big) = p(n-1)^{-1}\mathbb{I}_{n-1}.
\end{equation*}
Given an index $j\in[p]$ we let $\mathbb{U}^{-j}$ denote matrix $\mathbb{U}$ with the $j$-th column removed.  In view of Assumption A2, and because the magnitude of each entry of the matrix $\mathbb{U}$ is bounded by one, we then have
\begin{equation}
\label{bnd.tu}
\big(\mathbb{U}^{-j} (\mathbb{U}^{-j})^\top\big)^{-1}\mathbb{U}^{-j} = (n-1)(p-1)^{-1}\big[\mathbb{I}_{n-1}+\mathbb{O}_p(\delta_p)\big]\mathbb{U}^{-j},
\end{equation}
where $\mathbb{O}_p(\delta_p)$ is an $(n-1)\times (n-1)$ matrix whose entries are $O_p(\delta_p)$, uniformly over~$j$.
Consequently,
\begin{eqnarray}
\label{bnd.u3}(\mathbb{U}^{-j})^\top\big(\mathbb{U}^{-j} (\mathbb{U}^{-j})^\top\big)^{-2}\mathbb{U}^{-j} =(n-1)^2(p-1)^{-2}\big[(\mathbb{U}^{-j})^\top\mathbb{U}^{-j}+\mathbb{O}_p(\delta_p)\big],
\end{eqnarray}
where the bound~$\mathbb{O}_p(\delta_p)$ holds uniformly over $j\le p$.   Let $\mathbb{D}_{j}$ denote a diagonal matrix whose diagonal equals to the one of the matrix $(\mathbb{U}^{-j})^\top\big(\mathbb{U}^{-j} (\mathbb{U}^{-j})^\top\big)^{-2}\mathbb{U}^{-j}$.  It follows from~(\ref{bnd.u3}) that, uniformly over~$j$,
\begin{equation}
\label{bnd.D}
\mathbb{D}^{-1/2}_j =(n-1)^{-1}(p-1)\big[\mathbb{I}_p+\mathbb{O}_p(\delta_p)\big].
\end{equation}
Because $\tilde{\mathbb{U}}^{-j} = \big(\mathbb{U}^{-j} (\mathbb{U}^{-j})^\top\big)^{-1}\mathbb{U}^{-j} \mathbb{D}^{-1/2}_j$,
we can combine bounds~(\ref{bnd.tu}) and~(\ref{bnd.D}) to derive
\begin{eqnarray*}
\max_{j\le p} \big\|\tilde{\mathbb{U}}^{-j} - \mathbb{U}^{-j}\big\|_{\infty} = O_p\big(\delta_p\big),
\end{eqnarray*}
as $p\rightarrow\infty$. Consequently, if we define
\begin{eqnarray*}
\Delta_p=\max_{j\le p} \big\| (\tilde{\mathbb{U}}^{-j})^\top\mathbf{U}_j - ({\mathbb{U}}^{-j})^\top\mathbf{U}_j \|_{\infty},
\end{eqnarray*}
then $\Delta_p=O_p\big(\delta_p\big)$.

Let $N^C_{\rho}$ denote the number of discoveries corresponding to screening level~$\rho$ when screening the elements of the sample correlation matrix~$\mathbf{R}$.  Because $\mathbf{R}=\mathbb{U}^\top\mathbb{U}$, vector $(\mathbb{U}^{-j})^\top U_j$ contains the off-diagonal elements in the $j$-th row of~$\mathbf{R}$.  It follows that $N^C_{\rho_{p}+\Delta_{p}} \le N_{\rho_{p}} \le N^C_{\rho_{p}-\Delta_{p}}$.  Furthermore, we have an analogous relationship for the tail probabilities:
\begin{equation*}
P(N^C_{\rho_{p}+\Delta_{p}}>k) \le P(N_{\rho_{p}}>k)\le P(N^C_{\rho_{p}-\Delta_{p}}>k).
\end{equation*}

Given an arbitrarily small positive~$\epsilon$, we take~$M$ large enough to ensure $P(\Delta_p>M\delta_p)\le\epsilon$ for all sufficiently large~$p$.  It follows that
\begin{eqnarray}
\label{M.eps.bound}
\sup_{k\ge0} \big|P(N_{\rho_{p}}\ge k) - P(N^*_{p}\ge k)\big|&\le&\epsilon+\sup_{k\ge0} \big|P(N^C_{\rho_{p}+M\delta_{p}}\ge k) - P(N^*_{p}\ge k)\big|\nonumber\\
&&~+\sup_{k\ge0} \big|P(N^C_{\rho_{p}-M\delta_{p}}\ge k) - P(N^*_{p}\ge k)\big|.
\end{eqnarray}

We start with the second term on the right-hand side of display~(\ref{M.eps.bound}) and note that
\begin{eqnarray}
\big|E[N^C_{\rho_{p}+M\delta_{p}}]-E[N^*_{p}]\big|&\le&
\big|E[N^C_{\rho_{p}+M\delta_{p}}]-E[N_{\rho_{p}}]\big|+\big|E[N_{\rho_{p}}]-E[N^*_{p}]\big|\nonumber\\
\label{th1.prf.totvar.terms}
&\le&E[N^C_{\rho_{p}-M\delta_{p}}]-E[N^C_{\rho_{p}+M\delta_{p}}]+\big|E[N_{\rho_{p}}]-E[N^*_{p}]\big|.
\end{eqnarray}
Using the bound
\begin{equation}
\label{P0.approx}
P_0=(n-2)^{-1}a_n(1-\rho^2)^{(n-2)/2}\big(1+O(1-\rho^2)\big),
\end{equation}
which is stated in the appendix of \textcite{HeroRajaratnam:2011}, we derive that
\begin{equation}
\label{th1.prf.exp.diff1}
E[N^C_{\rho_{p}-M\delta_{p}}]-E[N^C_{\rho_{p}+M\delta_{p}}]=O\Big(p^2 {P}_0\delta_p(1- {\rho}_{p})^{-1}\Big).
\end{equation}
Consequently, combining bounds~(\ref{th1.prf.totvar.terms}), (\ref{th1.prf.exp.diff1}) and~(\ref{th1.prf.exp.bnd.Nst}), we conclude that
\begin{equation}
\label{th1.prf.totvar.bnd1}
E[N^C_{\rho_{p}+M\delta_{p}}]-E[N^*_{p}]=O\Big(p^2 {P}_0\big[\sqrt{1-\rho_p}+\delta_p(1- {\rho}_{p})^{-1}\big]\Big).
\end{equation}

We define $\tilde{N}^*_{p}$ as a Poisson random variable with the rate equal to $E[N^C_{\rho_{p}+M\delta_p}]$.  By Corollary~3.1 in Adell and Lekuona (2005), which bounds the total variation distance between two Poisson distributions, we have
\begin{equation*}
\sup_{k\ge0} \big|P(\tilde{N}^*_{p}\ge k) - P(N^*_{p}\ge k)\big|=O\left(\frac{\big|E[N^C_{\rho_{p}+M\delta_p}]-E[N^*_{p}]\big|}{\big(1+E[N^*_{p}]\big)^{1/2}}\right).
\end{equation*}
Combining the last two bounds, we then derive
\begin{equation}
\label{Nstar.tilde.bnd}
\sup_{k\ge0} \big|P(\tilde{N}^*_{\rho_{p}}\ge k) - P(N^*_{p}\ge k)\big|=O\Big(p^2 {P}_0\big[(1- {\rho}_{p})^{1/2}+\delta_p(1- {\rho}_{p})^{-1}\big]\big(1+E[N^*_{p}]\big)^{-1/2}\Big).
\end{equation}
Finally, we control the total variation distance between $N^C_{\rho_{p}+M\delta_{p}}$ and $\tilde{N}^*_{\rho_{p}}$.  Using the bound\footnote{Inequality~(A.29) and the subsequent bounds on the terms~$b_1$, $b_2$, $b_3$.} in the proof of Proposition~1 in \textcite{HeroRajaratnam:2011} on the total variation distance between $N^C_{\rho}$ and a Poisson distribution with rate~$E[N^C_{\rho}]$, we derive
\begin{equation}
\label{Nstar.NC.bnd}
\sup_{k\ge0} \big|P(N^C_{\rho_{p}+M\delta_{p}}\ge k) - P(\tilde{N}^*_{\rho_{p}}\ge k)\big|=O\Big(p^2 {P}_0\big[l_p^2P_0+\|\Delta_{p,l_p}\|_1\big]\Big).
\end{equation}
Combining bounds~(\ref{Nstar.tilde.bnd}) and~(\ref{Nstar.NC.bnd}), we arrive at
\begin{eqnarray*}
\sup_{k\ge0} \big|P(N^C_{\rho_{p}+M\delta_{p}}\ge k) - P(N^*_{p}\ge k)\big|&=&O\Big(p^2P_0\big[l_p^2P_0+\|\Delta_{p,l_p}\|_1\big]\Big)\\
&&+\,O\Big(p^2P_0\big[\sqrt{1-\rho_{p}}+\delta_p(1-{\rho}_{p})^{-1}\big]\big(1+E[N^*_p])^{-1/2}\Big).
\end{eqnarray*}
Thus, we have bounded the second term on the right-hand side of display~(\ref{M.eps.bound}). The corresponding bound for third term, involving $N^C_{\rho_{p}}-M\delta_p$, follows by analogous arguments. Consequently, we can rewrite inequality~(\ref{M.eps.bound}) as follows:
\begin{eqnarray*}
&&\sup_{k\ge0} \big|P(N_{\rho_{p}}\ge k) - P(N^*_{p}\ge k)\big|
\,\le\\
&&~~~~~~~~~~~~~~\epsilon+O\Big(\eta_p\big[l_p^2P_0+\|\Delta_{p,l_p}\|_1+\big(\sqrt{1-\rho_{p}}+\delta_p(1-{\rho}_{p})^{-1}\big)\big(1+E[N^*_p])^{-1/2}\big]\Big).
\end{eqnarray*}
Because the last term is $o(1)$ by assumption, we have $\sup_{k\ge0} \big|P(N_{\rho_{p}}\ge k) - P(N^*_{p}\ge k)\big|\le 2\epsilon$ for all sufficiently large~$p$.  As the above argument can be repeated for every given positive~$\epsilon$,  we conclude that
\begin{equation*}
\sup_{k\ge0} \big|P(N_{\rho_{p}}\ge k) - P(N^*_{p}\ge k)\big|=o(1).
\end{equation*}

\subsection{Proof of Theorem~\ref{thm.block.sparse}}
We partition matrix $\mathbb{U}$ as $\mathbb{U}=[\mathbb{U}_1,\mathbb{U}_2]$, where $\mathbb{U}_1$ is a matrix comprised of the columns of~$\mathbb{U}$ that contribute to the ``dependent''  block of the block sparse covariance matrix~$\mathbf{\Sigma}$,  and $\mathbb{U}_2$ is  a matrix comprised of the columns of~$\mathbb{U}$ that contribute to the ``independent'' block.  We note that the dimension of $\mathbb{U}_1$ is $(n-1)\times q_p$ and the dimension of $\mathbb{U}_2$ is $(n-1)\times (p-q_p)$.

For concreteness, we will assume, without loss of generality, that the ``dependent'' block of the covariance matrix~$\mathbf{\Sigma}$ is located in the top left corner.  It follows that
\begin{eqnarray*}
E[N_{\rho_{p}}]&=&\sum_{j=1}^{p-1}\sum_{k=j+1}^p P\big(|\mathbf{U}_j\tilde{\mathbf{U}}^{-j}_k|>\rho_p \big)\\
&=&\sum_{j=1}^{q_p}\sum_{k=j+1}^p P\big(|\mathbf{U}_j\tilde{\mathbf{U}}_k^{-j}|>\rho_p \big)+\sum_{j=q_p+1}^{p-1}\sum_{k=j+1}^p P\big(|\mathbf{U}_j\tilde{\mathbf{U}}_k^{-j}|>\rho_p \big).
\end{eqnarray*}
By the approximation~(\ref{cap.prob.appr}), we have $P\big(|\mathbf{U}_j\tilde{\mathbf{U}}_k^{-j}|>\rho_p \big)=O(P_0)$, uniformly over all~$j$ and~$k$.  Furthermore, repeating the argument used in the proof of Theorem~\ref{thm.pval} below, we deduce that $P\big(|\mathbf{U}_j\tilde{\mathbf{U}}_k^{-j}|>\rho_p \big)=P_0$ when $j>q_p$.  Consequently,
\begin{equation*}
E[N_{\rho_{p}}] = \frac{p(p-1)}{2}P_0\big[1+O(q_p/p)\big],
\end{equation*}
which establishes the stated bound for the expected number of discoveries.

Because the magnitude of each entry of the matrix $\mathbb{U}$ is bounded by one, we have
\begin{equation}
\label{bnd.u1}
\mathbb{U}_1 \mathbb{U}_1^\top = \mathbb{O}\big(q_p\big),
\end{equation}
where we interpret the expression on the right-hand side as an $(n-1)\times(n-1)$ matrix whose entries are $O(q_p/p)$.
As noted in \textcite{HeroRajaratnam:2011}, the columns of the matrix~$\mathbb{U}_2$ are i.i.d. and uniform over the unit sphere~$S_{n-2}$. Thus, by the central limit theorem, we have
\begin{equation}
\label{CLT.apr.U2}
\mathbb{U}_2 \mathbb{U}_2^\top = E\big[\mathbb{U}_2 \mathbb{U}_2^\top\big] + \mathbb{O}_p\big(p^{1/2}\big).
\end{equation}
Combining bounds~(\ref{bnd.u1}) and~(\ref{CLT.apr.U2}), we conclude that
\begin{equation}
\label{U.apr.stoch.bnd}
\mathbb{U} \mathbb{U}^\top - E\big[\mathbb{U} \mathbb{U}^\top\big] = \mathbb{U}_2 \mathbb{U}_2^\top - E\big[\mathbb{U}_2 \mathbb{U}_2^\top\big] + \mathbb{U}_1 \mathbb{U}_1^\top - E\big[\mathbb{U}_1 \mathbb{U}_1^\top\big] =\mathbb{O}_p\big(p^{1/2}\big)+\mathbb{O}\big(q_p\big).
\end{equation}
Consequently, assumption A1 holds with $s_p=p^{1/2}+q_p$ and Poisson approximation~(\ref{Bl.sp.Pois.apr1}) follows from Theorem~\ref{thm.gen} by taking $\delta_p=p^{-1/2}+q_p/p$ and $l=q_p$, after noting that the average dependency coefficient, $\|\Delta_{p,q_p}\|_1$, is equal to zero.

It is only left to establish
\begin{equation}
\label{Bl.sp.Pois.apr2}
\sup_{k\in\mathbb{N}}\left|P(N_{\rho_{p}}>k)-P(N^*_p>k)\right| = O\left(\frac1{p}+\eta_p^2\Big(\frac{q_p}{p}\Big)^2+{\sqrt{\eta_p}\Big[\frac{\sqrt{\log(p)}}{\sqrt{p}}+\frac{q_p}{p}\Big]}(1-{\rho}_{p})^{-1}\right).
\end{equation}
Applying Hoeffding's inequality, we convert the stochastic bound~(\ref{CLT.apr.U2}) into a non-stochastic bound that holds with high probability.  More specifically, there exists a constant~$C$, such that for all sufficiently large~$p$ inequality
\begin{equation}
\label{CLT.apr.U2.Hoef}
\Big|\mathbb{U}_2 \mathbb{U}_2^\top - E\big[\mathbb{U}_2 \mathbb{U}_2^\top\big]\Big| \le C\sqrt{p\log(p)}
\end{equation}
holds with probability at least $1-1/p$. As a consequence, stochastic bound~(\ref{U.apr.stoch.bnd}) becomes non-stochastic:
\begin{equation*}
\Big|\mathbb{U} \mathbb{U}^\top - E\big[\mathbb{U} \mathbb{U}^\top\big]\Big| = \mathbb{O}\big(\sqrt{p\log(p)}+q_p\big),
\end{equation*}
with probability at least $1-1/p$. Repeating the argument in the proof of Theorem~\ref{thm.gen} that starts with equation~(\ref{M.eps.bound}), but taking $\epsilon=1/p$, $l_p=q_p$ and $\delta_p=\sqrt{\log(p)/p}+q_p/p$, we arrive at the stated approximation~(\ref{Bl.sp.Pois.apr2}).

\subsection{Result 3: Limiting behavior of $\rho_p$} \label{appendix:thm.finite.lim}

The next theorem corresponds to Result 3 described in Section \ref{sec:theoretical_results}.  Here, we focus on the
setting where the expected number of discoveries converges to a finite limit as $p$ goes to infinity and specify the required limiting behavior of the screening level~$\rho_p$.

\begin{theorem}
\label{thm.finite.lim}
Let~$\kappa_n = a_n/[n-2]$.  Suppose that $n>10$, assumption~$A1$ holds, $\mathbf{\Sigma}$ is block-sparse of degree~$q_p=o(p)$, and $p^2(1-\rho^2_{p})^{(n-2)/2}\rightarrow e_n$ as $p\rightarrow\infty$.  Then,
\begin{equation*}
E[N_{\rho_{p}}]= e_n\kappa_n/2+o(1).
\end{equation*}
\noindent Now, let~$N^*$ denote a Poisson distributed random variable with rate $E[N^*_{n}]=e_n\kappa_n/2$ for some finite constant~$e_n$.  If $q_p=O(\sqrt{p})$, then
\begin{equation*}
\sup_{k\in\mathbb{N}}\left|P(N_{\rho_{p}}>k)-P(N^*>k)\right| = o(1).
\end{equation*}
\noindent In particular,
\begin{equation*}
P(N_{\rho_{p}}>0)\rightarrow 1-\exp\big(-e_n\kappa_n/2\big) \quad \text{as}\quad p\rightarrow\infty.
\end{equation*}
\end{theorem}
\begin{remark}
The result in Theorem~\ref{thm.finite.lim} subsumes the case where~$\mathbf{\Sigma}$ is diagonal, which is the setting of Corollary~\ref{thm.ell.cont}.  However, Corollary~\ref{thm.ell.cont} allows the expected number of discoveries to go off to infinity, while the conditions in Theorem~\ref{thm.finite.lim} require it to converge to a finite constant.
\end{remark}
\begin{remark}
Our methods of proof in this section go well beyond a simple application of the union bound together with the error control for the row-wise screening of the elements of matrix~$\mathbf{H}$. In fact, we demonstrate in Appendix~\ref{append.union} that the error control corresponding to the latter approach is overly conservative.
\end{remark}

\subsubsection{Proof of Theorem~\ref{thm.finite.lim}}

We now provide the proof of Theorem~\ref{thm.finite.lim}.  In view of bound~(\ref{P0.approx}), the imposed assumption on~$\rho_p$ and $q_p$ imply $\eta_p=e_n\kappa_n/2+o(1)$.  Hence, an application of Theorem~\ref{thm.block.sparse} yields
$E[N_{\rho_p}]=\eta_p[1+o(1)]=e_n\kappa_n/2+o(1)$, which establishes the stated limiting result for the expected number of discoveries.

The imposed assumptions also imply that $(p-1)P_0\le1$ and
\begin{equation*}
\eta_p^2(q_p/p)^2+\sqrt{\eta_p}\big[1/\sqrt{p}+q_p/p\big](1-{\rho}_{p})^{-1}=o(1).
\end{equation*}
Hence, applying Theorem~\ref{thm.block.sparse} again we derive
\begin{equation*}
\sup_{k\in\mathbb{N}}\left|P(N_{\rho_{p}}>k)-P(N^*_p>k)\right| = o(1),
\end{equation*}
where~$N^*_{p}$ is a Poisson random variable with $E[N^*_{p}]=\eta_p=e_n\kappa_n/2+o(1)$.  Because $E[N^*_{p}]=E[N^*]+o(1)$, an application of the bound on the total variation distance between two Poisson distributions yields
\begin{equation*}
\sup_{k\in\mathbb{N}}\left|P(N_{\rho_{p}}>k)-P(N^*>k)\right| = o(1).
\end{equation*}

\subsection{Proof of Theorem~\ref{thm.pval}}

As noted in \textcite{HeroRajaratnam:2011}, the diagonal form of~$\mathbf{\Sigma}$ implies that U-scores $\mathbf{U}_j$ are i.i.d uniform on the sphere~$S_{n-2}$. Because $\tilde{\mathbf{U}}^{-j}_k$ is a function of the U-scores other than $\mathbf{U}_j$, we deduce that partial U-score $\tilde{\mathbf{U}}^{-j}_k$ and U-score~$\mathbf{U}_j$ are independent. Writing, as before, $B(\rho,{\bf v})$ for the union of the spherical cap regions on~$S_{n-2}$ centered at~${\bf v}$ and~$-{\bf v}$ with radius $\sqrt{2(1-\rho)}$, we recall that $|B(\rho,{\bf v})|=P_0(\rho,n)|S_{n-2}|$, for every~${\bf v}\in S_{n-2}$. Consequently,
\begin{eqnarray*}
P\big(|H_{jk}|>\rho \big) &=&P\big(|\mathbf{U}_j\tilde{\mathbf{U}}_k^{-j}|>\rho \big)=E\left[P\Big(\mathbf{U}_j\in B(\rho,\tilde{\mathbf{U}}_k^{-j})\Big|\tilde{\mathbf{U}}_k^{-j}\Big)\right]=|B(\rho,{\bf v})|/|S_{n-2}|\\
&=&P_0(\rho,n).
\end{eqnarray*}

\subsection{Error control for the union-bound approach}
\label{append.union}

The next result uses Proposition~1 in \textcite{FirouziEtAl:2017}, together with the union bound approach, to establish an asymptotic upper bound on the FWER as $p$ tends to infinity.
\begin{proposition}
Consider a fixed $\alpha\in(0,1)$.  Suppose that $p^2(1-\tilde\rho^2_p)^{(n-2)/2}\rightarrow -\log(1-\alpha)/\kappa_n$ as $p\rightarrow\infty$.  If assumptions $A1$ and $A2$ hold, then:
$$
\limsup_{p\rightarrow\infty} P(N_{\tilde\rho_p}>0)\le\alpha.
$$
\end{proposition}

We now apply Theorem~\ref{thm.finite.lim} to derive the exact limiting behavior for the FWER corresponding to the screening level $\tilde\rho_p$ in Proposition~1.
\begin{proposition}
Under the assumptions of Theorem~\ref{thm.finite.lim},
$$
P(N_{\tilde\rho_p}>0)\rightarrow 1 - \exp(-\alpha/2) \quad \text{as}\quad p\rightarrow\infty.
$$
\end{proposition}
\noindent We note that $1 - \exp(-\alpha/2) < \alpha/2$ for $\alpha\in(0,1)$, so the bound in Proposition~1 is overly conservative.

\subsection{Numerical validation and phase transitions} \label{appendix:phase_transitions}

We now proceed to obtain a better understanding of the accuracy of the theoretical expressions in Section~\ref{sec:theoretical_results}, by comparing them to the quantity they aim to approximate: the true number of false discoveries.  The true number of false discoveries in the null model can be obtained through examining the empirical behavior of the number of PARSEC discoveries as we vary the screening level~$\rho$.  More specifically, we generate samples  from the null model where $\Sigma=\mathbf{I}$, performing 1000 replications for each setting of~$n$ and~$p$. Figure~\ref{fig:hub_transitions} in Section \ref{sec:theoretical_results} plots the (empirical) median proportion (over the replicates) of the features with at least one falsely identified partial correlation, together with the corresponding theoretical approximation implied by Theorem 3 for various values of~$n$ and~$p$.  It is abundantly clear that the theoretical approximations almost exactly mirror the truth, even when $p$ is as small as 50.  These figures thus provide a compelling numerical validation of PARSEC's theoretical properties.  Note also that both curves exhibit an interesting phase transition.

\newpage

\section{Algorithms} \label{appendix:algs}

\subsection{Base PARSEC algorithm} \label{appendix:base_algorithm}

{\SetKwProg{myproc}{Procedure}{}{end}
\small
\begin{algorithm}[h]
 \caption{{\bf Base PARSEC algorithm (direct evaluation of $\mathbf{H}$)} \label{alg:directeval_H}}
\KwIn{$\mathbb{X}_{n\times p}$}
\KwOut{$\mathbf{H}_{p\times p}$}
\Begin{
\For{$j=1$ \KwTo p}{
$ \displaystyle \mathbf{Z}_j = \dfrac{\mathbf{X}_j - \bar{X}_j\mathbf{1}}{\sqrt{\mathbf{S}_{jj}(n-1)}}$\;}
Define $\mathbb{T}= [n^{-1/2}\mathbf{1}, \mathbf{T}_{2:n}]$, where $\mathbf{1}^\top \mathbf{T} = [\sqrt{n},0,...,0]$ and $\textbf{T}_{2:n}^\top \textbf{T}_{2:n}=\textbf{I}_{n-1}$\;
 Obtain $ \displaystyle \mathbb{U}_{n-1 \times p} =\mathbf{T}^\top_{2:n} \mathbb{Z}$\;
 \For{$j=1$ \KwTo p}{
 Define $\displaystyle \tilde{\mathbb{U}}^{-j}=(\mathbb{U}^{-j}(\mathbb{U}^{-j})^\top)^{-1} \mathbb{U}^{-j} \mathbf{D}^{-\frac{1}{2}}_{ {\mathbb{U}^{-j}}^\top[ {\mathbb{U}^{-j}} {\mathbb{U}^{-j}}^\top]^{-2} {\mathbb{U}^{-j}}}$\;
 Estimate $\displaystyle \big(H_{j1},\dots,H_{j(j-1)},H_{j(j+1)},\dots,H_{jp}\big)^\top = (\tilde{\mathbb{U}}^{-j})^\top \mathbf{U}_j$\; }}
  \end{algorithm}}

\subsection{Scalable FDR algorithm} \label{appendix:scalable_FDR_algorithm}

Recall, we observe that the FDR control method described in Section \ref{uncertainty.subsec} requires the calculation and ordering of $\binom{p}{2}$ p-values, and thus can be computationally expensive in ultra-high dimensional settings.  To address this challenge, we now present a scalable iterative procedure to determine the BH/BY rejection level.  First, we set an initial screening level~$\tilde{\rho}_{(0)}$, which we determine by equating~$P_0$ to the given FDR level~$\alpha$. Next, we pre-screen $\mathbf{H}$ using screening level~$\tilde{\rho}_{(0)}$, which serves to significantly reduce the number of p-values under consideration. As the initial screening level is too liberal, it is easily shown that this pre-screening procedure retains all the partial correlation estimates that exceed the FDR level (for either the BH or BY procedure). This process is then repeated.  In particular, we iteratively decrease the BH/BY rejection level (which corresponds to increasing the partial correlation screening level); reducing the number of p-values under consideration until the BH or BY rejection level is found.  We note that this procedure is equivalent to implementing the FastLSU approach of \textcite{MadarBatista:2016} in the partial correlation screening setting.
 Algorithm~\ref{alg:fdr_proc} provides the specific implementation details. Table~\ref{tbl:timing_fdr_rank_1} (b) in Section~\ref{sec:simulated_data} reports wall-times, demonstrating significant improvements from using this approach.
{\begin{algorithm}[h]
\small
 \caption{{\bf FDR iterative procedure} \label{alg:fdr_proc}}
\KwIn{$\mathbf{H}_{p\times p}$, FDR method, FDR level $\alpha$ }
\KwOut{Screened scaled partial correlation matrix $\tilde{\mathbf{H}}_{p\times p}$}
\Begin{
\uIf{FDR method = FDR-BH}{
   $\displaystyle m_0=p(p-1)/2$\;
   }\ElseIf{FDR method = FDR-BY}{
   $\displaystyle m_0=(\sum^m_{k=1}\dfrac{1}{k})p(p-1)/2$\;
  }
Solve for $\tilde{\rho}_{(0)}$ where $\displaystyle \alpha = a_n \int^1_{\tilde{\rho}_{(0)}}(1-u^2)^{(n-4)/2}du$ and $\displaystyle a_n= \dfrac{2\Gamma ((n-1)/2)}{\sqrt{\pi}\Gamma((n-2)/2)}$\; \label{lst:line:spherical_cap}
Set $\displaystyle m_{1} = \sum_{k,j : j>k} \mathbf{1}(|H_{kj}| > \tilde{\rho}_{(0)})$\;
Set $t=1$\;
\While{$m_t \neq m_{t-1}$}{
Set $\tilde{\rho}_{(t)}$ by solving $\displaystyle \dfrac{m_{t}}{m_0}\alpha = a_n \int^1_{\tilde{\rho}_{(t)}}(1-u^2)^{(n-4)/2}du$ (using $a_n$ defined in line \ref{lst:line:spherical_cap})\;
Set $\displaystyle m_{t+1}= \sum_{k,j : j>k} \mathbf{1}(|H_{kj}| > \tilde{\rho}_{(t)})$\;
Update $t=t+1$\;
}
Set $\tilde{\mathbf{H}} = \mathbf{I}_p$\;
Update $\tilde{H}_{kj} = H_{kj}\mathbf{1}(|H_{kj}| > \tilde{\rho}_{(t)}) \quad \forall k,j$\;}
\end{algorithm}}

\clearpage
\newpage

\subsection{Methods for inverse covariance estimation} \label{appendix:inv_cov_estim}

{
\begin{algorithm}[h]
\small
 \caption{{\bf Inverse covariance estimation via CONCORD-based coordinate-wise descent \parencite{KhareOhRajaratnam:2015}} \label{alg:concord_coordinatewise}}
\KwIn{$\hat{\mathbb{E}}_{p\times p}$, screened partial correlation structure}
\KwIn{$\mathbf{S}$, sample covariance matrix}
\KwIn{$\epsilon$, convergence threshold}
\KwOut{$\hat{\Sigma}$}
\KwOut{$\hat{\Sigma}^{-1}$}
\Begin{
Initialize $\textbf{W}=\mathbf{S}$\;
 \For{$i=1$ \KwTo p}{
 \If{$\displaystyle \sum^p_{j=1}\hat{E}_{ij}=1$}{
 	Remove $i^{th}$ row and $i^{th}$ column from $\mathbf{W}$\;
 }}
 Initialize $\displaystyle \textbf{W}^0=\textbf{W}$ and $\textbf{W}^1=\mathbf{W}$\;
 \While{$\displaystyle \sum^p_{i=1}\sum^p_{j=1} \text{max} (| W^0_{ij}-W^1_{ij} |) > \epsilon$}{
Set $\textbf{W}^0=\textbf{W}^1$\;
 \For{$i=1$ \KwTo p}{
    \For{$j=1$ \KwTo p}{
    \If{$e_{ij}>0$}{
        Set $\omega^1_{ij} = \dfrac{-(\sum_{j'\neq j} \omega^1_{ij'} s_{jj'} e_{ij'} + \sum_{i'\neq i} \omega^1_{ij'} s_{ii'} e_{ij'}}{s_{ii}+s_{jj}}$ \; }
        Set $\omega^1_{ii} = \dfrac{-\sum_{j \neq i} \omega^1_{ij} s_{ij} e_{ij} + \sqrt{(\sum_{j\neq i} \omega^1_{ij} s_{ij} e_{ij})^2 +4s_{ii}}}{2s_{ii}}$ \;
    }
}
}
Set $\hat{\Sigma} = \textbf{W}^1$ and $\hat{\Sigma}^{-1} = \hat{\Theta}$ by re-padding $\hat{\Sigma}$ and $\hat{\Sigma}^{-1}$ with zeros corresponding to inactive edges in $\hat{\mathbb{E}}$}
\end{algorithm}}

{
\begin{algorithm}[h]
\small
\caption{{\bf Covariance and inverse covariance estimation via coordinate-wise descent \parencite{HastieTibshiraniFriedman:2009}} \label{alg:esl_coordinatewise}}
\KwIn{$\hat{\mathbb{E}}_{p\times p}$}
\KwIn{$\mathbf{S}$, sample covariance matrix}
\KwIn{$\epsilon$, convergence threshold}
\KwOut{$\hat{\Sigma}$}
\KwOut{$\hat{\Sigma}^{-1}$}
\Begin{
Initialize $\textbf{W}=\mathbf{S}$\;
\For{$i=1$ \KwTo p}{
\If{$\displaystyle \sum^p_{j=1}\hat{E}_{ij}=1$}{
Remove $i^{th}$ row and $i^{th}$ column from $\mathbf{W}$\;
}}
Initialize $\displaystyle \textbf{W}^0=\textbf{W}$ and $\textbf{W}^1=\mathbf{W}$\;
\While{$\displaystyle \sum^p_{i=1}\sum^p_{j=1} | W^0_{ij}-W^1_{ij} | > \epsilon$}{
Set $W^0=W^1$\;
\For{$j=1$ \KwTo p}{
Partition matrix $\textbf{W}^0$ into part 1; comprising all but the $j$th row and column ($\textbf{W}_{11}$), and part 2; the $j$th row and column, ($\textbf{W}_{22}$)\;
Solve $\displaystyle \textbf{W}_{11}^*\beta^* - s_{12}^* = 0$ for unconstrained edge parameters $\beta^*$, where $\textbf{W}^*_{11}$ and $s_{12}^*$ correspond to active edges in $\hat{\mathbb{E}}^*$\;
Update $\textbf{W}^1_{12}=\textbf{W}_{11}\hat{\beta}$\;
Update inverse covariance matrix estimate $\hat{\Theta}$, by solving $\displaystyle \hat{\theta}_{12} = -\hat{\beta}.\hat{\theta}_{22}$ where $\displaystyle 1/\hat{\theta}_{22} = s_{22} - w^\top_{12}\hat{\beta}$\;}
}
Set $\hat{\Sigma} = \textbf{W}^1$ and $\hat{\Sigma}^{-1} = \hat{\Theta}$ by re-padding $\hat{\Sigma}$ and $\hat{\Sigma}^{-1}$ with zeros corresponding to inactive edges in $\hat{\mathbb{E}}$}
\end{algorithm}
}

\clearpage
\newpage

\section{Computational Complexity} \label{appendix:comp_complexity}

\noindent \textbf{Proof of Theorem~\ref{thm:comp_complexity}}.  Given any matrix $\mathbb{M}$, we will write $\mathbf{M}_j$ for the $j$-th column of $\mathbb{M}$, and we will write $M_{j,k}$ for the element of $\mathbb{M}$ located in $j$-th row and $k$-th column. The off-diagonal elements in the $j$-th row of matrix $\mathbf{H}$ are given by the vector $\mathbf{H}^{j}=\big(\tilde{\mathbb{U}}^{-j}\big)^\top\mathbf{U}_j$, where $\tilde{\mathbb{U}}^{-j}=\mathbb{C}_j\mathbb{D}_j^{-1/2}$, with $\mathbb{C}_j=\big(\mathbb{U}^{-j} (\mathbb{U}^{-j})^\top\big)^{-1}\mathbb{U}^{-j}$ and $\mathbb{D}_j$ denoting a diagonal matrix whose diagonal equals to that of the matrix $(\mathbb{U}^{-j})^\top\big(\mathbb{U}^{-j} (\mathbb{U}^{-j})^\top\big)^{-2}\mathbb{U}^{-j}$.  Here, $\mathbb{U}^{-j}$ is the $U$-score matrix $\mathbb{U}$ without the $j$-th column, and $\mathbf{U}_j$ is the $j$-th column of $\mathbb{U}$. Matrix $\mathbb{U}$ is computed by excluding the first row of the matrix $\mathbb{L}^\top\mathbb{Z}$, where $\mathbb{Z}$ is the standardized data matrix~$\mathbb{X}$, and $\mathbb{L}$ is an orthogonal $n\times n$ matrix, whose first column is a vector with equal positive elements.

We first focus on the PARSEC implementation that uses the rank-one updates. Using the Sherman–Morrison formula, and defining  $\mathbf{A}=\big(\mathbb{U} \mathbb{U}^\top\big)^{-1}$, we derive
\begin{equation*}
\Big(\mathbb{U}^{-j} (\mathbb{U}^{-j})^\top\Big)^{-1} = \Big(\mathbb{U} \mathbb{U}^\top - \mathbf{U}_j{\mathbf{U}_j}^\top \Big)^{-1} = \mathbf{A}+\Big[\frac1{1-\mathbf{U}_j^\top\mathbf{A}\mathbf{U}_j}\Big]\mathbf{A}\mathbf{U}_j\mathbf{U}_j^\top\mathbf{A}.
\end{equation*}
Recalling the notation $\mathbf{B} = \mathbb{U}^\top \mathbf{A} \mathbb{U}$ and $\mathbf{F} = \mathbf{A} \mathbb{U}$, we conclude that for $k<j$, the $k$-th diagonal element of $\mathbb{D}_j^{-1/2}$ is given by
\begin{equation*}
\Big\| \mathbf{F}_k + \dfrac{B_{k,j}}{1-B_{j,j}}\mathbf{F}_j \Big\|_2^{-1}.
\end{equation*}
When~$k\ge j$, the above formula needs to be adjusted by replacing~$k$ with $k-1$.
The off-diagonal elements in the $j$-th row of matrix $\mathbf{H}$ are given by
$\mathbf{H}^{j}=\big(\tilde{\mathbb{U}}^{-j}\big)^\top\mathbf{U}_j  = \mathbb{D}_j^{-1/2} \mathbb{C}_j^\top \mathbf{U}_j
$.
We rewrite the above equation as follows:
\begin{align*}
\mathbb{D}_j^{-1/2} \mathbb{C}_j^\top \mathbf{U}_j &= \mathbb{D}_j^{-1/2} \bigg( \big(\mathbf{A}+\Big[\frac1{1-\mathbf{U}_j^\top\mathbf{A}\mathbf{U}_j}\Big]\mathbf{A}\mathbf{U}_j\mathbf{U}_j^\top\mathbf{A}\big) \mathbb{U}^{-j} \bigg)^\top \mathbf{U}_j\\ &= \mathbb{D}_j^{-1/2} (\mathbb{U}^{-j})^\top \big(\mathbf{A}+\Big[\frac1{1-\mathbf{U}_j^\top\mathbf{A}\mathbf{U}_j}\Big]\mathbf{A}\mathbf{U}_j\mathbf{U}_j^\top\mathbf{A}\big) \mathbf{U}_j \\
&= \mathbb{D}_j^{-1/2} (\mathbb{U}^{-j})^\top\mathbf{A}\mathbf{U}_j +\mathbb{D}_j^{-1/2} \Big[\frac1{1-\mathbf{U}_j^\top\mathbf{A}\mathbf{U}_j}\Big](\mathbb{U}^{-j})^\top \mathbf{A}\mathbf{U}_j\mathbf{U}_j^\top\mathbf{A}\mathbf{U}_j
\end{align*}
We note that for $k<j$, the $k$-th element of the above vector is given by
\begin{equation*}
B_{k,j} + \dfrac{B_{k,j}B_{j,j}}{1-B_{j,j}} = \dfrac{B_{k,j}}{1-B_{j,j}}.
\end{equation*}
When~$k\ge j$, we again need to adjust the above formula by replacing~$k$ with $k-1$.
%
Thus, we can compute each off-diagonal element of the matrix~$\mathbf{H}$ via
\begin{equation*}
H_{jk} = \Big\|\frac{1-B_{j,j}}{B_{k,j}}\mathbf{F}_k+\mathbf{F}_j \Big\|^{-1}_2.
\end{equation*}

\color{black}

\noindent To derive PARSEC's computational complexity, we break down the above formula into nine steps:
\begin{enumerate}
\item We compute the matrix $\mathbb{Z}$ by standardizing each of the~$p$ columns of the $n\times p$ data matrix~$\mathbb{X}$.  Standardizing each column requires $O(n)$ operations, and hence we obtain $\mathbb{Z}$ in $O(np)$ operations.

\item We calculate $\mathbb{L}$ by the Gramm-Schmidt orthogonolization of an $n\times n$ matrix, which requires $O(n^3)$ operations.

\item To compute $\mathbb{L}^\top\mathbb{Z}$ we multiply an $n\times n$ matrix by an $n\times p$ matrix, which requires $O(n^2p)$ operations.

\item To compute $\mathbb{U}\mathbb{U}^\top$, we multiply an $(n-1)\times p$ matrix by its transpose, which is a $p\times (n-1)$ matrix.  This requires~$O(n^2p)$ operations.

\item To compute~$\mathbf{A}$, we invert an $(n-1)\times (n-1)$ matrix $\mathbb{U} (\mathbb{U})^\top$, which requires~$O(n^3)$ operations.

\item To compute $\mathbf{F}$, we multiply an $(n-1)\times (n-1)$ matrix~$\mathbf{A}$ by an $(n-1)\times p$ matrix~$\mathbb{U}$, which requires~$O(n^2p)$ operations.

\item To compute $\mathbf{B}$, we multiply a $p\times(n-1)$ matrix~$\mathbb{U}^\top$ by an $(n-1)\times p$ matrix~$\mathbf{F}$, which requires~$O(np^2)$ operations.

\item Once we have obtained matrixes~$\mathbf{B}$ and~$\mathbf{F}$, we compute each element of~$\mathbf{H}$ in~$O(1)$ operations.  Thus, we obtain the entire matrix~$\mathbf{H}$ in~$O(p^2)$ operations.

\item Finally, we reconcile the differences in the two estimates given by~$\mathbf{H}$ for each pairwise partial correlation and then screen the results at a given level~$\rho$.  This requires~$O(p^2)$ operations.
\end{enumerate}

Combining the operations in the steps listed above, we see that it takes $O(np^2+n^2p+n^3)$ operations to compute, reconcile and screen all the elements of matrix~$\mathbf{H}$.  If the computation of matrixes~$\mathbf{F}$ and~$\mathbf{B}$ in steps~6 and~7, respectively, is parallelized over the rows, then it takes $O(n^2p+n^3)$ operations to compute the $\mathbf{H}$ matrix.  Similarly, parallelizing the reconciliation and the screening steps reduces the corresponding order operations from~$O(p^2)$ down to~$O(p)$. \qed

\clearpage
\newpage

\section{Storage and Memory Complexity} \label{appendix:storage_complexity}

PARSEC is also amenable to alternative data structures, which provide storage gains (improved ``memory/storage complexity").
As a concrete example, the one row at a time implementation together with screening at estimation allows for scaled partial correlation coefficients to be instead stored in a compressed sparse matrix or, alternatively, as a list with row and column indices as keys.  As $p$ increases, corresponding memory gains can be considerable as we circumvent the need to store a large dense $p\times p$ matrix.  Such improvements in memory/storage complexity can be highly beneficial, especially when multiple partial correlation matrix estimates are required, for example, when re-sampling or leave-one-out influence/outlier detection is undertaken.


\section{Simulated Data in the Multivariate Normal setting} \label{appendix:simulated}

In this Appendix section, we provide additional results from our numerical investigations initially outlined in Section \ref{sec:simulation_inferential}.  Section \ref{appendix:pfdr_null} outlines PARSEC's positive false discovery rate (pFDR) control in the null setting.  Section \ref{appendix:type_2_error} then provides a detailed explanation and additional findings of Type II error assessments.

\vspace{-0.1in}

\subsection{Positive False Discovery Rate control in the null setting} \label{appendix:pfdr_null}

In addition to controlling the false discovery rate (FDR), PARSEC can control the pFDR using exact marginal p-values outlined in Theorem \ref{thm.pval}.  We provide assessments of PARSEC's pFDR error control in Table \ref{tbl:sim_pfdr_ec_null}.  PARSEC's pFDR control is consistent with the pre-specified significance level $\alpha$ in each setting.  PARSEC's ability to provide error control across multiple measures (including FWER, k-FWER, FDR and pFDR) again highlights its adaptability as a modern ultra-high dimensional inferential approach.

{
\begin{table}[!ht]
\centering
\small
\begin{tabular}{ll rr  rr}
\toprule
& &  \multicolumn{2}{c}{\textbf{p=$10^3$}} &  \multicolumn{2}{c}{\textbf{p=$10^4$}} \\ \cmidrule(r){3-4} \cmidrule(r){5-6}
 & &  PARSEC & PCS-Hub &  PARSEC & PCS-Hub \\
\midrule
 \multicolumn{6}{l}{\textbf{pFDR (averaged over all replications)}} \\ \cmidrule(r){2-6}
pFDR & $\alpha=0.01$ & 0.012 &  d.n.e  & 0.016 & d.n.e  \\
& $\alpha=0.05$ & 0.051 &  d.n.e  &  0.060 & d.n.e  \\
\bottomrule
\end{tabular}
\caption[Simulation performance, error control as p increases]{Positive False Discovery Rate (pFDR) control with fixed $n$ ($n=30$) and varying $p$ in null models. Performance measures are evaluated over 1000 replications for each dimension. ``d.n.e." denotes `does not exist' as the PCS-Hub approach does not readily provide FDR control.
\label{tbl:sim_pfdr_ec_null}}
\end{table}
}

\vspace{-0.2in}

\subsection{Type II error/statistical power assessments} \label{appendix:type_2_error}

In Table \ref{tbl:sim_p1000_MCC}, we compare PARSEC and PCS-Hub in terms of the MCC metric: a consolidated performance measure which integrates both the false positive and false negative rates.  We observe that PARSEC uniformly obtains higher MCC values across all four error control measures, across all models, and across all signal strengths.
Notably, PARSEC substantially outperforms PCS-Hub even when covariance signal strength is lower ($\sigma,\phi_1=0.6$).  PARSEC's stronger identification of true partial correlation signal even when the covariance signal is low demonstrates its improved consistency as a partial correlation screening approach.

{
\begin{table}[!t]
\centering
\small
\begin{tabular}{llllllll}
  \hline
\textbf{MCC} &   &  \multicolumn{2}{c}{$\sigma,\phi_1=0.6$} & \multicolumn{2}{c}{$\sigma,\phi_1=0.7$} & \multicolumn{2}{c}{$\sigma,\phi_1=0.8$}\\ \cmidrule(r){3-4} \cmidrule(r){5-6} \cmidrule(r){7-8}
 & \textbf{EC method} & {PARSEC} & {PCS-Hub} &  {PARSEC} & {PCS-Hub} &  {PARSEC} & {PCS-Hub} \\
  \hline
 \textbf{Block}  \\
a=20, n=40	& FDR-BH & 	\textbf{0.073} & 	N/A & 	\textbf{0.411} & 	N/A & \textbf{0.900} & N/A \\
& pFDR & 	\textbf{0.073} &  N/A & 	\textbf{0.428} & 	N/A & \textbf{0.904} & N/A \\
 & 	FWER  & 	\textbf{0.051} & 	0.000 & 	\textbf{0.162} & 	0.051 & \textbf{0.533} & 0.218 \\
 & 	k-FWER &	\textbf{0.161} &	0.098 & 	\textbf{0.192} & 	0.151 & \textbf{0.196} & 0.191 \\ \cmidrule(r){2-8}
\multicolumn{2}{l}{\textbf{AR(1) Block}} \\
a=50, n=30 & FDR-BH & 	\textbf{0.101} &  N/A & 	\textbf{0.431} & 	N/A & \textbf{0.771} & N/A \\
& pFDR & 	\textbf{0.000} &  N/A & 	\textbf{0.433} & 	N/A & \textbf{0.771} & N/A \\
 & 	FWER & 	\textbf{0.000} & \textbf{0.000} & 	\textbf{0.286} & 	0.247 & \textbf{0.590} & 0.535 \\
 & 	k-FWER & 	\textbf{0.083} &	0.079 & 	\textbf{0.098} &	0.096 & \textbf{0.100} & 0.099  \\ \cmidrule(r){3-8}
a=100, n=30	& 	FDR-BH & 	\textbf{0.142} & 	N/A & 	\textbf{0.505} & 	N/A & \textbf{0.787} & N/A \\
& pFDR & 	\textbf{0.142} &  N/A & 	\textbf{0.503} & 	N/A & \textbf{0.786} & N/A \\
 & 	FWER & 	\textbf{0.100} & 	\textbf{0.100} & 	\textbf{0.268} & 	0.246 & \textbf{0.586} & 0.532 \\
 & 	k-FWER & 	\textbf{0.118} &	0.112 & 	\textbf{0.139} &	0.134 & \textbf{0.143} & 0.139 \\ \cmidrule(r){3-8}
a=50, n=50	& 	FDR-BH & \textbf{0.554} & N/A & \textbf{0.843} & N/A & \textbf{0.768} & N/A \\
 & pFDR & \textbf{0.554} & N/A & \textbf{0.840} & N/A & \textbf{0.770} & N/A \\
 & 	FWER & \textbf{0.378} & 0.286 & \textbf{0.700} & 0.589 & \textbf{0.885} & 0.865 \\
 & 	k-FWER  & \textbf{0.098} & 0.095 & \textbf{0.101} & 0.099 & \textbf{0.101} & 0.099\\ \cmidrule(r){3-8}
a=100, n=50	& 	FDR-BH & \textbf{0.620} & N/A & \textbf{0.853} & N/A & \textbf{0.749} & N/A \\
 & pFDR & \textbf{0.621} & N/A & \textbf{0.852} & N/A & \textbf{0.750} & N/A \\
 & 	FWER & \textbf{0.376} & 0.301 & \textbf{0.696} & 0.589 & \textbf{0.883} & 0.860 \\
 & 	k-FWER & \textbf{0.140} & 0.134	& \textbf{0.145} & 0.140 & \textbf{0.146} & 0.137 \\ 
 \bottomrule
 \end{tabular}
   \caption[Simulation performance, p=1000]{Median MCC values for  AR(1) block structures and block covariance structures, where $p=1000$.  ``EC method" denotes the error control method employed.  We use $\alpha$=0.05 for all EC methods trialled, and set $k=4995$ when using k-FWER.  We highlight in bold the best method (with highest MCC values) in each setting. As the PCS-Hub approach does not readily provide FDR and pFDR control, hence N/A for this error control method.
\label{tbl:sim_p1000_MCC}}
\end{table}}

 {
\begin{table}[!t]
\centering
\small
\begin{tabular}{llllll}
  \hline
 &  & \multicolumn{2}{c}{\textbf{Sensitivity}} &  \multicolumn{2}{c}{\textbf{Screening level}} \\ \cmidrule(r){3-4} \cmidrule(r){5-6}
 & \textbf{EC measure} & {PARSEC} & {PCS-Hub} &  {PARSEC} & {PCS-Hub} \\
  \hline
  \textbf{AR(10) Block}  \\

a=15, n=20 & 	FDR=0.05 & \textbf{0.168} & N/A & 0.915 & N/A   \\
& pFDR=0.05  & \textbf{0.168} & N/A & 0.915 & N/A \\
 & 	FWER, k=0 & \textbf{0.084} & \textbf{0.084 }& 0.940 & 0.940 \\
 & 	k-FWER, k=100 &  \textbf{0.432} & 0.411 & 0.854 & 0.854 \\
 & 	k-FWER, k=1000 &  \textbf{0.663} & 0.642 & 0.805 & 0.804 \\
 & 	k-FWER, k=4995 &  \textbf{0.832} & 0.811 & 0.761 & 0.761 \\ \cmidrule(r){3-6}

a=20, n=20 & 	FDR=0.05  &  \textbf{0.179} & N/A & 0.915 & N/A  \\ 	
 & pFDR=0.05 &  \textbf{0.179} & N/A & 0.915 & N/A \\
 & 	FWER, k=0 &  \textbf{0.083} &  \textbf{0.083} & 0.941 & 0.941 \\
 & 	k-FWER, k=100 &  \textbf{0.441} & 0.428 & 0.855 & 0.855 \\
 & 	k-FWER, k=1000 &  \textbf{0.686} & 0.662 & 0.805 & 0.805 \\
 & 	k-FWER, k=4995 &  \textbf{0.841} & 0.828 & 0.761 & 0.761 \\  \cmidrule(r){3-6}


a=15, n=30 & 	FDR=0.05  & \textbf{0.568} & N/A & 0.820 & N/A  \\
 & 	pFDR=0.05 & \textbf{0.568} & N/A & 0.820 & N/A  \\
 & 	FWER, k=0 & \textbf{0.253} & \textbf{0.253} & 0.888 & 0.885 \\
 & 	k-FWER, k=100 & \textbf{0.874} & 0.847 & 0.753 & 0.753 \\
 & 	k-FWER, k=1000 & \textbf{0.979} & 0.968 & 0.697 & 0.697 \\
 & 	k-FWER, k=4995 & \textbf{1.000} & \textbf{1.000} & 0.651 & 0.651 \\  \cmidrule(r){3-6}

a=20, n=30  & 	FDR=0.05  & \textbf{0.572} & N/A & 0.825 & N/A   \\
 & 	pFDR=0.05 & \textbf{0.572} & N/A & 0.825 & N/A  \\
 & 	FWER, k=0 & \textbf{0.193} & 0.186 & 0.905 & 0.904 \\
 & 	k-FWER, k=100 & \textbf{0.859} & 0.828 & 0.760 & 0.759 \\
 & 	k-FWER, k=1000 & \textbf{0.972} & 0.966 & 0.697 & 0.697 \\
 & 	k-FWER, k=4995 & \textbf{1.000} & 0.993 & 0.651 & 0.651 \\  


 \bottomrule
 \end{tabular}
   \caption[Simulation performance, p=1000]{Type II error assessment of AR(10) block structures, when $p=10000$ and $\phi_1=0.8$. Type II error performance is assessed using the median sensitivity obtained over 1000 replications in each setting, when fixing Type I error to a given error control measure (EC measure).  Note, in all EC measures observed, we set the empirical probability of Type I error in the relevant EC measure to be 0.05 (i.e. $\alpha=0.05$). We also report the screening level corresponding to the Type I error control for additional comparison.  We highlight in bold the best method (with the highest median sensitivity) in each setting. As the PCS-Hub approach does not readily provide FDR and pFDR control, ``N/A" is reported for this error control method.
\label{tbl:sim_p10000_Power_slow_decay_phi08}}
\end{table}}
Note that the theory for PCS-Hub screening relies on approximating the large-scale partial correlation matrix with the large-scale marginal correlation matrix as $p$ tends to infinity.  PARSEC, however, is less reliant on this approximation as the inferential procedure for FDR and pFDR error control, for instance, does not employ marginal correlation coefficients.  Regardless, it is not completely clear whether methods such as PCS-Hub and PARSEC perform well when the underlying model is a sparse partial correlation model but not a sparse marginal correlation model.  This setting can be observed in slow decaying covariance structures (such as AR).  These models were already considered in Table \ref{tbl:sim_p1000_various_aucs} (in Section \ref{sec:simulation_inferential}) and Table \ref{tbl:sim_p1000_MCC}, and PARSEC was shown to have good performance in terms of the consolidated performance metrics: AUC, AUC when FPR$<0.1$, and MCC.  Though these composite measures provide an overall assessment of performance, they do not readily provide assessments of statistical power.
 To further unpack the trade-off between Type I and Type II errors in such sparse partial correlation models, Table \ref{tbl:sim_p10000_Power_slow_decay_phi08} explicitly compares Type II error (sensitivity in this case) when an error metric (such as FWER/k-FWER/FDR/pFDR) is fixed at a pre-specified level $\alpha$.  We implement FWER by identifying the screening level at which no false discoveries are made in $(1-\alpha)*100\%$ of replications.  Equivalently, we implement k-FWER by identifying the screening level at which k false discoveries are made in $(1-\alpha)*100\%$ of replications.  FDR is implemented by determining the screening level at which the false discovery proportion measure (FDP=(FP)/(FP+TP)) averages a pre-specified significance level $\alpha$.  Lastly, pFDR is implemented by finding the screening level at which FDP averages a pre-specified significance level $\alpha$, while restricting the screening levels considered to be only those where there is at least one discovery made across all replications. These approaches provide a means to empirically determine the screening levels which provide the desired level of error control.
 Thereafter, sensitivity is calculated following the application of the respective screening level for each replication, and the median is reported.  We also report the screening level applied.
 Note that in Table \ref{tbl:sim_p10000_Power_slow_decay_phi08} PARSEC achieves consistently higher sensitivity rates, establishing that PARSEC has good statistical power and overall performance.  Although PCS-Hub is competitive when applying FWER error metrics, it should be noted that FWER is often deemed inappropriate in large-scale problems given its stringency \parencite{Storey:2002}.  When we move to more realistic (higher) values of $k$ for k-FWER, which are more suitable for large-$p$ problems, PCS-Hub is not competitive.

\begin{figure}[t!]
\centering
\subfloat[FPR $\leq 1$]{%
	 \hspace*{-1cm}{
	 \includegraphics[width=80mm]{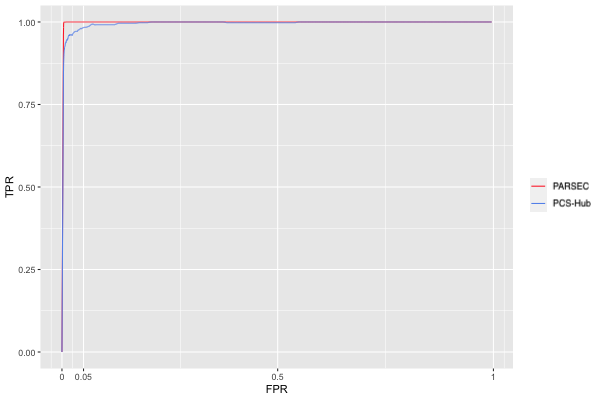}} } \quad \quad \quad
\subfloat[FPR $< 0.01$]{%
	 \hspace*{-1cm}{
	 \includegraphics[width=80mm]{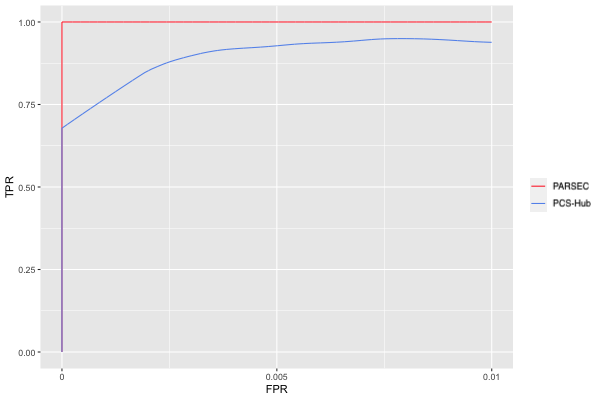}
             }}
\caption{ROC curves of an AR(10) setting where p=1000, a=50, $\phi_1=0.8$, n=30 model. We report the median values calculated over 1000 replications. \label{fig:AR10_roc}}
\end{figure}

\begin{figure}[t!]
\centering
\subfloat[PARSEC, stratified by coefficient lag]{%
	 \hspace*{-1cm}{
	 \includegraphics[width=120mm]{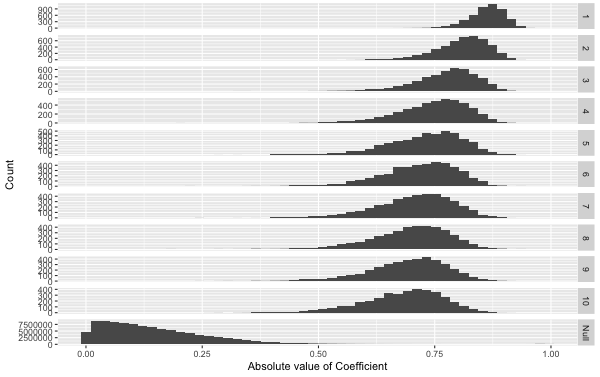}} } \\
\subfloat[PCS-Hub, stratified by coefficient lag]{%
	 \hspace*{-1cm}{
	 \includegraphics[width=120mm]{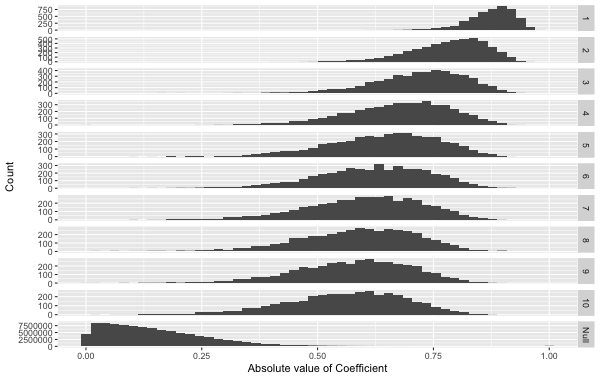}
             }}
\caption{Distribution of coefficients for an AR(10), p=1000, a=50, $\phi_1=0.8$, n=30 model, over 10 replications.  It is clear that PARSEC's estimates of non-null coefficients are further from the distribution of null coefficients.  \label{fig:AR10_lag_analysis}}
\end{figure}

\begin{figure}[!t]
\centering
\subfloat[$\rho=0.7$ (in $\Sigma$)]{%
	 \hspace*{-1cm}{
	 \includegraphics[width=85mm]{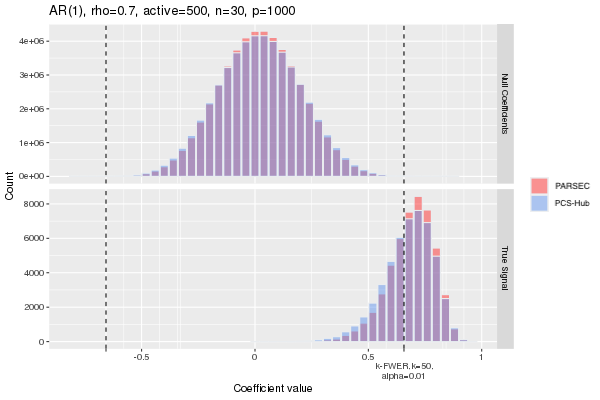}
            \includegraphics[width=85mm]{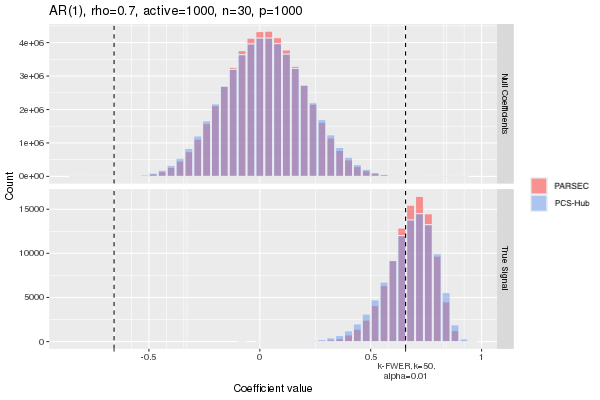} }} \\
\subfloat[$\rho=0.8$ (in $\Sigma$)]{%
	 \hspace*{-1cm}{
	 \includegraphics[width=85mm]{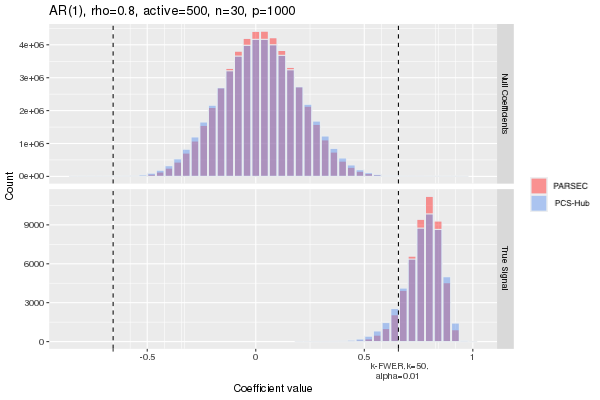}
            \includegraphics[width=85mm]{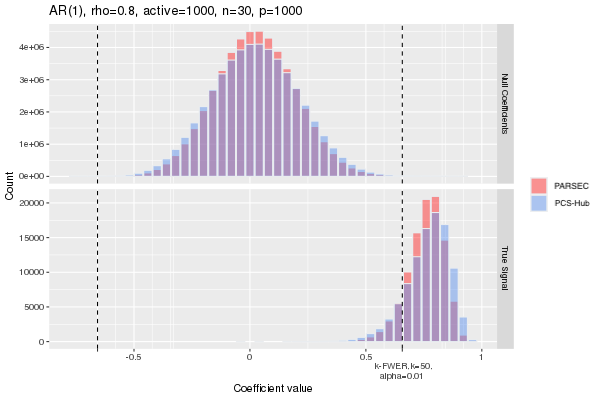}   }}    \\
\subfloat[$\rho=0.9$ (in $\Sigma$)]{%
	 \hspace*{-1cm}{
	 \includegraphics[width=85mm]{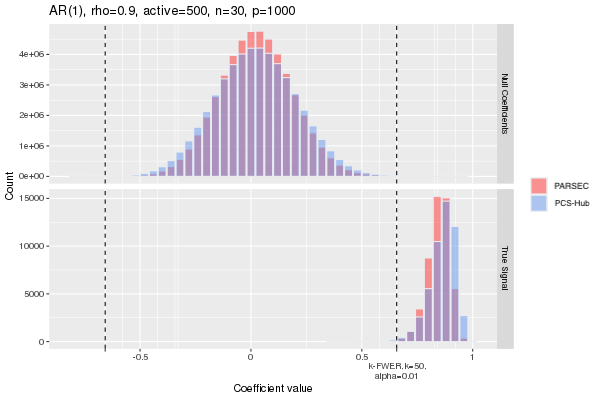}
            \includegraphics[width=85mm]{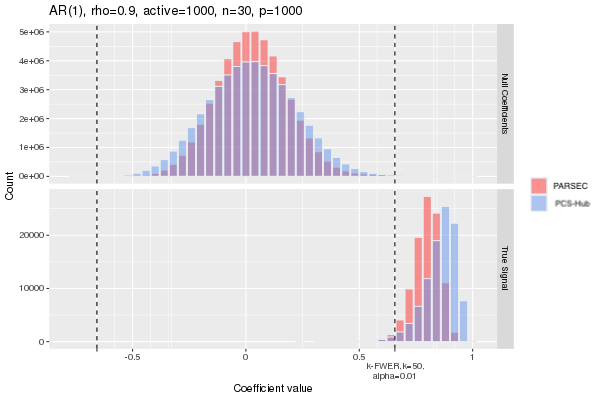} }}
\caption{Distribution of coefficients for an AR(1), p=1000, n=30 model. Block size varies from $a=500$ (left) and $a=1000$ (right), and signal strength of the first lag coefficient also increases from $\rho=0.7,0.8,0.9$.  Notably, PARSEC better identifies the true coefficients.  It also favorably shrinks null coefficients to zero.
 \label{fig:coeffs_AR_comparison_shrinkage_n30}}
\end{figure}

\begin{figure}[!htbp]
\centering
\subfloat[$\rho=0.7$ (in $\Sigma$)]{%
	 \hspace*{-1cm}{
	 \includegraphics[width=85mm]{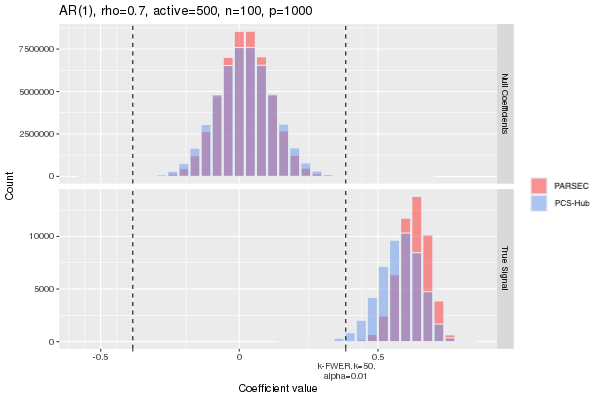}
            \includegraphics[width=85mm]{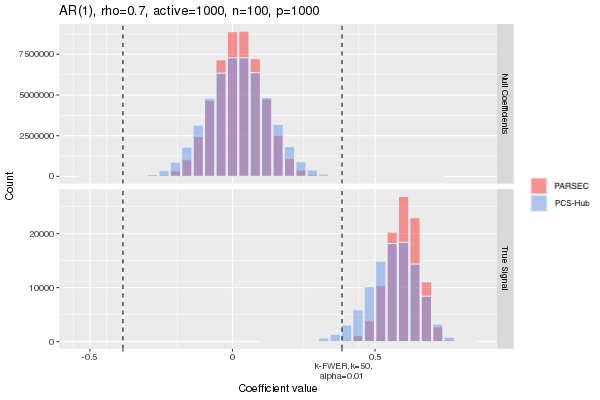} }} \\
\subfloat[$\rho=0.8$ (in $\Sigma$)]{%
	 \hspace*{-1cm}{
	 \includegraphics[width=85mm]{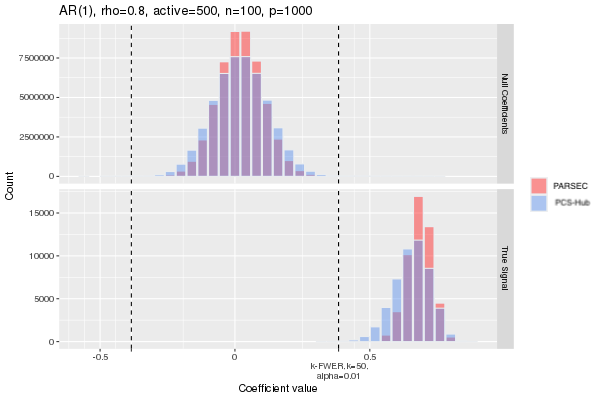}
            \includegraphics[width=85mm]{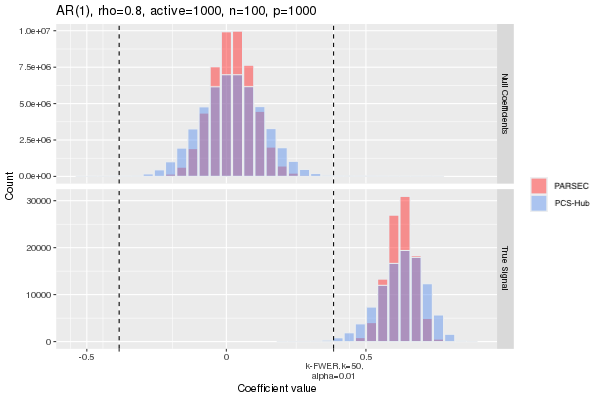}   }}    \\
\subfloat[$\rho=0.9$ (in $\Sigma$)]{%
	 \hspace*{-1cm}{
	 \includegraphics[width=85mm]{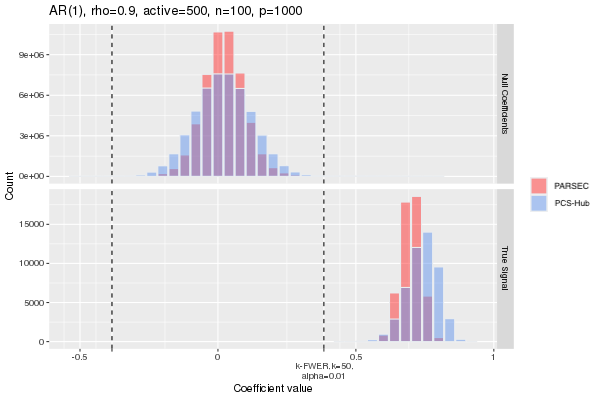}
            \includegraphics[width=85mm]{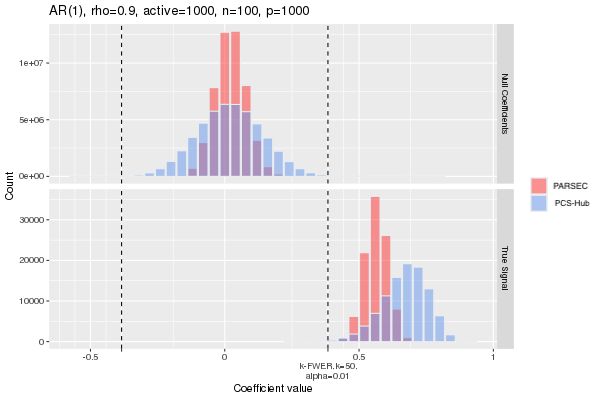} }}
\caption{Distribution of coefficients for an AR(1), p=1000, n=100 model.  Block size varies from $a=500$ (left) and $a=1000$ (right), and signal strength of the first lag coefficient also increases from $\rho=0.7,0.8,0.9$.  Notably, PARSEC better identifies the true coefficients.  It also favorably shrinks null coefficients to zero. \label{fig:coeffs_AR_comparison_shrinkage_n100}}
\end{figure}

We now undertake a comprehensive investigation to understand why PCS-Hub performs comparably with PARSEC when using the FWER error metric.  An example of a Receiver Operating Characteristic (ROC) curve from an AR(10) block setting illustrated in Figure \ref{fig:AR10_roc}, shows that PARSEC provides a higher true positive rate across the range of false positive rates. Further analysis of the AR(10) setting when $p=1000$ in Figure \ref{fig:AR10_lag_analysis}, shows that while PCS-Hub estimates strong partial correlation coefficients for the first and second lag in the AR(10) model, the remaining lag coefficients are closer to the distribution of null coefficients. However, PARSEC estimates higher partial correlation coefficients across all lags.  Hence, while PCS-Hub works well with highly conservative error control measures, its performance deteriorates rapidly as we permit more false discoveries.  In Figure \ref{fig:coeffs_AR_comparison_shrinkage_n30} and Figure \ref{fig:coeffs_AR_comparison_shrinkage_n100} we provide additional illustrations of the distribution of PARSEC vs. PCS-Hub coefficients for various AR(1) structured models.  The distribution of PCS-Hub estimates for null coefficients are notably heavier in the tails, again indicating that Type II error control deteriorates as additional false discoveries are permitted. In ultra-high dimensional problems, allowing for a higher number of false discoveries (equivalent to higher $k$) is more realistic to enable novel discoveries and findings.  Hence, PCS-Hub is not competitive in the large $p$ setting targeted in this work.


{
\begin{table}[!t]
\centering
\small
\begin{tabular}{llllll}
  \hline
$\alpha=0.05$ &  & \multicolumn{2}{c}{\textbf{Sensitivity}} &  \multicolumn{2}{c}{\textbf{Screening level}} \\ \cmidrule(r){3-4} \cmidrule(r){5-6}
 & \textbf{EC measure} & {PARSEC} & {PCS-Hub} &  {PARSEC} & {PCS-Hub} \\
  \hline
  \textbf{AR(10) Block}  \\

a=15, n=20 & 	FDR=0.05 & \textbf{0.337} & N/A & 0.845 & N/A \\
& pFDR=0.05 & \textbf{0.337} & N/A & 0.845 & N/A\\
 & 	FWER, k=0 & \textbf{0.147} & 0.137 & 0.904 & 0.896 \\
 & 	k-FWER, k=100 & \textbf{0.747} & 0.611 & 0.745 & 0.744 \\
 & 	k-FWER, k=1000 & \textbf{0.958} & 0.874 & 0.650 & 0.650 \\
 & 	k-FWER, k=4995 & \textbf{1.000} & 0.979 & 0.562 & 0.562 \\  \cmidrule(r){3-6}

a=20, n=20 & 	FDR=0.05 & \textbf{0.324} & N/A & 0.846 & N/A  \\
& pFDR=0.05 & \textbf{0.324} & N/A & 0.846 & N/A \\
 & 	FWER, k=0 & 0.097 & \textbf{0.103} & 0.913 & 0.901 \\
 & 	k-FWER, k=100 & \textbf{0.690} & 0.545 & 0.753 & 0.748 \\
 & 	k-FWER, k=1000 & \textbf{0.938} & 0.821 & 0.651 & 0.651 \\
 & 	k-FWER, k=4995 & \textbf{0.993} & 0.952 & 0.562 & 0.562 \\  \cmidrule(r){3-6}

a=15, n=30 & 	FDR=0.05 & \textbf{0.674} & N/A & 0.745 & N/A  \\
& pFDR=0.05 & \textbf{0.674} & N/A & 0.745 & N/A \\
 & 	FWER, k=0 & \textbf{0.242} & \textbf{0.242} & 0.854 & 0.826 \\
 & 	k-FWER, k=100 & \textbf{0.968} & 0.821 & 0.636 & 0.635 \\
 & 	k-FWER, k=1000 & \textbf{1.000} & 0.968 & 0.543 & 0.543 \\
 & 	k-FWER, k=4995 & \textbf{1.000} & \textbf{1.000} & 0.463 & 0.463 \\  \cmidrule(r){3-6}

a=20, n=30 & 	FDR=0.05 & \textbf{0.572} & N/A & 0.760 & N/A  \\
& pFDR=0.05 & \textbf{0.572} & N/A & 0.760 & N/A  \\
 & 	FWER, k=0 & 0.172 & \textbf{0.207} & 0.865 & 0.826 \\
 & 	k-FWER, k=100 & \textbf{0.924} & 0.724 & 0.648 & 0.641 \\
 & 	k-FWER, k=1000 & \textbf{1.000} & 0.924 & 0.544 & 0.544 \\
 & 	k-FWER, k=4995 & \textbf{1.000} & 0.986 & 0.463 & 0.463 \\  

 \bottomrule
 \end{tabular}
   \caption[Simulation performance, p=1000]{Type II error assessment of AR(10) block structures, when $p=1000$ and $\phi_1=0.8$. Type II error performance is assessed using the median sensitivity obtained over 1000 replications in each setting, when fixing Type I error to a given error control measure (EC measure).  Note, in all EC measures observed, we set the empirical probability of Type I error in the relevant EC measure to be 0.05 (i.e. $\alpha=0.05$). We also report the screening level corresponding to the Type I error control for additional comparison.  We highlight in bold the best method (with the highest median sensitivity) in each setting. As the PCS-Hub approach does not readily provide FDR and pFDR control, ``N/A" is reported for this error control method.
\label{tbl:sim_p1000_Power_slow_decay}}
\end{table}}

We also provide Type II error control assessments for the $p=1,000$ setting in Table \ref{tbl:sim_p1000_Power_slow_decay} to accompany the setting analyzed in Figure \ref{fig:AR10_lag_analysis}. Note that PARSEC performs well across all error control metrics, whereas PCS-Hub only does well in some of the cases in the FWER setting.  This limited performance does not extend upon moving to the large-$p$ setting where $p=10,000$ (already observed in Table \ref{tbl:sim_p10000_Power_slow_decay_phi08}).

\clearpage
\newpage

\section{Simulated Data in the Multivariate T setting} \label{appendix:heavier_simulated}

We now assess the partial correlation screening performance of PARSEC and PCS-Hub in the setting of heavier tail distributions.  This study investigates error control and signal recovery of both partial correlation approaches in a setting outside of the Multivariate Normal discussed in Section \ref{sec:simulated_data}. Specifically, we sample from the Multivariate T distribution, $vt_\nu(\mu,\Sigma)$ where $\nu$ denotes the degrees of freedom, $\mu$ denotes the mean and $\Sigma$ denotes the covariance matrix.  In all settings, we let $\nu=3$, and set $\mu=0$.  We vary the covariance matrix $\Sigma$ to sample from different structured settings.

Table \ref{tbl:sim_p1000_various_aucs_mvt_t} reports median overall AUC values and median AUC values when we restrict the false positive rate to be less than 0.1.  We evaluate these values over various structured models, including Auto-regressive (AR) block models, block covariance models and star structures, over 100 replications each.  We set the dimension $p=1000$ and the sample size $n=30$.  Extensive details of the data generating process for each of these structures is provides in Section \ref{sec:simulated_data}.  Across both AUC measures, we observe that PARSEC provides competitive if not superior values.  Hence, it is clear that even in heavier tail distributions outside the Multivariate Normal setting, PARSEC provides improved inferential performance for a range of structures.

Table \ref{tbl:sim_p1000_MCC_mvt_t} provides further comparison of PARSEC and PCS-Hub's signal recovery ability in other heavier-tail settings with varying signal strength.  We report the median MCC over 100 replications.  These values are determined following the application of critical screening levels derived from theory outlined in Section \ref{sec:theory}.  PARSEC's stronger MCC illustrates its statistical power even when working with heavy tail distributions.  These results are similar to PARSEC's strong performance demonstrated in the Multivariate Normal setting.

\begin{figure}[ht!]
\centering
\subfloat[$\phi_1=0.7$, $n=30$]{%
	 \includegraphics[width=110mm]{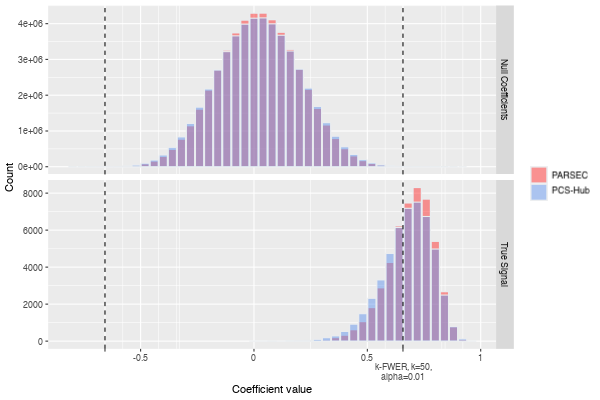}} \\
\subfloat[$\phi_1=0.7$, $n=100$]{%
            \includegraphics[width=110mm]{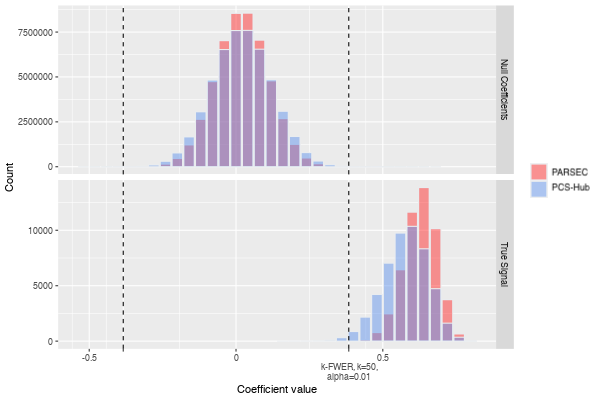}}
\caption{Distribution of coefficients for an AR(1) model where $p=1000$ and the block size $a=500$.  We sample from the Multivariate T distribution with $\nu=3$.  Even in heavier tail distributions, we observe that PARSEC favorably shrinks null coefficients and effectively identifies true coefficients. \label{fig:coeffs_AR_comparison_shrinkage_mvt_t}}
\end{figure}

We also observe the distribution of estimated partial correlation coefficients in the slow decay AR(10) setting, when sampling again from the Multivariate T distribution in Figure \ref{fig:coeffs_AR_comparison_shrinkage_mvt_t}.  A comparison of Figure \ref{fig:coeffs_AR_comparison_shrinkage_mvt_t} with the distribution of estimated partial correlation coefficients from the Multivariate Normal setting illustrated in Figure \ref{fig:coeffs_AR_comparison_shrinkage_n30} and Figure \ref{fig:coeffs_AR_comparison_shrinkage_n100}, reveals that PARSEC's estimation procedure demonstrates consistent behavior. PARSEC again favorably shrinks null coefficients, while providing stronger identification of true non-null coefficients.  There is, thus, conclusive evidence that PARSEC's signal recovery ability thus extends to the wider class of vector-elliptical distributions.

{
\begin{table}[!t]
\centering
\small
\begin{tabular}{llllllll}
  \hline
\textbf{MCC} &   &  \multicolumn{2}{c}{$\sigma,\phi_1=0.6$} & \multicolumn{2}{c}{$\sigma,\phi_1=0.7$} & \multicolumn{2}{c}{$\sigma,\phi_1=0.8$}\\ \cmidrule(r){3-4} \cmidrule(r){5-6} \cmidrule(r){7-8}
 & \textbf{EC method} & {PARSEC} & {PCS-Hub} &  {PARSEC} & {PCS-Hub} &  {PARSEC} & {PCS-Hub} \\
  \hline
 \textbf{Block}  \\
a=20, n=40	& FDR-BH & \textbf{0.073} & N/A & \textbf{0.451} & N/A  & \textbf{0.894} & N/A  \\
& pFDR & \textbf{0.073} & N/A  & \textbf{0.451} & N/A  & \textbf{0.894} & N/A  \\
 & 	FWER  & \textbf{0.073} & 0.000 & \textbf{0.178} & 0.073 & \textbf{0.508} & 0.178 \\
 & 	k-FWER  & \textbf{0.165} & 0.101 & \textbf{0.192} & 0.152 & \textbf{0.196} & 0.190 \\ \cmidrule(r){2-8}
\multicolumn{2}{l}{\textbf{AR(1) Block}} \\
a=50, n=30 & FDR-BH & \textbf{0.101} & N/A & \textbf{0.418} & N/A & \textbf{0.760} & N/A  \\
& pFDR & \textbf{0.101} & N/A & \textbf{0.418} & N/A & \textbf{0.760} & N/A \\
 & 	FWER & \textbf{0.122} & 0.000 & \textbf{0.286} & 0.247 & \textbf{0.572} & 0.516 \\
 & 	k-FWER & \textbf{0.082} & 0.078 & \textbf{0.096} & 0.095 & \textbf{0.100} & 0.099 \\  \cmidrule(r){3-8}
a=100, n=30	& 	FDR-BH &  \textbf{0.142} & N/A & \textbf{0.507} & N/A & \textbf{0.788} & N/A \\
& pFDR & \textbf{0.142} & N/A & \textbf{0.507} & N/A & \textbf{0.788} & N/A \\
 & 	FWER & \textbf{0.100} & \textbf{0.100} & \textbf{0.266} & 0.246 & \textbf{0.579} & 0.532 \\
 & 	k-FWER & \textbf{0.117} & 0.111 & \textbf{0.138} & 0.134 & \textbf{0.144} & 0.139 \\  \cmidrule(r){3-8}
a=50, n=50	& 	FDR-BH & \textbf{0.536} & N/A & \textbf{0.840} & N/A & \textbf{0.767} & N/A \\
 & pFDR & \textbf{0.536} & N/A & \textbf{0.840} & N/A & \textbf{0.767} & N/A \\
 & 	FWER & \textbf{0.350} & 0.286 & \textbf{0.700} & 0.572 & \textbf{0.883} & 0.872 \\
 & 	k-FWER  & \textbf{0.099} & 0.096 & \textbf{0.101} & 0.099 & \textbf{0.101} & 0.099 \\  \cmidrule(r){3-8}
a=100, n=50	& 	FDR-BH & \textbf{0.629} & N/A & \textbf{0.848} & N/A & \textbf{0.754} & N/A \\
 & pFDR & \textbf{0.629} & N/A & \textbf{0.848} & N/A & \textbf{0.754}  & N/A \\
 & 	FWER & \textbf{0.389} & 0.301 & \textbf{0.689} & 0.586 & \textbf{0.883} & 0.855 \\
 & 	k-FWER & \textbf{0.140} & 0.134 & \textbf{0.145} & 0.140 & \textbf{0.146} & 0.139 \\
 \bottomrule
 \end{tabular}
   \caption[Simulation performance, p=1000]{Median MCC values for  AR(1) block structures and block covariance structures, where $p=1000$ in the Multivariate T setting.  ``EC method" denotes the error control method employed.  We use $\alpha$=0.05 for all EC methods trialed, and set $k=4995$ when using k-FWER.  We highlight in bold the best method (with highest MCC values) in each setting. As the PCS-Hub approach does not readily provide FDR and pFDR control, hence N/A for this error control method.
\label{tbl:sim_p1000_MCC_mvt_t}}
\end{table}}

{
\begin{table}[ht!]
\small
\centering
\begin{tabular}{llcccc}
\toprule
 &  & \multicolumn{2}{c}{\textbf{PARSEC}} &  \multicolumn{2}{c}{\textbf{PCS-Hub}}\\ \cmidrule(r){3-4} \cmidrule(r){5-6}
\textbf{Structure} & \textbf{Dimensions} & \textbf{AUC} & \textbf{FPR$<$.1}  & \textbf{AUC} & \textbf{FPR$<$.1} \\
\midrule
 \multicolumn{5}{l}{\textbf{AR(1) Block, $\phi_1=0.7$}} \\
& a=50, n=20 & \textbf{0.9990} & \textbf{0.9896} & 0.9979 & 0.9788 \\
& a=100, n=50 & \textbf{0.9986} & \textbf{0.9861} & 0.9978 & 0.9787 \\
& a=500, n=100 & \textbf{0.9984} & \textbf{0.9854} & 0.9968 & 0.9735 \\ \cmidrule(r){2-6}
 \multicolumn{5}{l}{\textbf{AR(2) Block, $\phi_1=0.7$}} \\
& a=50, n=20 & 0.9999 & 0.9992 & \textbf{1.0000} & \textbf{0.9998} \\
& a=100, n=50 & 0.9998 & 0.9980 & \textbf{0.9999} & \textbf{0.9994} \\
& a=500, n=100 & 0.9982 & 0.9823 & \textbf{0.9996} & \textbf{0.9964} \\ \cmidrule(r){2-6}
 \multicolumn{5}{l}{\textbf{AR(5) Blocks, $\phi_1=0.7$}} \\
& a=50, n=20 & \textbf{0.9999} & 0.9989 & \textbf{0.9999} & \textbf{0.9995} \\
& a=100, n=50 & 0.9997 & 0.9970 & \textbf{0.9998} & \textbf{0.9984} \\
& a=500, n=100 & 0.9972 & 0.9723 & \textbf{0.9988} & \textbf{0.9879} \\ \cmidrule(r){2-6}
 \multicolumn{5}{l}{\textbf{AR(10) Blocks, $\phi_1=0.7$}} \\
& a=50, n=20 &\textbf{0.9998} & \textbf{0.9983} & \textbf{0.9998} & 0.9980 \\
& a=100, n=50 & \textbf{0.9994} & \textbf{0.9940} & 0.9982 & 0.9820 \\
& a=500, n=100 & \textbf{0.9923} & \textbf{0.9235} & 0.9826 & 0.8838 \\ \cmidrule(r){2-6}
 \multicolumn{5}{l}{\textbf{Block, $\sigma=0.7$}} \\
& a=5, n=20 & \textbf{0.9983} & \textbf{0.9829} & 0.9972 & 0.9720 \\
& a=30, n=60 & \textbf{0.9980} & \textbf{0.9804} & 0.9142 & 0.5898 \\
& a=50, n=100 & \textbf{0.9678} & \textbf{0.7372} & 0.5877 & 0.1242 \\ \cmidrule(r){2-6}
 \multicolumn{5}{l}{\textbf{Star separator, $c=-0.35$, $n=30$}} \\
& k=5, e=2 & \textbf{0.9137} & \textbf{0.6343} & 0.8967 & 0.5929 \\
& k=10, e=4 & \textbf{0.9750} & \textbf{0.8411} & 0.9686 & 0.8056 \\   \cmidrule(r){2-6}
 \multicolumn{5}{l}{\textbf{Star Block, $c=-0.35$, $n=30$}} \\
& k=5, e=2 & \textbf{0.8639} & \textbf{0.4911} & 0.8526 & 0.4573 \\
& k=20, e=2 & \textbf{0.8707} & \textbf{0.4965} & 0.8550 & 0.4703 \\
& k=20, e=4 & \textbf{0.9215} & \textbf{0.6354} & 0.9089 & 0.6009 \\
& k=50, e=4 &\textbf{ 0.9223} & \textbf{0.6318} & 0.9086 & 0.6022 \\
\bottomrule
\end{tabular}
\caption[Simulation performance, p=1000]{AUC values for the various covariance structures where $p=1000$ and varying $n$, in the Multivariate T setting.  Each setting is replicated 100 times.  As per the notation used in the Multivariate Normal setting, we let $\sigma$ denote the coefficients for non-zero elements simulated from a covariance matrix in block settings, $\phi_1$ is the coefficient of the first order lag in AR block settings and $a$ provides the block size. $c$ provides the signal for non-zero elements in the inverse covariance matrix in star structures. Performance is measured using median AUC, and median AUC where the FPR range is limited to less than 0.1.   We highlight in bold the best method (highest AUC) in each setting.
\label{tbl:sim_p1000_various_aucs_mvt_t}}
\end{table}
}

\clearpage
\newpage

\section{Computational performance} \label{appendix:computational_performance}

PARSEC's scalable implementation permits further computational advantages which extend beyond faster wall-times.  As established in Section \ref{sec:computational_complexity}, PARSEC's scalable implementation provides storage gains as its simultaneous estimation and screening procedure circumvents the need to store a large $p \times p$ matrix.  For the same reason, PARSEC's efficient implementation also requires less in-memory computational resources, resulting in lower RAM requirements. In the ultra-high dimensional setting, RAM is an important consideration given computational resource constraints and associated costs.  More specifically, recall that PARSEC evaluates partial correlation coefficients and then immediately performs screening.
As such, we note that the scalable implementation of PARSEC requires approximately 400GB of RAM (on average) in the large-scale setting where $p=100,000$. In contrast, PCS-Hub requires approximately 550GB of RAM (on average), or $37.5\%$ more computational resources than PARSEC for the same problem.  In summary, PARSEC is competitive and is often vastly superior compared to existing methods in all three aspects of computing: i) computational speed (i.e. processing speed), ii) storage needs, and iii) memory allocation requirements.

\clearpage
\newpage

\section{Breast Cancer Gene Screening} \label{sec:appendix_gene_screening}

\subsection{Data preparation and cleansing}

The sample of gene expression levels and patients included was reached by eliminating any gene expression which was not identifiable from its probe sequence ID, or for which more than 5\% of patients had missing records. We then excluded any patient with missing data for the remaining 15,220 gene expression levels.

\subsection{Identified genes following screening} \label{sec:genes_parsec}

{
\setstretch{1.2}
\footnotesize
\begin{longtable}{ p{1.5cm} p{1cm} p{1cm} p{1cm}  p{6.75cm} p{2.8cm}}
\toprule
\textbf{Gene} &  \rotatebox{90}{\parbox{3cm}{\textbf{FWER, \\ $\alpha = 0.05$} }} & \rotatebox{90}{\parbox{3cm}{\textbf{k-FWER, \\ $\alpha = 0.05, k = 1\%$} }} & \rotatebox{90}{\parbox{3cm}{\textbf{FDR-BH, \\ $\alpha = 0.01$} }}    & \textbf{Summary of Relationship} & \textbf{Reference} \\
\midrule
\endhead
PTGER3 (EP3) & + &  + & + & EP3 is a prostaglandin receptor identified as a prognostic factor for improved progression and survival of breast cancer. & \textcite{SemmlingerEtAl:2018} \\ \cmidrule(r){2-6}
CD53 & + &  + &    & CD53 has been found to regulate immune cell function.  In particular, extensive research has shown that high expressions of CD53 is present in radio-resistant tumor cells, as ligation of CD53 induces a survival response for cells that otherwise enter apoptosis. & \textcite{Dunlock:2020}, \textcite{YuntaLazo:2003}, \textcite{VoehringerEtAl:2000} \\ \cmidrule(r){2-6}
HSPG2 & + & + &   & HSPG2 is a glycosylated protein and has been recently identified as a promising target for breast cancer therapy due to confirmed expression in breast cancer cells.  Higher levels of HSPG2 in patients diagnosed with Triple Negative Breast Cancer (TNBC) in particular, is correlated with poor survival. & \textcite{KalscheuerEtAl:2019}, \textcite{HuEtAl:2013} \\ \cmidrule(r){2-6}
CRYBB3 & + &   &  + & CRYBB3 has not been identified as a prognostic or associative factor directly, but over-expression of the CRYBB2 gene within the same cluster has been linked with accelerated cell growth. & \textcite{BarrowEtAl:2019} \\ \cmidrule(r){2-6}
IFIT3 & + & & + & IFIT3 is an Interferon‐induced protein (IFN) that has been identified as a prognostic marker which regulates resistance to radiation in cancer cells and chemotherapy. In particular, IFIT3 is a marker of breast cancer cell sensitivity to immuno-stimulating therapeutics. & \textcite{NushtaevaEtAl:2018} \\
PCBP3 & + & & + & Identification of PCBP3 has been identified as a prognostic biomarker for pancreatic cancer. & \textcite{GerEtAl:2018} \\ \cmidrule(r){2-6}
LOR & + &  &  + \\ \cmidrule(r){2-6}
PHYHIPL & + &  & + \\ \cmidrule(r){2-6}
MX1 & + & & + & MX1 is over-expressed in various cancers, and has been identified as  a prognostic marker for shorter survival in breast cancer patients - further clinical trials have been recommended to verify the therapeutic use of MX1 further.  & \textcite{AljohaniEtAl:2020} \\ \cmidrule(r){2-6}
MEF2C & + & &   & MEF2 (including MEF2C) are transcription factors which play a role in the regulation, interaction and binding of co-repressors in cancer cells. In breast cancer however, MEF2C has also been linked to $\beta$-catenin through its interaction with miR-223, which promotes breast cancer invasion.  Hence, MEF2C is a likely candidate for targeted treatment.  & \textcite{GiorgioHancockBrancolini:2018} \\ \cmidrule(r){2-6}
ORC4L & + &  &   & \\ \cmidrule(r){2-6}
PLIN & + &  &   & PLIN was identified as an independent prognostic marker - within an in vitro study on mice, the exogenous expression of PLIN inhibits cell proliferation and invasion.  & \textcite{ZhouEtAl:2016} \\ \cmidrule(r){2-6}
EYA4 & + & &   & EYA4 has been associated with several cancer types. The prevalence of EYA4 in breast cancer tumors though has produced conflicting research findings whereby some studies have identified the gene as oncogenic, while others have asserted that it is a potential suppressor gene. & \textcite{LouEtAl:2018}, \textcite{Pandey:2010}, \textcite{FacklerEtAl:2011} \\ \cmidrule(r){2-6}
WAS & & + &   & The WAS gene encodes a family of proteins which include the WASp (Wiskott-Aldrich syndrome protein) and WIP (WASp interacting protein) that have been identified as biomarkers for breast cancer progression, and potential targets for therapeutic treatment. & \textcite{FrugtnietEtAl:2015}, \textcite{GarciaEtAl:2016}  \\ \cmidrule(r){2-6}
UMOD & & + &    & \\ \cmidrule(r){2-6}
NEUROD1 & & + &   & NEUROD1 is a marker for chemosensitivity as patients with high NEUROD1 methylation were found to be 10.8 times more likely to respond with a complete pathologic response to chemotherapy. & \textcite{FieglEtAl:2008} \\ \cmidrule(r){2-6}
PPAPDC1A (PLPP4) & & + &   & A number of studies have identified PLPP4 as a novel gene that is over-expressed in numerous cancers, including breast cancer.  In breast cancer, PLPP4 is a potential contributor to cancer development due to its role of signal transduction. & \textcite{DahlEtAl:2006}, \textcite{ZhangEtAl:2017}, \textcite{YangEtAl:2018}  \\ \cmidrule(r){2-6}
IL13RA2 & & + &   & Overexpression of IL13RA2 has been linked to metastasis in several cancer types - in breast cancer, patients with increased IL13RA2 had worse prognosis. Several studies recommend IL13RA2 as a target for gene silencing. & \textcite{PapageorgisEtAl:2015}, \textcite{ZhaoWangXu:2015}, \textcite{OkamotoEtAl:2019} \\ \cmidrule(r){2-6}
COL5A2 & & + &   & COL5A2 has been reported in the pathological process and evolution of a variety of cancers, including breast cancer, where it has been found to be increased in invasive breast cancer. & \textcite{Chai:2016}, \textcite{Vargas:2012}  \\ \cmidrule(r){2-6}
CLDN16 & & + &   & CLDN16 is a tight junction protein that contributes to the maintenance of cell polarity, adhesion and arrangement.  It has been suggested as a target for therapeutic treatment for this reason. & \textcite{KuoEtAl:2010} \\ \cmidrule(r){2-6}
CYP2B6 & & + &   & CYP2B6 is involved with the metabolism of testosterone and is associated with greater risk of breast cancer when CYP2B6 exhibit functional changes. CYP2B6 has also been found to have an interaction effect with the clinical efficacy of chemotherapeutic treatment on metastatic breast cancer patients. & \textcite{JustenhovenEtAl:2014}, \textcite{SongEtAl:2015}    \\ \cmidrule(r){2-6}
AKAP11 (AKAP220) & & + &   & A study has found that silencing of AKAP220 alters the rate of cell migration in cancer cells. & \textcite{LogueEtAl:2011} \\ \cmidrule(r){2-6}
CSMD1 & & & + & CSMD1 has been identified as a suppressor in breast cancer development, with a number of studies linking decreased expression of the gene to lower survival. &  \textcite{GialeliEtAl:2021}, \textcite{Escudero-EsparzaEtAl:2016}\\ \cmidrule(r){2-6}
KRT2 & & & + & Changes in Keratin expression (including KRT2) have found to demonstrate complex patterns in breast cancer patients, and is of interest for further study. & \textcite{JoosseEtAl:2012} \\ \cmidrule(r){2-6}
DLST & & & + & A recent study found that DLST plays an important role in the aggression of tumor growth in TNBC patients.  DLST depletion suppresses tumor growth and burden, while high expression of DLST is linked with decreased survival.  & \textcite{ShenEtAl:2021} \\ \cmidrule(r){2-6}
PPP1R3A & & & + & Mutations in PPP1R3A has been associated with interval cancer (primary breast cancers identified during mammogram screening). & \textcite{LiEtAl:2017} \\ \cmidrule(r){2-6}
ACRV1 & & & + \\
\bottomrule
\caption[Breast cancer gene screening]{Comparison of genes screened using FWER ($\alpha=0.05$) and k-FWER ($\alpha=0.05$, $k=1\%$ of $p^2$).  Both error control metrics produced a similar set of genes; indicating consistency in the topology of networks as we vary the error control approach. \label{tbl:breastCancerGenes}}
\end{longtable}}

\subsection{Analysis of genes identified by previous studies}

\begin{figure}[!htbp]
\includegraphics[width=12.5cm]{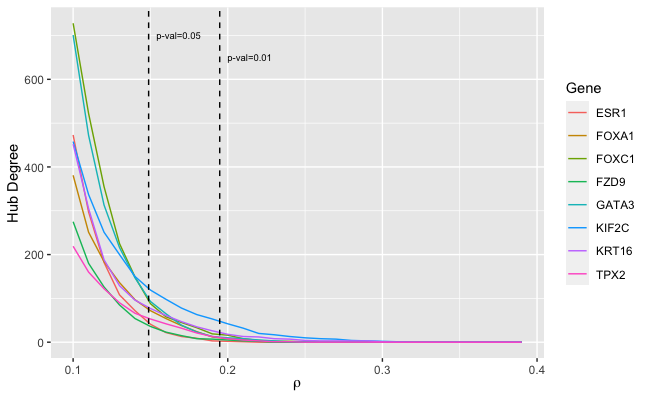}
\centering
\caption[Degree of hub genes previously identified by CONCORD]{Degree of hub genes previously identified by CONCORD \parencite{KhareOhRajaratnam:2015}, as we vary the critical screening level $\rho$, corresponding to different error control metrics. We illustrate the FWER screening levels corresponding to the significance levels of $\alpha=0.01$ and $\alpha=0.05$. These p-value cut-offs are derived from the spherical cap probability, from PARSEC's inferential framework.  \label{fig:hubs_concord}}
\end{figure}

\begin{figure}[H]
\includegraphics[width=12.5cm]{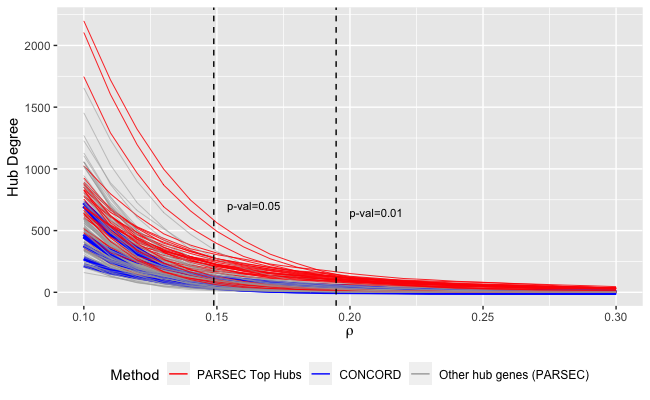}
\centering
\caption[Degree of top hub genes identified by PARSEC vs. CONCORD hub genes]{Degree of hub genes previously identified by CONCORD \parencite{KhareOhRajaratnam:2015} (marked in blue) vs. the top hub genes identified by PARSEC as we vary the critical screening level $\rho$ (marked in red).  Other hub genes which are connected to CONCORD's identified hub genes are also illustrated (in grey).  We illustrate the FWER screening levels corresponding to the significance levels of $\alpha=0.01$ and $\alpha=0.05$. These p-value cut-offs are derived from the spherical cap probability, from PARSEC's inferential framework.  \label{fig:connected_to_hubs_concord_parsec}}
\end{figure}

We compare the degree of hubs previously identified by CONCORD \parencite{KhareOhRajaratnam:2015}, to further understand the performance of PARSEC in contrast to competing $\ell_1$ methods.  Figure \ref{fig:hubs_concord} illustrates the hub degree of these genes as we vary the critical screening level ($\rho$) used by PARSEC.  We note that PARSEC still identifies these genes as hubs.  However, given PARSEC's inclusion of a wider gene expression set, we note that the degree of these ($\ell_1$-based) hub genes has been eclipsed by other hub genes outlined in Section \ref{sec:genes_parsec} of this appendix section.
We also note that although the degree of these hub genes are lower, they are connected to other strongly connected hub genes.  Figure \ref{fig:connected_to_hubs_concord_parsec} compares the node degree of previously identified $\ell_1$-based hub genes, the node degree of their strong connections, and the node degree of the top hub genes identified by PARSEC as the critical screening level $\rho$ is again varied. Table \ref{tbl:hub_gene_table_summary} also outlines the degree of these sets of genes at specific screening levels.  It is clear that PARSEC's results concur with previous studies, while also permitting further sensitivity analysis within an inferential setting.  Note in addition that limited comparative sample sizes and heterogeneity amongst cancer patients and tumor types present significant barriers to developing targeted and personalized therapies using existing methods \parencite{Campbell:2020}.
PARSEC, however, allows the identification of critical gene relationships for specific diagnoses or sub-groups due to the theoretical safeguards it provides in the small-$n$ inferential screening framework.  Hence, the sensitivity analysis demonstrated in this section can also be extended to the same specific diagnoses or sub-groups.

{
\setstretch{1.2}
\footnotesize
\begin{longtable}{ p{1.5cm} p{1.8cm} p{2.5cm} p{6cm}  p{1cm} p{1.5cm}}
\centering
& & & & \multicolumn{2}{c}{\textbf{Hub Degree}} \\ \cmidrule(r){5-6}
Method & Gene Name & Accession \# & Description & \rotatebox{90}{\parbox{1.8cm}{\textbf{FWER, \\ $\alpha = 0.05$} }} &
\rotatebox{90}{\parbox{1.8cm}{\textbf{k-FWER, \\ $\alpha = 0.05$ \\ $k=5000$} }} \\
  \midrule
\endhead
CONCORD & TPX2 & AB024704 & TPX2, microtubule-associated, homolog (Xenopus laevis) &   0 &  11 \\
  CONCORD & ESR1 & NM\_000125 & estrogen receptor 1 &   0 &   2 \\
  CONCORD & FOXC1 & NM\_001453 & forkhead box C1 &   0 &  18 \\
  CONCORD & GATA3 & NM\_002051 & GATA binding protein 3 &   0 &  11 \\
  CONCORD & FZD9 & NM\_003508 & frizzled homolog 9 (Drosophila) &   0 &   6 \\
  CONCORD & FOXA1 & NM\_004496 & forkhead box A1 &   0 &  11 \\
  CONCORD & KRT16 & NM\_005557 & keratin 16 (focal non-epidermolytic palmoplantar keratoderma) &   0 &  23 \\
  CONCORD & KIF2C & NM\_006845 & kinesin family member 2C &   0 &  50 \\ \cmidrule(r){2-6}
   & LOR & NM\_000427 & loricrin &  12 & 101 \\
   & HIST1H2BG & NM\_003518 & histone cluster 1, H2bg &  12 &  21 \\
   & CRYBB3 & NM\_004076 & crystallin, beta B3 &  12 & 124 \\
   & PHYHIPL & Contig1558\_RC & phytanoyl-CoA 2-hydroxylase interacting protein-like &  10 & 106 \\
   & PLIN & NM\_002666 & perilipin &  10 &  64 \\
   & PPP1R3A & NM\_002711 & protein phosphatase 1, regulatory (inhibitor) subunit 3A (glycogen and sarcoplasmic reticulum binding subunit, skeletal muscle) &  10 & 100 \\
   & TRAF4 & NM\_004295 & TNF receptor-associated factor 4 &  10 &  41 \\
   & KRT2 & NM\_000423 & keratin 2 (epidermal ichthyosis bullosa of Siemens) &   9 &  96 \\
   & DLST & NM\_001933 & dihydrolipoamide S-succinyltransferase (E2 component of 2-oxo-glutarate complex) &   9 &  52 \\
   & HPD & NM\_002150 & 4-hydroxyphenylpyruvate dioxygenase &   9 &  54 \\
   & CLCNKA & NM\_004070 & chloride channel Ka &   9 & 103 \\
   & ACRV1 & NM\_020107 & acrosomal vesicle protein 1 &   9 & 105 \\
   & HIST1H2BK & AJ223352 & histone cluster 1, H2bk &   8 &  23 \\
   & IFIT3 & NM\_001549 & interferon-induced protein with tetratricopeptide repeats 3 &   8 &  88 \\
   & CLDN16 & NM\_006580 & claudin 16 &   8 & 126 \\
   & PCBP3 & NM\_020528 & poly(rC) binding protein 3 &   8 & 114 \\
   & NUT & AL137416 & nuclear protein in testis &   7 &  48 \\
   & INTS4 & Contig3396\_RC & integrator complex subunit 4 &   7 &  58 \\
   & ARMETL1 & Contig39875 & arginine-rich, mutated in early stage tumors-like 1 &   7 &  69 \\
   & CLNS1A & NM\_001293 & chloride channel, nucleotide-sensitive, 1A &   7 &  24 \\
   & PDC & NM\_002597 & phosducin &   7 & 120 \\
   & ASH2L & NM\_004674 & ash2 (absent, small, or homeotic)-like (Drosophila) &   7 &  32 \\
   & HIST1H1C & NM\_005319 & histone cluster 1, H1c &   7 &  20 \\
   & KRT1 & NM\_006121 & keratin 1 (epidermolytic hyperkeratosis) &   7 &  65 \\
   & LSM1 & NM\_014462 & LSM1 homolog, U6 small nuclear RNA associated (S. cerevisiae) &   7 &  29 \\
   & ALG8 & AJ224875 & asparagine-linked glycosylation 8 homolog (S. cerevisiae, alpha-1,3-glucosyltransferase) &   6 &  23 \\
   & RAB11FIP1 & Contig1682\_RC & RAB11 family interacting protein 1 (class I) &   6 &  22 \\
   & CATSPER3 & Contig22546\_RC & cation channel, sperm associated 3 &   6 &  87 \\
   & SPFH2 & Contig2811\_RC & SPFH domain family, member 2 &   6 &  20 \\
   & AQP11 & Contig34222\_RC & aquaporin 11 &   6 &  17 \\
   & SASP & Contig43806\_RC & skin aspartic protease &   6 &  41 \\
   & C6orf165 & Contig48907\_RC & chromosome 6 open reading frame 165 &   6 &  67 \\
   & NARS2 & Contig51414\_RC & asparaginyl-tRNA synthetase 2 (mitochondrial)(putative) &   6 &  23 \\
   & TSPAN9 & Contig54824 & tetraspanin 9 &   6 & 116 \\
   & MX1 & NM\_002462 & myxovirus (influenza virus) resistance 1, interferon-inducible protein p78 (mouse) &   6 &  66 \\
   & TULP2 & NM\_003323 & tubby like protein 2 &   6 &  77 \\
   & HIST3H3 & NM\_003493 & histone cluster 3, H3 &   6 &  50 \\
   & NDUFC2 & NM\_004549 & NADH dehydrogenase (ubiquinone) 1, subcomplex unknown, 2, 14.5kDa &   6 &  44 \\
   & DGCR2 & NM\_005137 & DiGeorge syndrome critical region gene 2 &   6 &  52 \\
   & LIPE & NM\_005357 & lipase, hormone-sensitive &   6 &  54 \\
   & B3GALT5 & NM\_006057 & UDP-Gal:betaGlcNAc beta 1,3-galactosyltransferase, polypeptide 5 &   6 &  89 \\
   & RSF1 & NM\_016578 & remodeling and spacing factor 1 &   6 &  29 \\
   & C17orf63 & NM\_018182 & chromosome 17 open reading frame 63 &   6 &  29 \\
   & CA10 & NM\_020178 & carbonic anhydrase X &   6 &  52 \\
   & TNFAIP1 & NM\_021137 & tumor necrosis factor, alpha-induced protein 1 (endothelial) &   6 &  40 \\
   & KIAA1189 & AB033015 & KIAA1189 &   5 &  66 \\
   & IFT20 & AF070643 & intraflagellar transport 20 homolog (Chlamydomonas) &   5 &  34 \\
   & UNQ467 & Contig51687\_RC & KIPV467 &   5 &  71 \\
   & CSMD1 & Contig63051 & CUB and Sushi multiple domains 1 &   5 & 102 \\
   & TNFAIP1 & M80783 & tumor necrosis factor, alpha-induced protein 1 (endothelial) &   5 &  31 \\
   & ADH1A & NM\_000667 & alcohol dehydrogenase 1A (class I), alpha polypeptide &   5 &  51 \\
   & H3F3A & NM\_002107 & H3 histone, family 3A &   5 &  36 \\
   & MEF2C & NM\_002397 & MADS box transcription enhancer factor 2, polypeptide C (myocyte enhancer factor 2C) &   5 &  90 \\
   & HIST1H3E & NM\_003532 & histone cluster 1, H3e &   5 &  50 \\
   & HIST1H3J & NM\_003535 & histone cluster 1, H3j &   5 &  46 \\
   & GPAA1 & NM\_003801 & glycosylphosphatidylinositol anchor attachment protein 1 homolog (yeast) &   5 &  40 \\
   & NCR2 & NM\_004828 & natural cytotoxicity triggering receptor 2 &   5 &  42 \\
   & ISG15 & NM\_005101 & ISG15 ubiquitin-like modifier &   5 &  59 \\
   & UNC119 & NM\_005148 & unc-119 homolog (C. elegans) &   5 &  35 \\
   & CCL27 & NM\_006664 & chemokine (C-C motif) ligand 27 &   5 &  60 \\
   & SPFH2 & NM\_007175 & SPFH domain family, member 2 &   5 &  21 \\
   & POLDIP2 & NM\_015584 & polymerase (DNA-directed), delta interacting protein 2 &   5 &  35 \\
   & WHSC1L1 & NM\_017778 & Wolf-Hirschhorn syndrome candidate 1-like 1 &   5 &  27 \\
   & BRF2 & NM\_018310 & BRF2, subunit of RNA polymerase III transcription initiation factor, BRF1-like &   5 &  27 \\
   & AICDA & NM\_020661 & activation-induced cytidine deaminase &   5 &  66 \\
   & HIST1H3F & NM\_021018 & histone cluster 1, H3f &   5 &  54 \\
   & HIST1H2BD & AJ223353 & histone cluster 1, H2bd &   4 &  21 \\
   & BRD1 & AL080149 & bromodomain containing 1 &   4 &  46 \\
   & TMEM97 & AW139198\_RC & transmembrane protein 97 &   4 &  31 \\
   & C17orf32 & Contig1804\_RC & chromosome 17 open reading frame 32 &   4 &  37 \\
   & C16orf78 & Contig25830\_RC & chromosome 16 open reading frame 78 &   4 &  80 \\
   & DPPA3 & Contig34895\_RC & developmental pluripotency associated 3 &   4 &  69 \\
   & C17orf63 & Contig39950\_RC & chromosome 17 open reading frame 63 &   4 &  30 \\
   & FCRL4 & Contig41826\_RC & Fc receptor-like 4 &   4 &  48 \\
   & C12orf50 & Contig42142\_RC & chromosome 12 open reading frame 50 &   4 &  44 \\
   & PERLD1 & Contig56503\_RC & per1-like domain containing 1 &   4 &  29 \\
   & LPL & NM\_000237 & lipoprotein lipase &   4 &  54 \\
   & PTGER3 & NM\_000957 & prostaglandin E receptor 3 (subtype EP3) &   4 & 126 \\
   & AQP7 & NM\_001170 & aquaporin 7 &   4 &  62 \\
   & FOXJ1 & NM\_001454 & forkhead box J1 &   4 &  43 \\
   & DSG1 & NM\_001942 & desmoglein 1 &   4 &  46 \\
   & E2F3 & NM\_001949 & E2F transcription factor 3 &   4 &  67 \\
   & IFI6 & NM\_002038 & interferon, alpha-inducible protein 6 &   4 &  52 \\
   & NR4A1 & NM\_002135 & nuclear receptor subfamily 4, group A, member 1 &   4 &  47 \\
   & NEUROD1 & NM\_002500 & neurogenic differentiation 1 &   4 & 134 \\
   & PIP5K1B & NM\_003558 & phosphatidylinositol-4-phosphate 5-kinase, type I, beta &   4 &  54 \\
   & MTMR2 & NM\_003912 & myotubularin related protein 2 &   4 &  59 \\
   & FLOT2 & NM\_004475 & flotillin 2 &   4 &  23 \\
   & DDX3Y & NM\_004660 & DEAD (Asp-Glu-Ala-Asp) box polypeptide 3, Y-linked &   4 &  63 \\
   & GRB7 & NM\_005310 & growth factor receptor-bound protein 7 &   4 &  25 \\
   & GRAP & NM\_006613 & GRB2-related adaptor protein &   4 &  35 \\
   & NRG1 & NM\_013956 & neuregulin 1 &   4 &  25 \\
   & ANGPTL3 & NM\_014495 & angiopoietin-like 3 &   4 &  52 \\
   & ORMDL3 & NM\_016471 & ORM1-like 3 (S. cerevisiae) &   4 &  26 \\
   & SEMA6A & NM\_020681 & sema domain, transmembrane domain (TM), and cytoplasmic domain, (semaphorin) 6A &   4 &  73 \\
   & ORC4L & NM\_002552 & origin recognition complex, subunit 4-like (yeast) &   3 & 107 \\
   & CYP2B6 & X13494 & cytochrome P450, family 2, subfamily B, polypeptide 6 &   3 & 120 \\
   & CYP2B6 & M29873 & cytochrome P450, family 2, subfamily B, polypeptide 6 &   2 &   9 \\
   & CYP2B6 & M29874 & cytochrome P450, family 2, subfamily B, polypeptide 6 &   2 &   9 \\
   & CYP2B6 & NM\_000767 & cytochrome P450, family 2, subfamily B, polypeptide 6 &   2 &  13 \\
   & IL13RA2 & U70981 & interleukin 13 receptor, alpha 2 &   2 & 127 \\
   & AKAP11 & AK002166 & A kinase (PRKA) anchor protein 11 &   1 & 121 \\
   & IL13RA2 & NM\_000640 & interleukin 13 receptor, alpha 2 &   1 &  20 \\
   & HSPG2 & NM\_005529 & heparan sulfate proteoglycan 2 (perlecan) &   1 & 137 \\
   & PTGER3 & AL050227 & prostaglandin E receptor 3 (subtype EP3) &   0 &   4 \\
   & CSMD1 & Contig27001\_RC & CUB and Sushi multiple domains 1 &   0 &  28 \\
   & PPAPDC1A & Contig39616\_RC & phosphatidic acid phosphatase type 2 domain containing 1A &   0 & 130 \\
   & CD53 & M37033 & CD53 molecule &   0 &  77 \\
   & WAS & NM\_000377 & Wiskott-Aldrich syndrome (eczema-thrombocytopenia) &   0 & 146 \\
   & COL5A2 & NM\_000393 & collagen, type V, alpha 2 &   0 & 127 \\
   & CD53 & NM\_000560 & CD53 molecule &   0 & 167 \\
   & ACRV1 & NM\_001612 & acrosomal vesicle protein 1 &   0 &  20 \\
   & UMOD & NM\_003361 & uromodulin (uromucoid, Tamm-Horsfall glycoprotein) &   0 & 140 \\
   & EYA4 & NM\_004100 & eyes absent homolog 4 (Drosophila) &   0 &  92 \\
   & AKAP11 & NM\_016248 & A kinase (PRKA) anchor protein 11 &   0 &  23 \\
   & ACRV1 & NM\_020069 & acrosomal vesicle protein 1 &   0 &  13 \\
   & ACRV1 & NM\_020110 & acrosomal vesicle protein 1 &   0 &  86 \\
   & ACRV1 & NM\_020111 & acrosomal vesicle protein 1 &   0 &  17 \\
   & ACRV1 & NM\_020113 & acrosomal vesicle protein 1 &   0 &  25 \\
   & ACRV1 & NM\_020115 & acrosomal vesicle protein 1 &   0 &  11 \\
   & DLST & S72422 & dihydrolipoamide S-succinyltransferase (E2 component of 2-oxo-glutarate complex) &   0 &  13 \\
   \bottomrule
\caption[Breast cancer gene screening, hub degree comparison]{Comparison of the degree of hub genes, following the application of FWER ($\alpha=0.05$) and k-FWER ($\alpha=0.05$, $k=5000$).  It is clear that the set of previously identified hub genes from CONCORD, are also identified by PARSEC when we lower the critical screening level to a k-FWER screening level. We limit the genes included in this table to the previously identified set of CONCORD hub genes, the PARSEC's top hub genes discussed in Table \ref{tbl:breastCancerGenes}, and all remaining hub genes that present a node degree greater than three when using FWER ($\alpha=0.05$). \label{tbl:hub_gene_table_summary}}
\end{longtable}
}

\pagebreak

\section{Minimum Variance Portfolio Selection} \label{appendix:finance}

\subsection{Extended Motivation, Data, Method and Results} \label{appendix:finance_summary}

We now demonstrate how PARSEC can be leveraged in yet another application domain: down-stream covariance screening and estimation for the purpose of portfolio selection in finance.  Broadly speaking, for a given selection of $p$ shares, the aim of portfolio selection is to determine the set of optimal portfolio weights subject to a minimum expected rate of return or an acceptable level of risk.  Within the setting of the Markowitz mean-variance portfolio theory, a stable estimate of the (inverse) covariance matrix is a critical input in determining optimal weights \parencite{Markowitz:1952}.  Hence, this application provides a suitable context to assess PARSEC's performance in comparison with other competing methods.

In order to compare the performance of covariance estimation methods, we adopt the minimum variance portfolio framework as implemented by \textcite{WonEtAl:2013}. The goal here it to minimize the volatility of a given portfolio. Let $\mathbf{\Sigma}_t$ denote the covariance matrix of the daily returns for period~$t$. The minimum variance portfolio problem is defined as:
\vspace{-0.2in}
\begin{equation*}
\text{min} \quad  \mathbf{w}^\top_t \mathbf{\Sigma}_t \mathbf{w}_t \quad \text{subject to} \; \textbf{1}^\top  \mathbf{w}_t = 1,
\vspace{-0.2in}
\end{equation*}
where $\mathbf{w}_t$ denotes portfolio weights.  The above optimization problem has an analytic solution, which is a function of the inverse covariance matrix, and is given by $\boldsymbol{\mathbf{w}}^*_{t} = (\textbf{1}^\top\mathbf{\Sigma}_t^{-1}\textbf{1})^{-1} \mathbf{\Sigma}_t^{-1} \textbf{1}$.  Since the data is non-stationary, a rolling covariance estimate is required.  We re-estimate~$\mathbf{\Sigma}_t$ repeatedly at the beginning of each investment period~$t$, using a sample size of $n$ daily (adjusted) returns preceding the period.  As discussed (and implemented) in \textcite{WonEtAl:2013}, this re-estimation procedure enables re-balancing of the portfolio weights using recent financial returns data.  Hence, the re-estimation strategy addresses the challenge of non-stationarity in financial returns.

Previous illustrations of the minimum-variance portfolio problem focus on the Dow-Jones index, which consists of only 30 securities.  PARSEC's scalability allows us to easily consider the same application on the broader S\&P500 index.  We use a 20-year investment horizon starting from January 1, 1995 and ending on January 1, 2015.  Recall that non-stationarity in financial returns necessitates a rolling covariance estimate.  Hence, we recast our long investment horizon problem as a multi-period problem with shorter monthly investment periods.  Note that we re-estimate $\Sigma_t$ using past data from the ``estimation horizon" period.  These estimates are then used to compute the portfolio weights, $\boldsymbol{\mathbf{w}}^*_{t}$, at the beginning of each monthly investment (or ``hold-out") period, with $\boldsymbol{\mathbf{w}}^*_{t}$ held constant until the next investment period.  As in \textcite{WonEtAl:2013}, different estimation horizons are used to examine the trade-off between sample size requirement and lack of stationarity (longer estimation horizons provide more samples whereas shorter estimation horizons better address stationarity but yields fewer samples).  We consider the following number of (adjusted) daily returns used to estimate~$\mathbf{\Sigma}_t$: 1, 2, 3, 6 and 12 months.  As constituents enter and drop out of the S\&P500 periodically, the number of stocks considered in each estimation and investment periods varies from month to month: from a minimum of 420 to a maximum of 492 stocks.  We note that the number of stocks is also limited by the availability of price data, which was obtained from Compustat and accessed via Wharton Research Data Services.

We employ the PARSEC and PCS-Hub approaches to identify significant partial correlations among the different stocks, thus specifying the underlying partial correlation graph structure.  The screening levels of the two approaches are determined by controlling either FWER, k-FWER or FDR.  The FDR screening level is specific only to PARSEC, as this option is not available for PCS-Hub.  Once the graph structure is determined, we obtain estimates of the non-zero elements of the inverse covariance matrix by employing either a likelihood-based or a pseudo-likelihood-based estimation approach (see Appendix Section \ref{appendix:inv_cov_estim} for the details of the algorithms).
We also compare PARSEC to CONCORD, which is a leading $\ell_1$-penalized pseudo-likelihood method (implemented using the R package \texttt{gconcord} available on CRAN). We use cross-validation to determine the value of the CONCORD penalty parameter, where the objective is to minimize prediction risk within each estimation horizon.  Further details on the cross-validation strategy are provided in \textcite{KhareOhRajaratnam:2015}.  Although neither PARSEC nor PCS-Hub require selection of a computationally expensive penalty parameter, for comparison purposes we also include results for both methods when using the same cross-validation strategy.

\begin{figure}[t!]
\includegraphics[width=15cm]{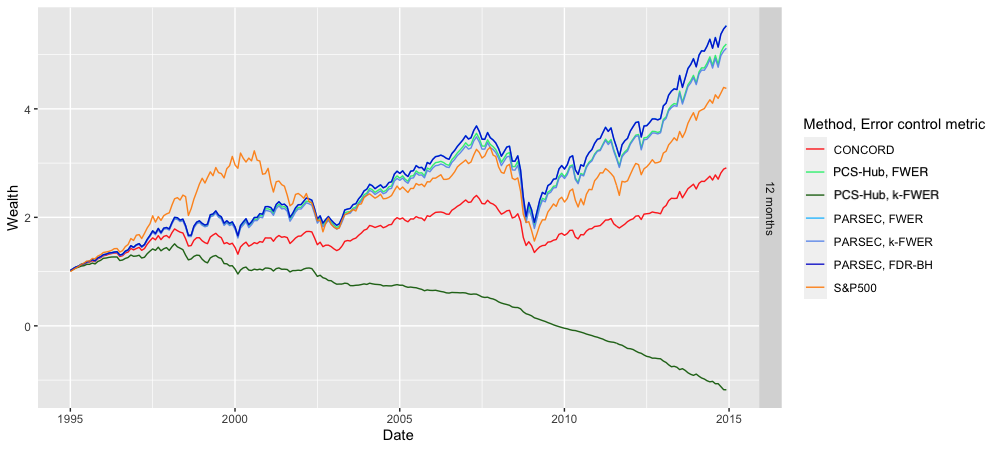}
\centering
\caption[Wealth growth for 12 month estimation horizon]{Wealth growth using a 12-month estimation horizon for various methods together with specific error control metrics.  Error control metrics for PARSEC and PCS-Hub use a) FWER where $\alpha=0.05$, b) k-FWER where $ \alpha=0.01, k\approx1000$; and c) FDR-BH screening where $\alpha=0.05$ for PARSEC.\label{fig:wealth_mvpt}}
\end{figure}

We apply back-testing to compare the behavior of each portfolio using the performance metrics implemented in \textcite{WonEtAl:2013}.  In particular, for a given investment period we use the optimal weights estimated from the preceding estimation horizon to track the performance of the corresponding portfolio at the end of that particular ``hold-out" period.  Doing so provides an objective approach to assess portfolio performance, as at any point only data from the past can be utilized for estimating optimal portfolio weights.  Figure~\ref{fig:wealth_mvpt} compares each method's wealth growth over the total investment horizon when using a 12-month estimation horizon to determine portfolio weights ($n\approx20$).  PARSEC's strong growth irrespective of the error control metric used demonstrates the consistency of its performance. Substantial additional details regarding the methods' performance using different error control metrics are provided in Appendix Section \ref{appendix:sensitivity_finance}, where we illustrate that PARSEC's superior normalized wealth is a consequence of yielding stable portfolios (due to lower turnover and transaction costs when compared to competing approaches).  For comparison purposes, the S\&P500 wealth is also included in Figure~\ref{fig:wealth_mvpt}. Here, the S\&P500 is calculated as the normalized change in the S\&P500 price for a unit investment in the index.
Hence, the S\&P500 wealth displayed in Figure~\ref{fig:wealth_mvpt} does not reflect additional transaction or management costs.  In spite of this, PARSEC has an overall higher wealth growth over the 20-year investment horizon, underscoring the fact that PARSEC can deliver better performance than passive trading strategies.

\begin{table}[t]
\small
\begin{tabular}{ p{1.8cm} p{1.05cm} p{1.05cm} p{1.05cm} p{1.05cm} p{1.05cm} p{1.05cm} p{1.05cm} p{1.05cm} p{1.05cm} p{1.05cm} }
\toprule
&  & \multicolumn{5}{c}{\textbf{PARSEC}} & \multicolumn{4}{c}{\textbf{PCS-Hub}}  \\ \cmidrule(r){3-7} \cmidrule(r){8-11}
$\textbf{N}_{est}$ &  \rotatebox{90}{\parbox{2cm}{\textbf{CONCORD}, \\ CV }}  &  \rotatebox{90}{\parbox{2cm}{\textbf{FWER, \\ $\alpha = 0.05$} }} & \rotatebox{90}{\parbox{2cm}{\textbf{k-FWER, \\ $\alpha = 0.01, \\ k=\approx1000$} }} &  \rotatebox{90}{\parbox{2cm}{\textbf{k-FWER, \\ $\alpha = 0.05, \\ k\approx5000$} }} &  \rotatebox{90}{\parbox{2cm}{\textbf{FDR-BH, \\ $\alpha = 0.05$}}} & \rotatebox{90}{\parbox{2cm}{\textbf{PARSEC} \\ CV }} &  \rotatebox{90}{\parbox{2cm}{\textbf{FWER, \\ $\alpha = 0.05$} }} & \rotatebox{90}{\parbox{2cm}{\textbf{k-FWER, \\ $\alpha = 0.01, \\ k\approx1000$} }} &  \rotatebox{90}{\parbox{2cm}{\textbf{k-FWER, \\ $\alpha = 0.05, \\ k\approx5000$}}} &  \rotatebox{90}{\parbox{2cm}{\textbf{PCS-Hub}\\ CV }}   \\
\midrule
1 month & -1.198 & \textbf{0.424} & 0.221 & -0.809 & \textbf{0.424} & 0.107 & 0.418 & -0.337 & -1.145 & -0.210 \\  \cmidrule(r){2-11}
2 months & 0.194 & \textbf{0.423} & 0.297 & -0.679 & \textbf{0.424} & 0.232 & \textbf{0.420} & -0.900 & -0.311 & -0.372 \\  \cmidrule(r){2-11}
3 months & 0.168 & \textbf{0.420} & 0.316 & -0.595 & \textbf{0.420} & 0.240 & \textbf{0.417} & -0.865 & -1.038 & -0.583 \\  \cmidrule(r){2-11}
6 months & 0.274 & \textbf{0.444} & 0.381 & 0.128 & \textbf{0.445} & 0.352 & \textbf{0.447} & -0.443 & -0.709 & 0.110 \\  \cmidrule(r){2-11}
12 months & 0.259 & \textbf{0.462} & 0.434 & 0.337 & \textbf{0.462} & 0.423 & 0.446 & -0.391 & -0.843 & -1.016 \\
\bottomrule
\end{tabular}
\caption[Minimum Variance Portfolio Adjusted Sharpe Ratios]{Comparison of Adjusted Sharpe Ratios across different methods.  Similar to \textcite{KhareOhRajaratnam:2015}, the highest Adjusted Sharpe Ratios for each estimation horizon, and values within 1\% of this maximum, are highlighted in bold.  Note that PARSEC with FWER ($\alpha=0.05$) and FDR-BH ($\alpha=0.05$) consistently produce the highest Adjusted Sharpe Ratio.   \label{tbl:MVP_Adjusted_SharpeRatio_corr}}
\end{table}

For a fairer and more meaningful comparison, we also use the industry-standard measure, the Sharpe Ratio, to assess the effect of turnover and portfolio stability over the entire investment horizon.  In particular, we calculate the adjusted Sharpe Ratio, $SR_{A}=(r_{p}-r_{f})/\sigma_{p}$, where $r_{p}$ is the average of the monthly wealth returns for a given portfolio, $\sigma_{p}$ is the realized risk of monthly wealth (i.e., the standard deviation of monthly wealth returns), and  $r_{f}$ is the monthly risk-free rate.  In each investment period, wealth is determined as the portfolio return less transaction and borrowing costs, allowing us to holistically assess the long-term net return of each portfolio.  We report the adjusted Sharpe Ratio for competing methods in Table~\ref{tbl:MVP_Adjusted_SharpeRatio_corr} under different estimation horizons.  It is clear PARSEC  provides uniformly competitive estimates across every estimation horizon.  This superior performance is attained when using FWER screening ($\alpha=0.05$) and FDR-BH screening ($\alpha=0.05$).  In contrast, the PCS-Hub method exhibits volatile performance.  This can be seen in a number of ways: i) the adjusted Sharpe Ratio values corresponding to PCS-Hub are unstable (see Table \ref{tbl:MVP_Adjusted_SharpeRatio_corr}), ii) as screening levels are varied, the portfolio performance corresponding to PCS-Hub fluctuates severely (see Figure \ref{fig:tuning_k_KFWER_1month} to Figure \ref{fig:adj_sr_tuning_k_KFWER_1month_4horizons} in Appendix Section \ref{appendix:sensitivity_finance}), iii) portfolio weights, turnover and borrowing costs corresponding to PCS-Hub oscillate over time (see Table \ref{tbl:MVP_Turnover_corr} and Figure \ref{fig:weights_analysis} in Appendix Section \ref{appendix:sensitivity_finance}).  We further investigate PARSEC's performance and its properties (see Appendix Section \ref{appendix:sensitivity_finance} for more details). In particular, we demonstrate PARSEC's lower turnover and borrowing costs.  We also illustrate financial networks obtained by PARSEC in Figure \ref{fig:fwer_network} in Appendix Section~\ref{appendix:sensitivity_finance}, underscoring its vast potential for portfolio selection.

\subsection{Detailed sensitivity analysis of PARSEC's performance} \label{appendix:sensitivity_finance}

Estimation of non-zero elements of the covariance matrix, $\Sigma_t$, is obtained using either a likelihood or pseudo-likelihood estimation approach.  Once the non-zero structure is determined using either the screened PARSEC or screened PCS-Hub approach, we employ one of the algorithms provided in Appendix \ref{appendix:inv_cov_estim}.  Algorithm \ref{alg:concord_coordinatewise} specifies the pseudo-likelihood approach motivated by CONCORD \parencite{KhareOhRajaratnam:2015}, and Algorithm \ref{alg:esl_coordinatewise} specifies the Gaussian likelihood approach \parencite{HastieTibshiraniFriedman:2009}.  Thus far, we have reported results obtained using the pseudo-likelihood approach as it requires fewer assumptions and is more robust to heavy tails commonly observed in the financial returns data.  We now report the corresponding results for the Gaussian likelihood approach.

\begin{figure}[t!]
\includegraphics[width=15cm]{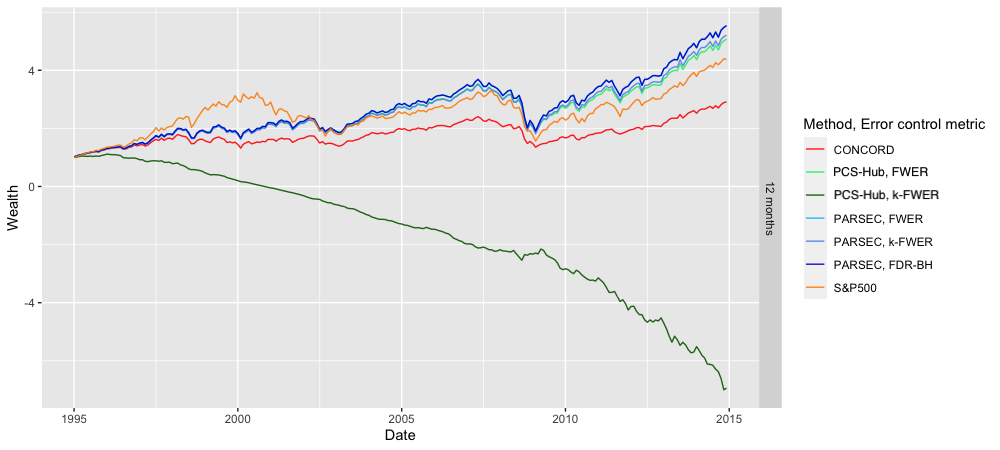}
\centering
\caption[Wealth for 12 month estimation horizon using the Gaussian likelihood approach for downstream covariance estimation]{Wealth using the 12-month estimation horizon and the Gaussian likelihood approach for downstream covariance estimation in screening methods.  Error control metrics for PARSEC and PCS-Hub use FWER where $\alpha=0.05$, and  $ \alpha=0.01, k\approx1000$ for k-FWER.  We also include Wealth from using FDR-BH control ($\alpha=0.05$) for PARSEC.  These results are consistent with the wealth growth obtained using the pseudo-likelihood approach for downstream covariance estimation illustrated in Figure \ref{fig:wealth_mvpt}. \label{fig:wealth_mvpt_gauss}}
\end{figure}

Table \ref{tbl:MVP_Adjusted_SharpeRatio_corr_gauss} reports the adjusted Sharpe ratios using the Gaussian likelihood approach for estimation of $\Sigma_t$.  We note that PARSEC screening still leads to stable results, which are consistent with the adjusted Sharpe ratios obtained using the pseudo-likelihood approach reported in Table \ref{tbl:MVP_Adjusted_SharpeRatio_corr}.  PARSEC adjusted Sharpe ratios also demonstrate greater consistency than those of other methods, even as we vary the screening level or increase~$k$ for k-FWER screening.  In contrast, the PCS-Hub approach exhibits high volatility once the screening level is lowered.  Figure \ref{fig:wealth_mvpt_gauss} illustrates the normalized wealth of competing methods using a 12-month estimation horizon and the Gaussian likelihood approach.  We note that PARSEC's normalized wealth again demonstrates greater stability even as we vary the screening level used.  PARSEC thus uniformly outperforms PCS-Hub across both down-stream covariance estimation approaches.

\begin{table}[ht!]
\small
\begin{tabular}{ p{1.8cm} p{1.05cm} p{1.05cm} p{1.05cm} p{1.05cm} p{1.05cm} p{1.05cm} p{1.05cm} p{1.05cm} p{1.05cm} p{1.05cm} }
\toprule
&  & \multicolumn{5}{c}{\textbf{PARSEC}} & \multicolumn{4}{c}{\textbf{PCS-Hub}}  \\ \cmidrule(r){3-7} \cmidrule(r){8-11}
$\textbf{N}_{est}$ &  \rotatebox{90}{\parbox{2cm}{\textbf{CONCORD}, \\ CV }}  &  \rotatebox{90}{\parbox{2cm}{\textbf{FWER, \\ $\alpha = 0.05$} }} & \rotatebox{90}{\parbox{2cm}{\textbf{k-FWER, \\ $\alpha = 0.01, \\ k=\approx1000$} }} &  \rotatebox{90}{\parbox{2cm}{\textbf{k-FWER, \\ $\alpha = 0.05, \\ k\approx5000$} }} &  \rotatebox{90}{\parbox{2cm}{\textbf{FDR-BH, \\ $\alpha = 0.05$}}} & \rotatebox{90}{\parbox{2cm}{\textbf{PARSEC} \\ CV }} &  \rotatebox{90}{\parbox{2cm}{\textbf{FWER, \\ $\alpha = 0.05$} }} & \rotatebox{90}{\parbox{2cm}{\textbf{k-FWER, \\ $\alpha = 0.01, \\ k\approx1000$} }} &  \rotatebox{90}{\parbox{2cm}{\textbf{k-FWER, \\ $\alpha = 0.05, \\ k\approx5000$}}} &  \rotatebox{90}{\parbox{2cm}{\textbf{PCS-Hub}\\ CV }}   \\
\midrule
1 month & -1.198 & \textbf{0.422} & 0.106 & -0.426 & \textbf{0.420} & \textbf{0.424}  & 0.406 & -0.951 & -0.225 & 0.413   \\  \cmidrule(r){2-11}
2 months & 0.194 & \textbf{0.425} & 0.257 & -1.076 & \textbf{0.421} & 0.403 & 0.415 & -0.239 & -0.308 & 0.249  \\  \cmidrule(r){2-11}
3 months & 0.168 & \textbf{0.418} & 0.282 & -1.140 & \textbf{0.417} & 0.154 & 0.411 & 0.075 & 0.070 & 0.171 \\  \cmidrule(r){2-11}
6 months & 0.274 & \textbf{0.443} & 0.369 & -0.622 & \textbf{0.441} & -0.096 & \textbf{0.444} & -0.304 & -1.175 & -0.500 \\  \cmidrule(r){2-11}
12 months & 0.259 & \textbf{0.462} & 0.436 & -0.133 & \textbf{0.462} & 0.348 & 0.407 & 0.063 & 0.028 & -0.854  \\
\bottomrule
\end{tabular}
\caption[Minimum Variance Portfolio Adjusted Sharpe Ratios using the Gaussian likelihood approach for downstream covariance estimation]{Comparison of Adjusted Sharpe Ratios across different methods using the realized return and risk of monthly wealth.  We use the Gaussian likelihood approach for estimating each estimation horizons covariance matrix, $\Sigma_t$, in this setting. PARSEC using FWER ($\alpha=0.05$), k-FWER ($\alpha=0.01, k\approx1000$) and FDR-BH control ($\alpha=0.05$)  consistently produces the highest Adjusted Sharpe Ratio, which is consistent with the results obtained when using the pseudo-likelihood approach to estimate $\Sigma_t$ reported in Table \ref{tbl:MVP_Adjusted_SharpeRatio_corr}.  The highest Adjusted Sharpe Ratios for each estimation horizon, and values within 1\% of this maximum, are highlighted in bold. \label{tbl:MVP_Adjusted_SharpeRatio_corr_gauss}}
\end{table}

\begin{figure}[!htbp]
\includegraphics[width=15cm]{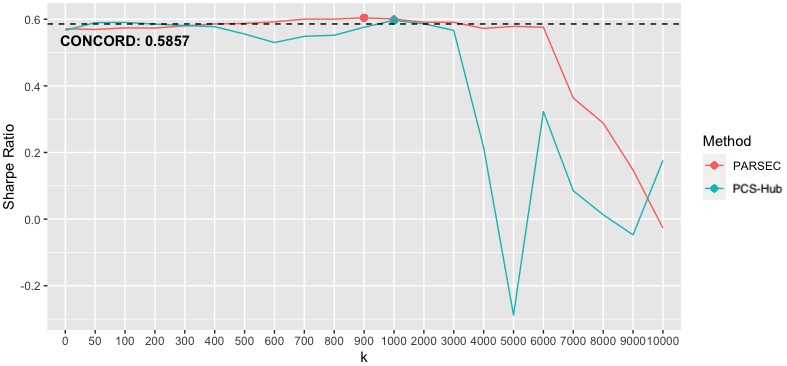}
\centering
\caption[Sharpe Ratio for 1 month estimation horizon]{Sharpe Ratios when using different $k$ in k-FWER screening and a 1-month estimation horizon.  The value of $\alpha$ when determining the screening level is held consistent at 0.01. \label{fig:tuning_k_KFWER_1month}}
\end{figure}

We report the additional metrics implemented by \textcite{KhareOhRajaratnam:2015} and \textcite{WonEtAl:2013}. We report results obtained using the pseudo-likelihood approach for down-stream covariance estimation in the remainder of this section, to match the initial results provided in Appendix Section \ref{appendix:finance_summary}.   Table \ref{tbl:MVP_SharpeRatio_corr} summarizes the Sharpe Ratio for $\textbf{N}_{est} \in \{1,2,3,6,12\}$ months.  Within each estimation horizon beyond 2 months, CONCORD's Sharpe Ratios generally outperforms both PARSEC and PCS-Hub.   However, when using an estimation horizon of 1 month which typifies a sample-staved ultra-high dimensional regime, PARSEC produces consistently competitive Sharpe Ratios across different screening levels.  Interestingly, there is also great variability in the Sharpe Ratio of PCS-Hub across different screening levels - PCS-Hub's variable performance when using k-FWER ($\alpha = 0.05$ and $k=5\%\approx5000$) provides evidence of inconsistent screening.  Additional analysis of the Sharpe Ratios obtained by PARSEC and PCS-Hub as $k$ is varied in k-FWER screening is provided in Figure \ref{fig:tuning_k_KFWER_1month} using a 1-month estimation horizon.

\begin{figure}[H]
\includegraphics[width=15cm]{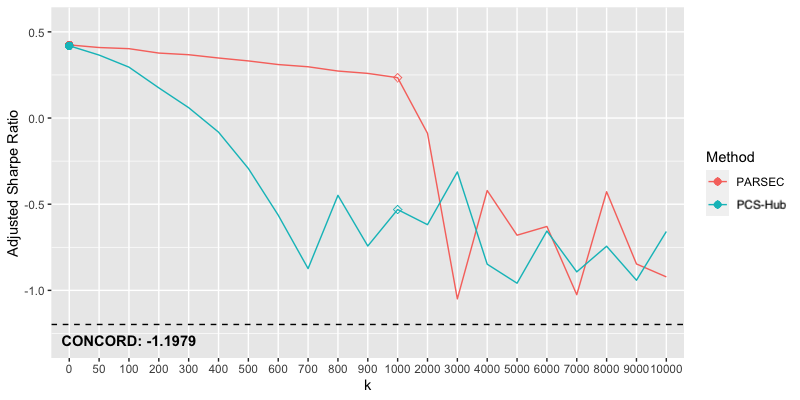}
\centering
\caption[Adjusted Sharpe Ratio for 1 month estimation horizon, varying $k$]{Adjusted Sharpe Ratios when using different $k$ in k-FWER screening and a 1-month estimation horizon.  The value of $\alpha$ when determining the screening level is held consistent at 0.01. \label{fig:adj_sr_tuning_k_KFWER_1month}}
\end{figure}

\begin{figure}[H]
\includegraphics[width=15cm]{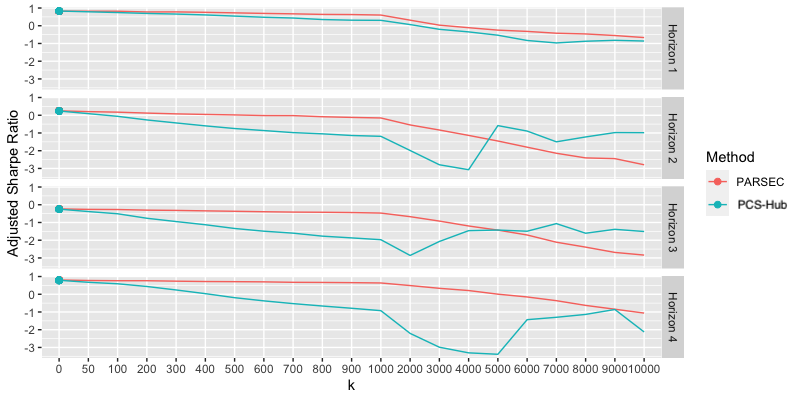}
\centering
\caption[Adjusted Sharpe Ratio for 1 month estimation horizon, varying $k$ over 4 horizons]{Adjusted Sharpe Ratios when using different $k$ in k-FWER screening and a 1-month estimation horizon.  In this setting, we have split the 20-year investment period into 5 years each, to assess the consistency of both methods when using different data.  The value of $\alpha$ when determining the screening level is held consistent at 0.01.  \label{fig:adj_sr_tuning_k_KFWER_1month_4horizons}}
\end{figure}

\begin{figure}[H]
\includegraphics[width=15cm]{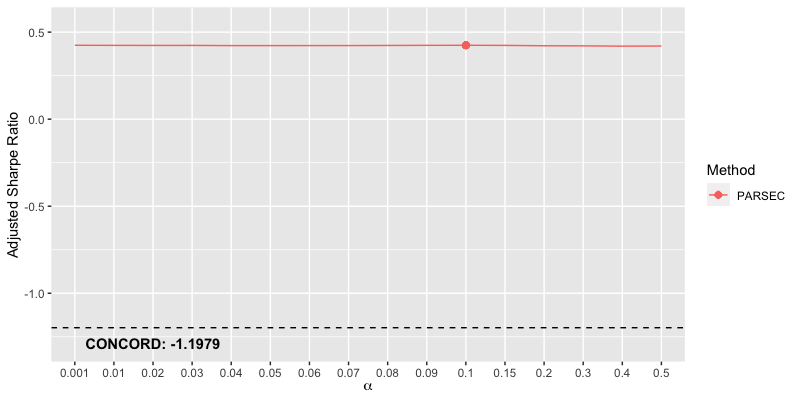}
\centering
\caption[Adjusted Sharpe Ratio for 1 month estimation horizon, varying $\alpha$ with FDR-BH control]{Adjusted Sharpe Ratios when using different $\alpha$ in FDR-BH control and a 1-month estimation horizon. \label{fig:adj_sr_tuning_alpha_fdr_bh_1month}}
\end{figure}

\begin{figure}[H]
\includegraphics[width=18cm]{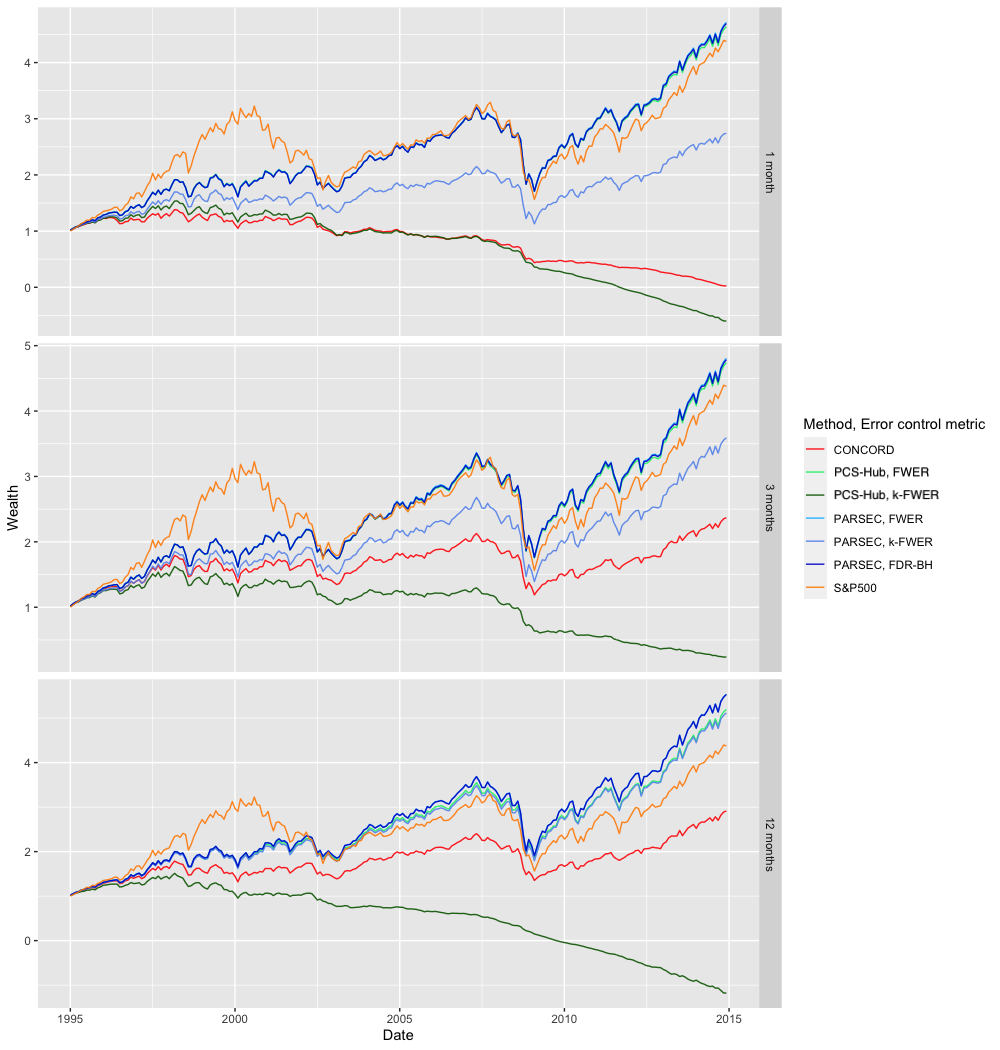}
\centering
\caption[Wealth over 1, 3 and 12 month horizons]{Wealth for 1 month (top), 3 month (middle) and 12 month (bottom) estimation horizons.  PARSEC and PCS-Hub estimates use $\alpha=0.05$ for FWER, and $\alpha=0.01, k\approx1000$ for k-FWER.  We set $\alpha=0.05$ for PARSEC FDR control. \label{fig:wealth_collated_1_3_12}}
\end{figure}

We also compare the adjusted Sharpe ratios for PARSEC and PCS-Hub as $k$ is varied using k-FWER screening in Figure \ref{fig:adj_sr_tuning_k_KFWER_1month}.  While PCS-Hub achieves higher adjusted Sharpe ratios at certain values of $k$, this performance appears to be an artifact of the data.  An additional comparison of performance when the investment horizon is split into four periods in Figure \ref{fig:adj_sr_tuning_k_KFWER_1month_4horizons}, reveals that the PCS-Hub method produces inconsistent adjusted Sharpe ratio values at the same values of $k$.   We also assess the performance of PARSEC as we vary $\alpha$ in the context of FDR-BH screening.  Figure \ref{fig:adj_sr_tuning_alpha_fdr_bh_1month} illustrates that the performance of PARSEC is very stable even as we significantly shift $\alpha$.  Hence, there is strong evidence that PARSEC provides superior and more stable uncertainty quantification as we adjust screening levels and/or vary the selected error control measure.  Note, PARSEC also demonstrates competitive performance when using different estimation horizons, as seen in the Wealth plots provided in Figure \ref{fig:wealth_collated_1_3_12}.  Even when using longer estimation horizons of 3 and 12 months, PARSEC still produces the strongest wealth growth indicating that the method is also flexible when moving to longer estimation horizons (or larger sample sizes).

{\begin{table}[htbp!]
\small
\begin{tabular}{ p{1.8cm} p{1.05cm} p{1.05cm} p{1.05cm} p{1.05cm} p{1.05cm} p{1.05cm} p{1.05cm} p{1.05cm} p{1.05cm} p{1.05cm} }
\toprule
&  & \multicolumn{5}{c}{\textbf{PARSEC}} & \multicolumn{4}{c}{\textbf{PCS-Hub}}  \\ \cmidrule(r){3-7} \cmidrule(r){8-11}
$\textbf{N}_{est}$ &  \rotatebox{90}{\parbox{2cm}{\textbf{CONCORD}, \\ CV }}  &  \rotatebox{90}{\parbox{2cm}{\textbf{FWER, \\ $\alpha = 0.05$} }} & \rotatebox{90}{\parbox{2cm}{\textbf{k-FWER, \\ $\alpha = 0.01, \\ k\approx1000$} }} &  \rotatebox{90}{\parbox{2cm}{\textbf{k-FWER, \\ $\alpha = 0.05, \\ k\approx5000$} }} &  \rotatebox{90}{\parbox{2cm}{\textbf{FDR-BH, \\ $\alpha = 0.05$}}} & \rotatebox{90}{\parbox{2cm}{\textbf{PARSEC} \\ CV }} &  \rotatebox{90}{\parbox{2cm}{\textbf{FWER, \\ $\alpha = 0.05$} }} & \rotatebox{90}{\parbox{2cm}{\textbf{k-FWER, \\ $\alpha = 0.01, \\ k\approx1000$} }} &  \rotatebox{90}{\parbox{2cm}{\textbf{k-FWER, \\ $\alpha = 0.05, \\ k\approx5000$}}} &  \rotatebox{90}{\parbox{2cm}{\textbf{PCS-Hub}\\ CV }}   \\
\midrule
1 month & 0.5857 & 0.5728 & 0.5969 & 0.6371 & 0.5722 & 0.5511 & 0.5701 & 0.6262 & 0.5895 & \textbf{0.6598} \\  \cmidrule(r){2-11}
2 months & 0.5668 & 0.5453 & 0.5451 & 0.5350 & 0.5462 & 0.5412 & 0.5465 & \textbf{0.5868} & 0.5612 & 0.5021 \\  \cmidrule(r){2-11}
3 months & \textbf{0.5426} & 0.5283 & 0.5204 & 0.5241 & 0.5282 & 0.5171 & 0.5283 & 0.5386 & \textbf{0.5477} & 0.4593 \\  \cmidrule(r){2-11}
6 months & 0.5521 & 0.5295 & 0.5173 & 0.5056 & 0.5298 & 0.5207 & 0.5334 & 0.5095 & 0.4605 & \textbf{0.6365} \\  \cmidrule(r){2-11}
12 months & \textbf{0.5653} & 0.5348 & 0.5300 & 0.5169 & 0.5347 & 0.5292 & 0.5369 & 0.5290 & 0.4906 & 0.5160 \\
\bottomrule
\end{tabular}
\caption[Minimum Variance Portfolio Sharpe Ratios]{Comparison of Sharpe Ratios across different methods.  For all  $\textbf{N}_{est} \in \{1,2,3,6\}$ months, CONCORD outperforms our screening based methods.  However, it is noticeable that despite different screening levels, covariance estimates using PARSEC provides consistently strong Sharpe Ratios in comparison to the PCS-Hub. Furthermore, cross-validation does not improve the performance of PARSEC, indicating that the use of a selected screening level provides robust and consistent estimates.  The highest Sharpe Ratios for each estimation horizon, and values within 1\% of this maximum, are highlighted in bold.  \label{tbl:MVP_SharpeRatio_corr}}
\end{table}}

\begin{figure}[H]
\includegraphics[width=18cm]{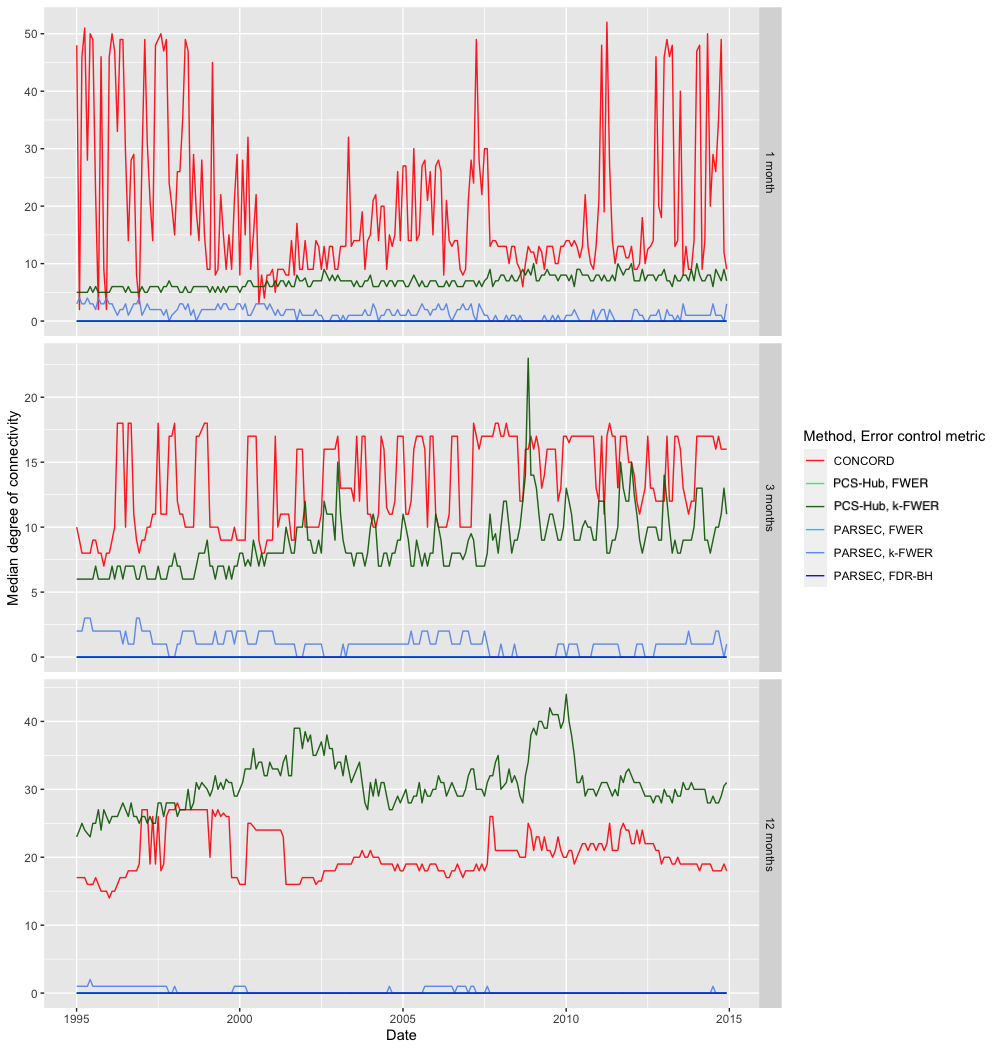}
\centering
\caption[Node distribution over the investment horizon]{Median degree of nodes (shares) over the investment horizon, using different estimation horizons.  PARSEC again provides the most stable degree distribution over time. We use a FWER screening level where $\alpha=0.05$ and k-FWER screening level where $k\approx1000$ and $\alpha=0.01$ for both PARSEC and PCS-Hub.  We set $\alpha=0.05$ for PARSEC FDR control.  \label{fig:node_degree_over_horizon}}
\end{figure}

\begin{figure}[H]
\includegraphics[width=18cm]{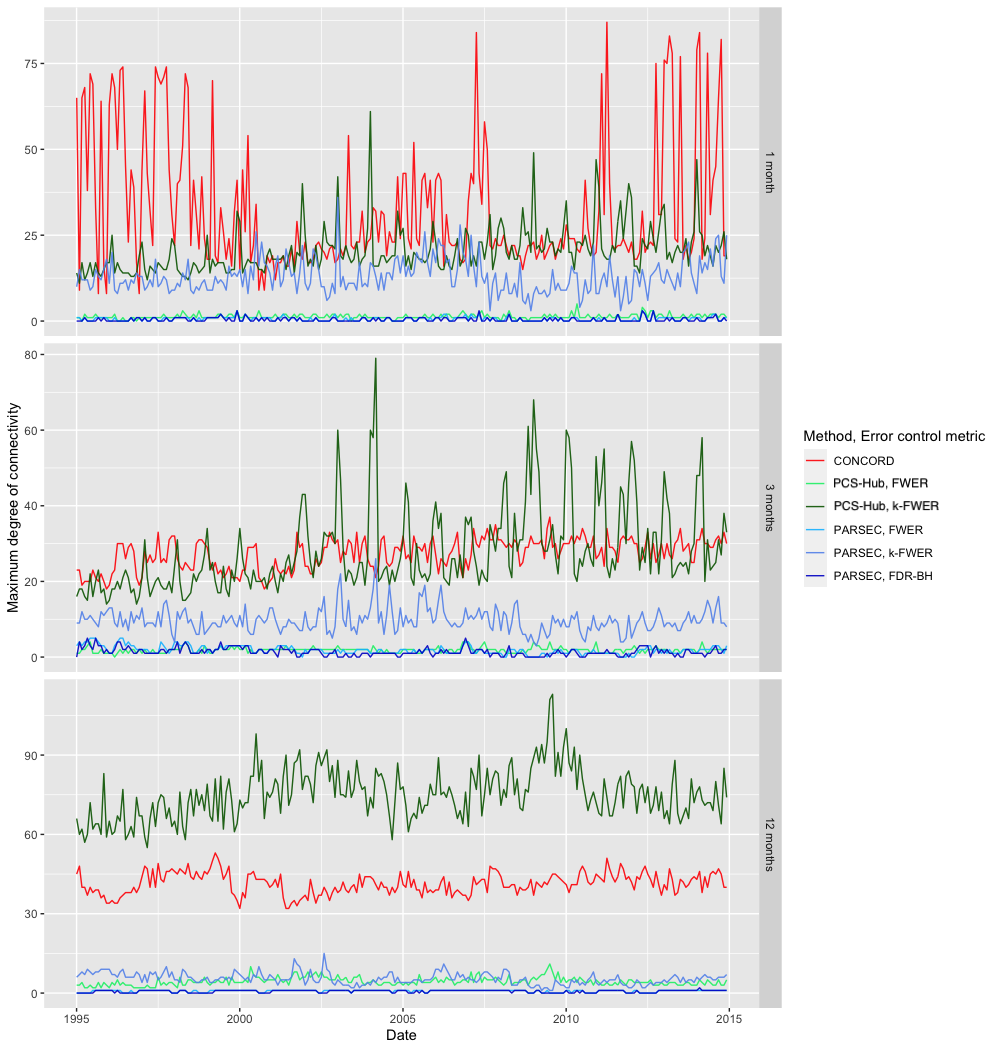}
\centering
\caption[Node distribution over the investment horizon]{Maximum degree of nodes (shares) over the investment horizon, using different estimation horizons.  PARSEC again provides the most stable degree distribution over time.  We use a FWER screening level where $\alpha=0.05$ and k-FWER screening level where $k\approx1000$ and $\alpha=0.01$ for both PARSEC and PCS-Hub.  We set $\alpha=0.05$ for PARSEC FDR control. \label{fig:node_degree_over_horizon_max}}
\end{figure}

PARSEC estimates also achieve greater sparsity than PCS-Hub.  This is evident in both Table \ref{tbl:MVP_estimation_density} which reports the mean density of the covariance matrix per estimation horizon and method, and Table \ref{tbl:average_median_degree} which reports the average and median degree of nodes over time (i.e. connectivity of shares).  The density of PARSEC's estimated covariance matrix, and its mean and median degree of nodes over the investment horizon is consistently lower than both PCS-Hub and CONCORD.  Figure \ref{fig:node_degree_over_horizon} and Figure \ref{fig:node_degree_over_horizon_max} depicts the median and maximum degree of nodes over the investment horizon.  It is illustrated that PARSEC estimates remain sparse even during financial crisis periods around 2001-2003 (the Dot-com bubble) and 2008-2010 (the Global Financial Crisis).  This is in contrast to PCS-Hub which fluctuates at these volatile times.

\begin{figure}[H]
\includegraphics[width=18cm]{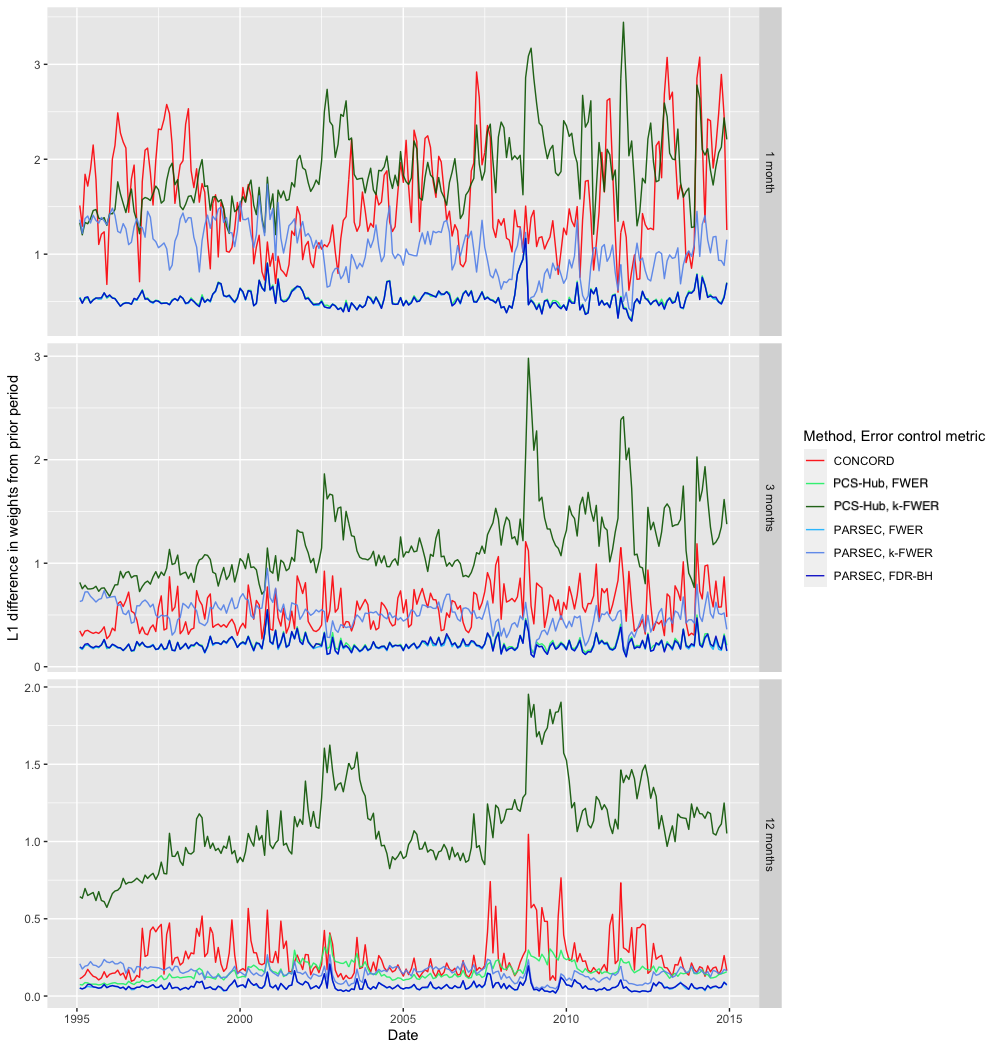}
\centering
\caption[Difference in portfolio weights]{L1 difference in consecutive portfolio weights over 1, 3 and 12 month estimation horizons, and using different error control metrics.  We use FWER  where $\alpha=0.05$ and k-FWER where $k\approx1000$ and $\alpha=0.01$ for both PARSEC and PCS-Hub.  We set $\alpha=0.05$ for PARSEC FDR control.  \label{fig:weights_analysis}}
\end{figure}

Table \ref{tbl:MVP_Turnover_corr} compares mean turnover per portfolio across the entire investment horizon.  These results further support the stability of PARSEC's error control properties across various estimation horizons.  PARSEC's portfolios consistently yield the lowest turnover, with less sensitivity to lower screening levels.  In contrast, PCS-Hub witnesses a drastic increase in turnover as we lower the screening level to k-FWER ($\alpha=0.05, k\approx5000$).  Figure \ref{fig:weights_analysis} compares the difference in portfolio weights in each consecutive estimation horizon over the investment horizon, illustrating the $\ell_1$ norm of the difference in weights between periods.  It is evident again that PARSEC provides the most stable estimates, with the lowest overall difference in weights from period to period.

\begin{figure}[!htbp]
\includegraphics[width=16cm]{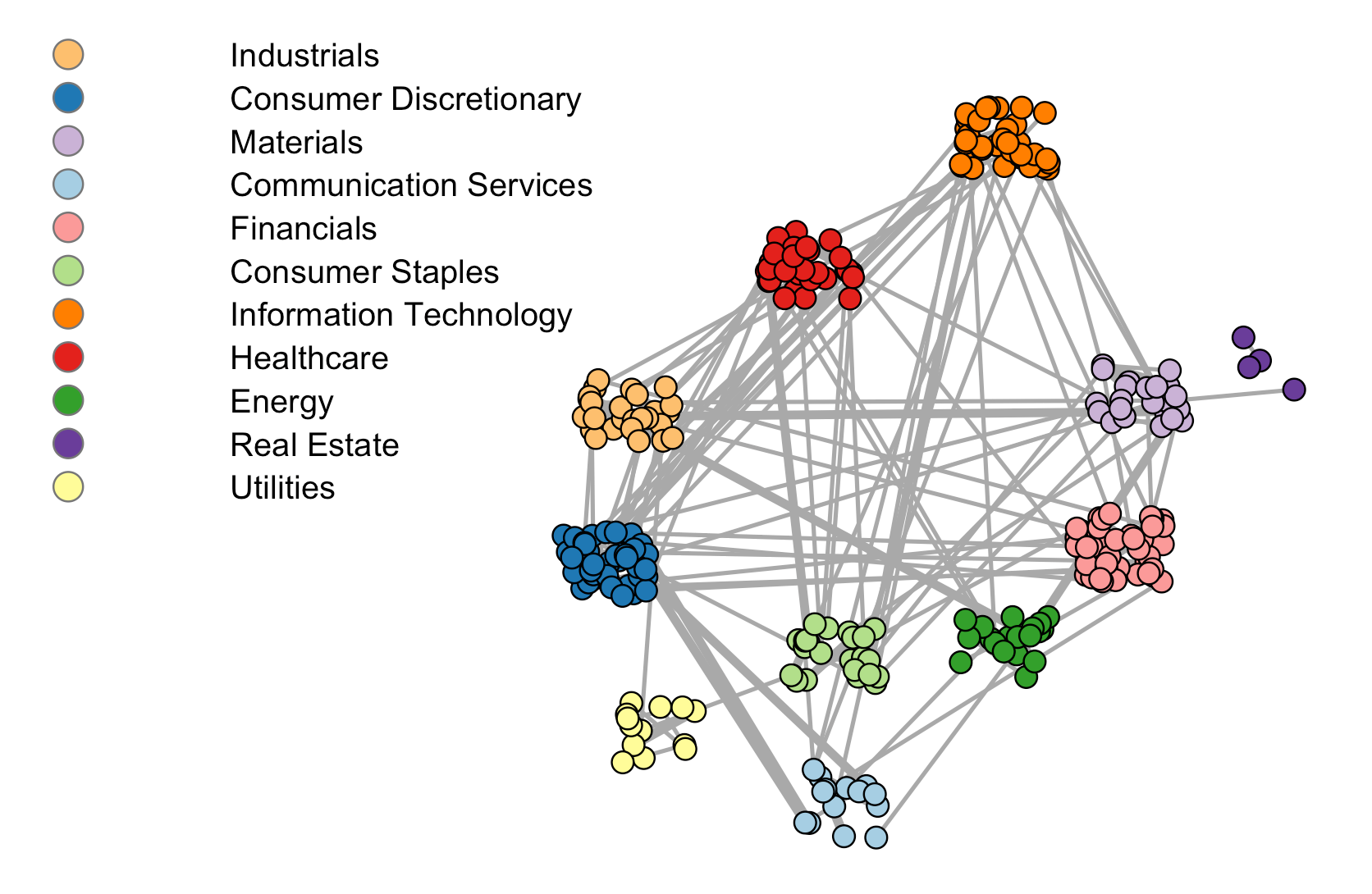}
\centering
\caption[Network of constituents determined by FWER]{Network of screened S\&P500 constituents over the investment period, by GICs sectors, using a FWER $\alpha=0.05$ screening level and 1 month estimation horizon.  Each node represents a screened constituent, with edge weights illustrating the relative volume of edges (or screened partial correlation coefficients) between the connected constituents.  We note a larger number of edges/connections within the same GICs sector, indicating that PARSEC independently identifies within-sector dependencies.  \label{fig:fwer_network}}
\end{figure}

\begin{figure}[H]
\includegraphics[width=16cm]{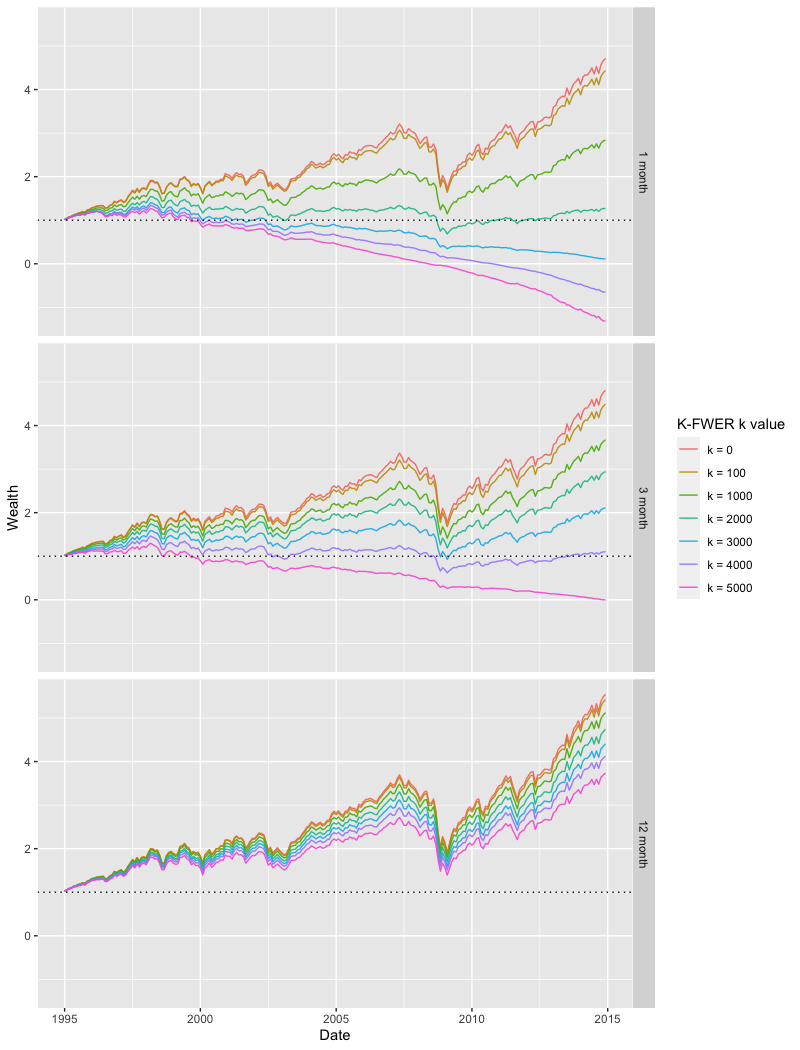}
\centering
\caption[Wealth for 1, 3 and 12 month estimation horizons]{PARSEC's normalized wealth for the 1 month (top), 3 month (middle) and 12 month (bottom) estimation horizons using different values of $k$ for k-FWER ($\alpha=0.01$).  It is clear that sparser estimates (achieved through conservative error control) produce more profitable portfolios. \label{fig:wealth_kfwer_threshold}}
\end{figure}

To further assess whether sparsity of the inverse covariance matrix contributes to more stable estimates and stronger wealth growth, Figure \ref{fig:wealth_kfwer_threshold} compares the normalized wealth in 1, 3 and 12 month estimation horizons.  Here, we vary the value of $k$ when using a k-FWER screening level ($\alpha=0.01$). From these results, it can be inferred that the sparsity of the inverse covariance matrix does contribute to portfolio performance, as the most stringent FWER screening level produces the strongest wealth growth over the investment horizon.  However, even with larger values of $k$ (up to $k=2000$ in the 1-month estimation horizon), we note that PARSEC demonstrates positive wealth growth.  The stable performance of PARSEC as $\alpha$ and $k$ vary, thus permits flexibility in the choice of PARSEC's screening level.  Portfolio or fund managers can select a screening level aligned with a preferred level of risk while still maintaining strong wealth growth.  Interestingly, an analysis of the financial networks estimated by PARSEC over the investment horizon reveals that PARSEC also independently identifies high partial correlations and covariances between industries.  Figure \ref{fig:fwer_network} illustrates that the node connectivity within industries is highest, when screening using a FWER screening level ($\alpha=0.05$) and 1 month estimation horizon. PARSEC captures these dependencies appropriately, and hence can guide diversification strategies.

{\begin{table}[htbp!]
\small
\begin{tabular}{ p{1.8cm} p{1.05cm} p{1.08cm} p{1.05cm} p{1.05cm} p{1.08cm} p{1.05cm} p{1.08cm} p{1.05cm} p{1.05cm} p{1.05cm} }
\toprule
&  & \multicolumn{5}{c}{\textbf{PARSEC}} & \multicolumn{4}{c}{\textbf{PCS-Hub}}  \\ \cmidrule(r){3-7} \cmidrule(r){8-11}
$\textbf{N}_{est}$ &  \rotatebox{90}{\parbox{2cm}{\textbf{CONCORD}, \\ CV }}  &  \rotatebox{90}{\parbox{2cm}{\textbf{FWER, \\ $\alpha = 0.05$} }} & \rotatebox{90}{\parbox{2cm}{\textbf{k-FWER, \\ $\alpha = 0.01, \\ k\approx1000$} }} &  \rotatebox{90}{\parbox{2cm}{\textbf{k-FWER, \\ $\alpha = 0.05, \\ k\approx5000$} }} &  \rotatebox{90}{\parbox{2cm}{\textbf{FDR-BH, \\ $\alpha = 0.05$}}} & \rotatebox{90}{\parbox{2cm}{\textbf{PARSEC} \\ CV }} &  \rotatebox{90}{\parbox{2cm}{\textbf{FWER, \\ $\alpha = 0.05$} }} & \rotatebox{90}{\parbox{2cm}{\textbf{k-FWER, \\ $\alpha = 0.01, \\ k\approx1000$} }} &  \rotatebox{90}{\parbox{2cm}{\textbf{k-FWER, \\ $\alpha = 0.05, \\ k\approx5000$}}} &  \rotatebox{90}{\parbox{2cm}{\textbf{PCS-Hub}\\ CV }}   \\
\midrule
1 month & 4656.03 (0.0428) & 1.09 (1E-5) & 450.27 (0.0042) & 2349.44 (0.0219) & 2.33 (2E-5) & 543.47 (0.0050) & 4.81 (4E-5) & 1684.44 (0.0154) & 6939.45 (0.0635) & 1547.60 (0.0142) \\  \cmidrule(r){2-11}
2 months & 2886.83 (0.0267) & 3.94 (4E-5) & 378.27 (0.0035) & 1903.01 (0.0179) & 8.42 (8E-5) & 525.01 (0.0049) & 10.06 (9E-5) & 1937.32 (0.0179) & 7543.04 (0.0697) & 1939.96 (0.0179) \\  \cmidrule(r){2-11}
3 months & 3173.4 (0.0296) & 5.78 (5E-5) & 333.05 (0.0031) & 1663.62 (0.0158) & 11.59 (0.0001) & 520.95 (0.0049) & 11.24 (0.0001) & 2173.68 (0.0202) & 8139.83 (0.0759) & 2395.35 (0.0224) \\  \cmidrule(r){2-11}
6 months & 3673.86 (0.0353) & 6.20 (6E-5) & 240.63 (0.0023) & 1268.50 (0.0124) & 11.09 (0.0001) & 387.43 (0.0038) & 8.8 (8E-5) & 2957.69 (0.0283) & 9872.93 (0.0948) & 2799.77 (0.0270) \\  \cmidrule(r){2-11}
12 months & 4570.51 (0.0468) & 1.82 (2E-5) & 134.78 (0.0014) & 898.08 (0.0093) & 2.61 (3E-5) & 217.16 (0.0023) & 70.53 (0.0007) & 7061.97 (0.0725) & 16818.92 (0.1722) & 4708.21 (0.0484) \\
\bottomrule
\end{tabular}
\caption[Mean density of estimates]{Comparison of the mean number of non-zero elements in the estimated inverse covariance matrix across all estimation horizons, as well as the percentage of non-zero elements (in brackets).  This provides a sparsity measure of the inverse covariance estimates for each method.  PARSEC provides sparser estimates than PCS-Hub in each estimation horizon and for each screening level, indicating the effectiveness of PARSEC's error control properties.  \label{tbl:MVP_estimation_density}}
\end{table}}

{\begin{table}[htbp!]
\small
\begin{tabular}{ p{1.8cm} p{1.05cm} p{1.05cm} p{1.05cm} p{1.05cm} p{1.05cm} p{1.05cm} p{1.05cm} p{1.05cm} p{1.05cm} p{1.05cm} }
\toprule
&  & \multicolumn{5}{c}{\textbf{PARSEC}} & \multicolumn{4}{c}{\textbf{PCS-Hub}}  \\ \cmidrule(r){3-7} \cmidrule(r){8-11}
$\textbf{N}_{est}$ &  \rotatebox{90}{\parbox{2cm}{\textbf{CONCORD}, \\ CV }}  &  \rotatebox{90}{\parbox{2cm}{\textbf{FWER, \\ $\alpha = 0.05$} }} & \rotatebox{90}{\parbox{2cm}{\textbf{k-FWER, \\ $\alpha = 0.01, \\ k\approx1000$} }} &  \rotatebox{90}{\parbox{2cm}{\textbf{k-FWER, \\ $\alpha = 0.05, \\ k\approx5000$} }} &  \rotatebox{90}{\parbox{2cm}{\textbf{FDR-BH, \\ $\alpha = 0.05$}}} & \rotatebox{90}{\parbox{2cm}{\textbf{PARSEC} \\ CV }} &  \rotatebox{90}{\parbox{2cm}{\textbf{FWER, \\ $\alpha = 0.05$} }} & \rotatebox{90}{\parbox{2cm}{\textbf{k-FWER, \\ $\alpha = 0.01, \\ k\approx1000$} }} &  \rotatebox{90}{\parbox{2cm}{\textbf{k-FWER, \\ $\alpha = 0.05, \\ k\approx5000$}}} &  \rotatebox{90}{\parbox{2cm}{\textbf{PCS-Hub}\\ CV }}   \\
\midrule
1 month  & 19.938  (14) & 0.005 (0) & 1.939 (1) & 10.122 (9) & 0.003 (0) & 2.333 (1) & 0.021 (0) & 7.189 (7) & 29.647 (29) & 6.614 (2) \\  \cmidrule(r){2-11}
2 months  & 12.402 (12.5) & 0.017 (0) & 1.636 (1) & 8.239 (7) & 0.018 (0) & 2.266 (1) & 0.043 (0) & 8.303 (7.75) & 32.370 (31) & 8.310 (3) \\  \cmidrule(r){2-11}
3 months & 13.676 (13)  & 0.025 (0) & 1.446 (1) & 7.237 (6) & 0.018 (0) & 2.267 (1) & 0.048 (0) & 9.358 (8) & 35.091 (33) & 10.346 (4) \\  \cmidrule(r){2-11}
6 months & 16.086 (15) & 0.0273 (0) & 1.059 (1) & 5.598 (5) & 0.030 (0) & 1.718 (0) & 0.038 (0) & 12.923 (12) & 43.176 (42) & 12.273 (5.5) \\  \cmidrule(r){2-11}
12 months & 20.634  (19) & 0.008 (0) & 0.611 (0) & 4.083 (3) & 0.007 (0) & 0.989 (0) & 0.322 (0) & 31.926 (30) & 75.930 (75) & 21.290 (5.5) \\
\bottomrule
\end{tabular}
\caption[Average and median degree]{Average degree (and median degree) of nodes, over all investment periods.  PARSEC again produces the lowest degree nodes compared to PCS-Hub and CONCORD, across all screening levels.  \label{tbl:average_median_degree}}
\end{table}}

{\begin{table}[htbp!]
\small
\begin{tabular}{ p{1.8cm} p{1.05cm} p{1.05cm} p{1.05cm} p{1.05cm} p{1.05cm} p{1.05cm} p{1.05cm} p{1.05cm} p{1.05cm} p{1.05cm} }
\toprule
&  & \multicolumn{5}{c}{\textbf{PARSEC}} & \multicolumn{4}{c}{\textbf{PCS-Hub}}  \\ \cmidrule(r){3-7} \cmidrule(r){8-11}
$\textbf{N}_{est}$ &  \rotatebox{90}{\parbox{2cm}{\textbf{CONCORD}, \\ CV }}  &  \rotatebox{90}{\parbox{2cm}{\textbf{FWER, \\ $\alpha = 0.05$} }} & \rotatebox{90}{\parbox{2cm}{\textbf{k-FWER, \\ $\alpha = 0.01, \\ k\approx1000$} }} &  \rotatebox{90}{\parbox{2cm}{\textbf{k-FWER, \\ $\alpha = 0.05, \\ k\approx5000$} }} &  \rotatebox{90}{\parbox{2cm}{\textbf{FDR-BH, \\ $\alpha = 0.05$}}} & \rotatebox{90}{\parbox{2cm}{\textbf{PARSEC} \\ CV }} &  \rotatebox{90}{\parbox{2cm}{\textbf{FWER, \\ $\alpha = 0.05$} }} & \rotatebox{90}{\parbox{2cm}{\textbf{k-FWER, \\ $\alpha = 0.01, \\ k\approx1000$} }} &  \rotatebox{90}{\parbox{2cm}{\textbf{k-FWER, \\ $\alpha = 0.05, \\ k\approx5000$}}} &  \rotatebox{90}{\parbox{2cm}{\textbf{PCS-Hub}\\ CV }}   \\
\midrule
1 month  & 1.8432  (0.0337) & 0.7433 (0.0102) & 1.2813 (0.0179) & 2.2702 (0.0239) & 0.7452 (0.0102) & 1.3079 (0.0272) & 0.7546 (0.0103) & 2.1530 (0.0311) & 3.3355 (0.0520) & 1.7864 (0.0520) \\  \cmidrule(r){2-11}
2 months  & 1.1182 (0.0189) & 0.6179 (0.0108) & 0.9844 (0.0137) & 1.9650 (0.0276) & 0.6244 (0.0109) & 1.1136 (0.0216) & 0.6330 (0.0111) & 1.8365 (0.0336) & 3.1918 (0.0608) & 1.7161 (0.0508) \\  \cmidrule(r){2-11}
3 months & 1.0598 (0.0216) & 0.5546 (0.0113) & 0.8503 (0.0136) & 1.6842 (0.0273) & 0.5621 (0.0114) & 1.0098 (0.0197) & 0.5686 (0.0115) & 1.667 (0.0340) & 3.064 (0.0600) & 1.7081 (0.0472) \\  \cmidrule(r){2-11}
6 months & 0.9105 (0.0261) & 0.4804 (0.0126) & 0.6632 (0.0132) & 1.1561 (0.0192) & 0.4859 (0.0126) & 0.7557 (0.0165) & 0.4888 (0.0127) & 1.4718 (0.0375) & 2.7501 (0.0620) & 1.4437 (0.0484) \\  \cmidrule(r){2-11}
12 months & 0.9455 (0.0281) & 0.4480 (0.0118) & 0.5443 (0.0119) & 0.8104 (0.0139) & 0.4493 (0.0118) & 0.5856 (0.0120) & 0.5198 (0.0123) & 1.9823 (0.0438) & 3.2766 (0.0687) & 1.6513 (0.0549) \\
\bottomrule
\end{tabular}
\caption[Average Turnover]{Average turnover (and standard error) over all investment periods.  PARSEC consistently produces the lowest turnover across all estimation horizons when using a FWER ($\alpha=0.05$) screening level.   \label{tbl:MVP_Turnover_corr}}
\end{table}}

\clearpage
\newpage

\printbibliography

\end{refsection}

\end{appendices}

\end{document}